\documentclass[12pt]{article}
\usepackage{mathpazo}
\usepackage{fancyhdr}
\usepackage[numbers,sort&compress,merge,round]{natbib}
\usepackage{soul}

\usepackage{longtable}
\usepackage{amsmath}
\usepackage{times}
\usepackage{rotating}
\usepackage{transparent}
\usepackage{amsthm}
\usepackage{color, colortbl}
\usepackage[dvipsnames,table]{xcolor}
\definecolor{lightgray}{rgb}{0.83, 0.83, 0.83}
\usepackage{amssymb}
\usepackage[hmargin=1in,vmargin=1in]{geometry}
\usepackage[hyphens]{url}
\usepackage[]{graphicx}
\usepackage{epstopdf}
\usepackage{blindtext}
\usepackage{array}
\usepackage{lscape} 
\usepackage{afterpage}
\usepackage{graphicx}  
\usepackage{setspace}
\usepackage{subcaption} 
\usepackage{longtable}
\usepackage{mathptmx}
\usepackage[final]{pdfpages}
\usepackage{tgtermes}
\usepackage[T1]{fontenc}
\usepackage{multicol}
\usepackage{authblk}
\usepackage{standalone}
\usepackage[dvipsnames]{xcolor}
\usepackage{booktabs,amsfonts,dcolumn}
\usepackage[hidelinks]{hyperref}

\usepackage[draft,inline,nomargin,index]{fixme}
\fxsetup{theme=colorsig,mode=multiuser,inlineface=\itshape,envface=\itshape}
\FXRegisterAuthor{rr}{arr}{Ron}
\FXRegisterAuthor{bn}{abn}{Brendan}
\FXRegisterAuthor{jr}{ajr}{Jason}
\FXRegisterAuthor{cw}{acw}{Christo}
\FXRegisterAuthor{ac}{aac}{Annie}

\DeclareUnicodeCharacter{2009}{~}

\usepackage{footmisc}

\makeatletter
\renewcommand*{\@fnsymbol}[1]{\ensuremath{\ifcase#1\or \dagger \else\@ctrerr\fi}}
\makeatother

\DeclareGraphicsExtensions{.eps,.pdf,.png}

\hypersetup{colorlinks=false}

\title{Subscriptions and external links help drive resentful users to alternative and extremist YouTube channels}

\author[1]{\large Annie Y. Chen}
\author[2]{\large Brendan Nyhan}
\author[3]{\large Jason Reifler}
\author[4,5]{\large Ronald E. Robertson}
\author[5]{\large Christo Wilson}

\affil[1]{CUNY Institute for State \& Local Governance}
\affil[2]{Dartmouth College}
\affil[3]{University of Exeter}
\affil[4]{Stanford University}
\affil[5]{Northeastern University}

\date{}  

\begin{document}
\thispagestyle{empty}
\setcounter{page}{0}

\maketitle
\thispagestyle{empty}
\begin{abstract}
\noindent Do online platforms facilitate the consumption of potentially harmful content? Using paired behavioral and survey data provided by participants recruited from a representative sample in 2020 (n=1,181), we show that exposure to alternative and extremist channel videos on YouTube is heavily concentrated among a small group of people with high prior levels of gender and racial resentment. These viewers often subscribe to these channels (prompting recommendations to their videos) and follow external links to them. In contrast, non-subscribers rarely see or follow recommendations to videos from these channels. Our findings suggest YouTube's algorithms were not sending people down ``rabbit holes'' during our observation window in 2020, possibly due to changes that the company made to its recommender system in 2019. However, the platform continues to play a key role in facilitating exposure to content from alternative and extremist channels among dedicated audiences. 

\begin{center}
\textbf{125-character teaser}
\end{center}
Exposure to extremist YouTube channels is driven by resentful users seeking out this content, not algorithmic recommendations
\end{abstract}

\newpage

\begin{doublespacing}

\section*{Introduction}

What role do technology platforms play in exposing people to dubious and hateful information and enabling its spread? Concerns have grown in recent years that online communication is exacerbating the human tendency to engage in preferential exposure to congenial information \citep{sunstein01,stroud2010polarization,gentzkow2011ideological}. Such concerns are particularly acute on social media, where people may be especially likely to view content about topics such as politics and health that is false, extremist, or otherwise potentially harmful. The use of algorithmic recommendations and platform affordances such as following and subscribing features may enable this process by helping people to find potentially harmful content and helping content creators build and monetize an audience for it. 

These concerns are particularly pronounced for YouTube, the most widely used social media platform in the U.S. \citep{pew21}. Critics highlight the popularity of extreme and harmful content such as videos by white nationalists on YouTube, which they often attribute to the recommendation system that the company itself says is responsible for 70 percent of user watch time \cite{cnet18}. Many fear that these algorithmic recommendations are an engine for radicalization. For instance, the sociologist Zeynep Tufecki wrote that the YouTube recommendation system ``may be one of the most powerful radicalizing instruments of the 21st century'' \cite{tufekci18}. These claims seem to be supported by reports that feature descriptions of recommendations to potentially harmful videos and accounts of people whose lives were upended by content they encountered online \cite{nicas18,roose19,mozilla21}.  

YouTube subsequently announced changes in 2019 to ``reduce the spread of content that comes close to---but doesn't quite cross the line of---violating our Community Guidelines'' \citep{yt2019a}. It claimed that these interventions resulted in a 50\% drop in watch time from recommendations for ``borderline content and harmful misinformation'' \citep{citeyt2019b} and a 70\% decline in watch time from non-subscribed recommendations \citep{yt2019}. However, these claims have not been independently evaluated using behavioral data, nor have the implications or caveats of ``non-subscribed recommendations'' been sufficiently explored. 

In general, questions remain about the size and composition of the audience for potentially harmful videos on YouTube following these changes, the manner in which people reach those videos, and the role of the recommendation system in that process. Studies show that sites like Twitter and Facebook can amplify tendencies toward extreme opinions or spread false information \citep{bail2018exposure,vosoughi2018spread}, though the extent of these effects and the prevalence of exposure is often overstated \citep{bakshy2015exposure,lazer2018science,guess2019less}. YouTube may operate differently, though, given its focus on video and the central role of its recommendation system \citep{munger2020right,yesilada-2022-ipr}. Browsing data has documented the existence of a sizeable audience of dedicated far-right news consumers on YouTube who often reach extremist videos via external links \citep{hosseinmardi2020evaluating}, but these data lack information on the recommendations shown to users by YouTube or the channels the users follow (a key source of recommendations). Random walk simulations conducted during and after 2019 found that problematic content was reachable, but its prevalence in recommendations fell during this period~\cite{faddoul2020longitudinal}. Research conducted after 2019 found that watching videos promoting misinformation still led to recommendations of similar videos on some topics, though their overall prevalence among recommendations was low \cite{hussein2020measuring,papadamou-2022-icwsm,ribeiro2020auditing,bisbee-2022-jots}. We build on these studies, seeking to determine the extent to which YouTube's goal of ``reduc[ing] recommendations of borderline content and harmful misinformation'' has been met using a novel measurement approach \citep{yt2019}. 

This study advances scientific understanding of the audience for potentially harmful content on YouTube and the manner in which people are exposed to it. We pair individual-level viewer histories and the associated video recommendations shown with survey data from a sample of 1,181 US respondents who were weighted to resemble the US adult population on key demographic traits. This research design allows us to examine the association between demographic and attitudinal variables, especially gender and racial resentment, and YouTube consumption behavior.
Using these data, we address three limitations of prior research in the field.
First, prior work has not taken YouTube users' channel subscriptions into account, a key indicator of user demand for specific types of content as well as a major factor in what recommendations are shown to users. We address this point by inferring the channels that our participants subscribe to and stratifying our analysis of recommendations along this axis.
Second, existing work has either relied on data from controlled experiments and random walks---which lack ecological validity---or browsing histories that lack data on video recommendations. Our dataset offers the ecological validity that comes from directly observing user behavior on YouTube, providing the first direct evidence of the extent to which real-world algorithmic recommendations push people toward potentially harmful content.
Third, prior fears about the frequency of ``rabbit holes'' are based on anecdotes and lack a precise definition. We address this problem by constructing a specific set of rules to define a ``rabbit hole'' event.
This definition builds on and reaffirms prior work \cite{ledwich2019algorithmic,faddoul2020longitudinal,ribeiro2020auditing,bisbee-2022-jots}, and we applied it to our dataset to measure the prevalence of radicalization rabbit holes among U.S. YouTube users in 2020.

Our sample of 1,181 participants is recruited from a sample of 4,000 YouGov panelists, including oversamples of two groups who we identified as especially likely to be exposed to potentially harmful video content: (1) people who previously expressed high levels of gender and/or racial resentment and (2) those who indicated they used YouTube frequently. Participants voluntarily agreed to install a custom browser extension in Chrome or Firefox that monitored their web browsing behavior. The study was conducted from July 21--December 31, 2020 (i.e., after the 2019 changes to YouTube's algorithm); respondents were enrolled in data collection for a median of 133 days. (See Methods below for further details on measurement. We provide descriptive statistics on study participants and their browser activity data availability and aggregate consumption patterns in the Supplementary Material [SM].)

We report two key findings. First, we replicate findings from \citet{hosseinmardi2020evaluating} concerning the overall size of the audience for alternative and extreme content and enhance their validity by examining participant's attitudinal variables. Though almost all participants use YouTube, videos from alternative and extremist channels are overwhelmingly watched by a small minority of participants with high levels of gender and racial resentment. Within this group, total viewership is heavily concentrated among a few individuals, a common finding among studies examining potentially harmful online content \cite{robertson2022uncommon}. Like prior work~\cite{hosseinmardi2020evaluating}, we observe that viewers often reach these videos via external links (e.g., from other social media platforms). Additionally, we find that viewers are often subscribers to the channels in question. These findings relative to existing work demonstrate the robustness of our study. They also highlight that YouTube remains a key hosting provider for alternative and extremist channels, which reinforces concerns about lax content moderation on the platform~\cite{locatelli-2022-ht} and enables these content creators to continue profiting from their audience~\cite{ballard-2022-www,factcheckers-2022-poynter}.

Second, we investigate the prevalence of ``rabbit holes'' in YouTube's recommendations during Fall 2020. We rarely observe recommendations to alternative or extremist channel videos being shown to, or followed by, non-subscribers. During our study period, only 3\% of participants who were not already subscribed to alternative or extremist channels viewed a video from one of these channels based on a recommendation. On one hand, this finding suggests that unsolicited exposure to potentially harmful content on YouTube in the post-2019 era is rare, in line with findings from prior work~\cite{ribeiro2020auditing,bisbee-2022-jots}. On the other hand, even low levels of algorithmic amplification can have damaging consequences when extrapolated over YouTube's vast user base and across time~\cite{hosseinmardi2020evaluating}. Further, it may be the case that the susceptible population was already radicalized during YouTube's pre-2019 era. Finally, given the limitations of our study, our results must be interpreted as a lower bound on ``rabbit hole'' events, which suggests that YouTube may still need to do more to remove ``borderline'' content from recommendations.

\section*{Materials and Methods}

\subsection*{Study participants}

Study participants completed a public opinion survey and installed a browser extension that recorded their browser activity ($n$=1,181). Specifically, we contracted with the survey company YouGov to conduct a public opinion survey with 4,000 respondents from three distinct populations: a nationally representative sample of 2,000 respondents who previously took part in the 2018 Cooperative Congressional Election Survey (CCES) when it was fielded by YouGov; an oversample of 1,000 respondents who expressed high levels of racial resentment \cite{kinder1996divided}, hostile sexism \cite{glick1997hostile}, and denial of institutional racism \cite{desante2020less} in their responses to the 2018 CCES; and an oversample of 1,000 respondents who did not take part in the 2018 CCES but indicated that they use YouTube ``several times per day'' or ``almost constantly'' in their survey response. (The prior measures of racial resentment and hostile sexism, which were collected as part of the 2018 CCES for 3,000 of our 4,000 respondents, are also used as independent variables in our analysis; see below for details on question wording.)
    
While completing the survey, participants who used an eligible browser (Chrome or Firefox) were offered the opportunity to download a browser extension that would record their browser activity in exchange for additional compensation. A total of 1,181 respondents did so (778 from the nationally representative sample, 97 from the high resentment oversample, and 306 from the high YouTube user oversample). 

All analyses we report below use survey weights created by YouGov to account for the fact that, in addition to a national sample, we have also specifically recruited participants who fall into one of two oversample groups: (1) those who previously expressed gender and/or racial resentment, or (2) those who are frequent YouTube users. When we apply these weights to all three samples, the total sample is weighted to be nationally representative. Applying these weights to the subset of participants who installed the browser extension helps us to best approximate the characteristics of a nationally representative sample, though the sample is of course not fully representative of the US adult population. We therefore report weighted estimates of the number of users or cases of a behavior as well as weighted percentages or proportions for maximum clarity. Additional details about respondent demographics and other characteristics are provided in the SM.

\subsection*{Ethics and privacy}

Our study methods were approved by the Institutional Review Boards (IRBs) at the authors' respective institutions (Dartmouth CPHS STUDY00032001, Northeastern IRB \#20-03-04, and University of Exeter Social Sciences and International Studies Ethics Committee \#201920-111).

All participants were asked to consent to data collection before completing our survey and again when they installed our browser extension. Participants were fully informed about the data collected by our extension when they were invited to install it and again during installation of the extension. The extension did not collect any data until consent was provided and participants were free to opt out at any time by uninstalling our extension. The extension automatically uninstalled itself from participants' browsers at the end of the study period. (See the SM for the full text of our informed consent notices.)

To protect participants' security and privacy, we adopted a number of best practices. Our participants are indexed by pseudonymous identifiers. Our browser extension used TLS to encrypt collected data while it was in transit. All participant data is stored on servers that are physically secured by key cards. We use standard remote access tools like SSH to access participant data securely. 

We have posted data and code on Dataverse that allows for the replication of all results in this article (linked in the "Data availability" section). All analysis code has also been posted. However, raw behavior data cannot be posted publicly to protect the privacy of respondents.

\subsection*{Data collection and measurement}

The browser extension passively logged user page views, including the full URL and a timestamp, and collected HTML snapshots when users viewed YouTube videos, allowing us to examine the video recommendations that participants received. This combination of passive monitoring and HTML snapshots provides us with the ability to measure not just what respondents watched but also what YouTube showed them prior to that action.
To account for duplicate data, we dropped additional page views of the same URL within one second of the prior page view on the assumption that the user refreshed the page \citep{guess2021almost}. 

Our data collection approach focuses on browser activity data, which provides important advantages relative to the history data that is provided by the web browser's WebExtension API. The browser APIs report the time when a given web page was first opened and the time when a user makes a transition from that page to another page (e.g., by clicking a link). However, the APIs do not report the total dwell time on a given web page taking into account changes in the active browser tab. For example, if someone opens web page $A$ in a tab, then opens web page $B$ in another tab, and then switches their browser tab back to $A$, the browser history APIs will not register this shift in attention, making it difficult to obtain accurate estimates of time spent on a given web page. Our passive monitoring records all changes in the active tab, allowing us to overcome this issue. (In the SM, we validate our browser activity data against browser history data from the extension.)

In this article, we describe YouTube ``views,'' ``consumption,'' and ``exposure'' using the browser activity data described above. As with any passive behavioral data, we cannot verify that every user saw the content that appeared on their device in every instance. 

We measured the amount of time a user spent on a given web page by calculating the difference between the timestamp of the page in question and the next one they viewed. This measure is imperfect because we do not have a measure of eye gaze or a proxy for active viewing. Though some participants might rewind and rewatch videos more than once, we are more concerned about our measure overstating watch time due to users leaving their browser idling. We therefore refine this measure by capping our measure of time spent at the length of the video in question (obtained from the YouTube API). 

We measure which channels users subscribed to by extracting additional information from the HTML snapshots of the videos they watched. Specifically, we parsed the subscribe button from each HTML snapshot, which reads ``Subscribe'' when the participant was not subscribed to the video channel at the time the video was watched and ``Subscribed'' when they were already subscribed. Because we must use this indirect method to infer channel subscriptions, we do not know the full set of channels to which participants subscribe. In particular, not all recommended videos in our dataset were viewed by participants. As a result, we could not determine the subscription status for all recommended videos.

We denote the web page that a participant viewed immediately prior to viewing a YouTube video as the ``referrer.'' We are unable to measure HTTP \texttt{Referrer} headers using our browser extension, so instead we rely on browser activity data to identify referrers to YouTube videos. Using prior browsing history is a common proxy used to analyze people's behavior on the web \citep{guess2020exposure,wojcieszak2021avenues}.

All analyses of the percentage of recommendations seen or followed are based on the full set of recommendations that we could extract from each video. The mean number of recommended videos captured was 17.9 and the median was 20, which aligns with the default number of recommendations shown on a YouTube video (20) at the time our study was conducted.

\subsection*{Channel definitions and measurement}

Following studies of information consumption online that rely on ratings of content quality at the domain level \citep{grinberg2019fake,guess2020exposure}, we construct a typology of YouTube channel types to measure participant exposure. Given that YouTube has tens of millions of channels and that the types of content we are interested in a relatively rare, it is necessary to rely on the judgement of experts to help us identify alternative, extremist, and mainstream media channels. We use the resulting channel lists to classify all videos to which our participants are exposed as coming from an alternative channel, an extremist channel, a mainstream media channel, or some other type of channel (``other''). The process by which these channel lists are described further below; the SM provides more detail on the procedures used by these experts to label channels. 

In our typology, alternative channels discuss controversial topics through a lens that attempts to legitimize discredited views by casting them as marginalized viewpoints (despite the channel owners often identifying as White and/or male). Our list combines the 223 channels classified by Ledwich and Zaitsev \cite{ledwich2019algorithmic} as Men's Rights Activists or Anti-Social Justice Warriors, the 141 Intellectual Dark Web and Alt-lite channels from Ribeiro et al. \cite{ribeiro2020auditing}, and the 24 channels from Lewis' Alternative Influence Network \cite{lewis18}. After removing duplicates, our alternative channel list contains 322 channels, of which 68 appeared on two source lists, and nine appeared on three. Example alternative channels in our typology include those hosted by Steven Crowder, Tim Pool, Laura Loomer, and Candace Owens. Joe Rogan's is the most prominent alternative channel in our typology (it appears on all three source lists), accounting for 11.8\% of all visits and 21.8\% of all time spent on alternative channel videos.

Our list of extremist channels consists of those labelled as white identitarian by Ledwich and Zaitsev (30 channels) \cite{ledwich2019algorithmic}, white supremacist by Charles (23 channels) \cite{charles2020main}, alt-right by Ribeiro et al. (37 channels) \cite{ribeiro2020auditing}, extremist or hateful by the Center on Extremism at the Anti-Defamation League (16 channels), and those compiled by journalist Aaron Sankin from lists curated by the Southern Poverty Law Center, the Canadian Anti-Hate Network, the Counter Extremism Project, and the white supremacist website Stormfront (157 channels) \cite{sankin-2019-gizmodo}. After removing duplicates, our extremist channel list contains 290 channels, of which 36.2\% appeared on two or more source lists.
Example extremist channels include those hosted by Stefan Molyneux, David Duke, Mike Cernovich, and Faith J. Goldy. 

As the examples above suggest, the potentially harmful alternative and extremist channels identified by scholarly and subject matter experts are predominantly from the (far) right in the U.S. Other forms of extremism exist, of course, especially outside the U.S. (e.g., Islamic extremism).

Following prior research, we define both alternative and extremist channels as potentially harmful \citep{ledwich2019algorithmic,ribeiro2020auditing,lewis18,charles2020main}. Of the 302 alternative and 213 extremist channels that were still available on YouTube as of January 2021 (i.e., they had not been taken down by the owner or by YouTube), videos from 208 alternative and 55 extremist channels were viewed by at least one participant in our sample. We are not making these lists publicly available to avoid directing attention to them but are willing to privately share them with researchers and journalists upon request.

To create our list of mainstream media channels, we collected news channels from Buntain et al. \cite{buntain2021youtube} (65 mainstream news sources), Ledwich et al. \cite{ledwich2019algorithmic} (75 mainstream media channels), Stocking et al. \cite{pew20} (81 news channels), Ribeiro et al. \cite{ribeiro2020auditing} (68 popular media channels), Eady et al. \cite{eady2020} (219 national news domains), and Zannettou et al. \cite{zannettou_web_2017} (45 news domains). We manually found the corresponding YouTube channels via YouTube search when authors only provided websites \cite{ribeiro2020auditing,eady2020,zannettou2019}. In cases where news organizations have multiple YouTube channels (e.g., Fox News and Fox Business), all YouTube channels under the parent organization were included. Any channels appearing in fewer than three of these sources were omitted. Finally, we also included channels that were featured on YouTube's \href{https://www.youtube.com/channel/UCYfdidRxbB8Qhf0Nx7ioOYw}{News} hub from February 10--March 5, 2021.

The resulting list of mainstream media channels was then checked to identify those that meet all of the following criteria:
\begin{enumerate}
    \item They must publish credible information, which we define as having a NewsGuard score greater than 60 (\url{https://www.newsguardtech.com}) and not being associated with any ``black'' or ``red'' fake news websites listed in \citet{grinberg2019fake}.
    \item They must meet at least one criteria for mainstream media recognition or distribution, which we define as having national print circulation, having a cable TV network, being part of the White House press pool, or having won or been nominated for a prestigious journalism award (e.g., Pulitzer Prize, Peabody Award, Emmy, George Polk Award, or Online Journalism Award).
    \item They must be a US-based organization with national news coverage. 
\end{enumerate} 

Our final mainstream media list consists of 127 YouTube channels. 

We then placed all YouTube channels in our dataset that did not fall into one of these three categories (alternative, extremist, or mainstream media) into a residual category that we call ``other.'' (These may include alternative, extremist, or mainstream media that were missed by the processes described above.)

\subsection*{Survey measures of racial resentment and hostile sexism}

We measure anti-Black animus with a standard four-item scale intended to measure racial resentment \citep{kinder1996divided}. For example, respondents were asked whether they agree or disagree with the statement ``It's really a matter of some people just not trying hard enough: if blacks would only try harder they could be just as well off as whites.'' Responses are provided on a five-point agree/disagree scale and coded such that higher numbers represent more resentful attitudes. Respondents' racial resentment score is the average of these four questions. Responses to these questions are taken from respondent answers to the 2018 Cooperative Congressional Election Survey (as noted above, participants were largely recruited from the pool of previous CCES respondents). %

We operationalized hostile sexism using two items from a larger scale that was also asked on the 2018 Cooperative Congressional Election Survey (CCES) \cite{glick1997hostile}. For example, one of the questions asks if respondents agree or disagree with the statement ``When women lose to men in a fair competition, they typically complain about being discriminated against.'' Responses are provided on a five-point agree/disagree scale and coded such that higher numbers represent more hostile attitudes. 

All other question wording is provided in the survey codebook in the SM. Racial resentment and hostile sexism measures were also included in our 2020 survey; responses showed a high degree of persistence over time ($r=.92$ for racial resentment, $r=.79$ for hostile sexism). The two measures, which we refer to as measuring ``resentment'' or identifying ``resentful'' users per, e.g., \citet{banda2022hostile} and \citet{schaffner2021optimizing}, were highly correlated with each other as well ($r=.84$). 

\section*{Results}

\subsection*{Exposure levels}

Though 91\% of participants visited YouTube, the vast majority of participants did not view any alternative or extremist channel videos. Just 15\% of the sample for whom we have browser activity data ($n$=1,181) viewed any video from an alternative channel and only 6\% viewed any video from an extremist channel. By comparison, 44\% viewed at least one video from a mainstream media channel. (See Methods for how channel types were defined and how view history and watch time were defined.) Videos from mainstream media channels account for 3.6\% of videos watched in our sample---a figure that falls between recent estimates that 2.9--11\% of videos watched on YouTube are news \cite{yang2021online,hosseinmardi2020evaluating}. The corresponding numbers for videos from alternative and extremist channels are 3.0\% and 0.5\%, respectively (similar to estimates from 2019~\cite{hosseinmardi2020evaluating}).

The audience for alternative and extremist channels is skewed toward people who subscribe to the channel in question or one like it, which we determine by inspecting whether the subscription button is activated when a participant views a video from that channel (see Methods for more details). Among the set of people who saw at least one extremist channel video during the study period, for instance, 52\% watched a video from an extremist channel to which they subscribed. Similarly, 39\% of alternative channel viewers watched at least one video from an alternative channel to which they subscribed. 

\begin{figure}[!tb]
    \begin{center}
    \caption{Distribution of video views by subscription status and channel type}
    \label{fig:subscriptions}
    \includegraphics[width = .99\textwidth]{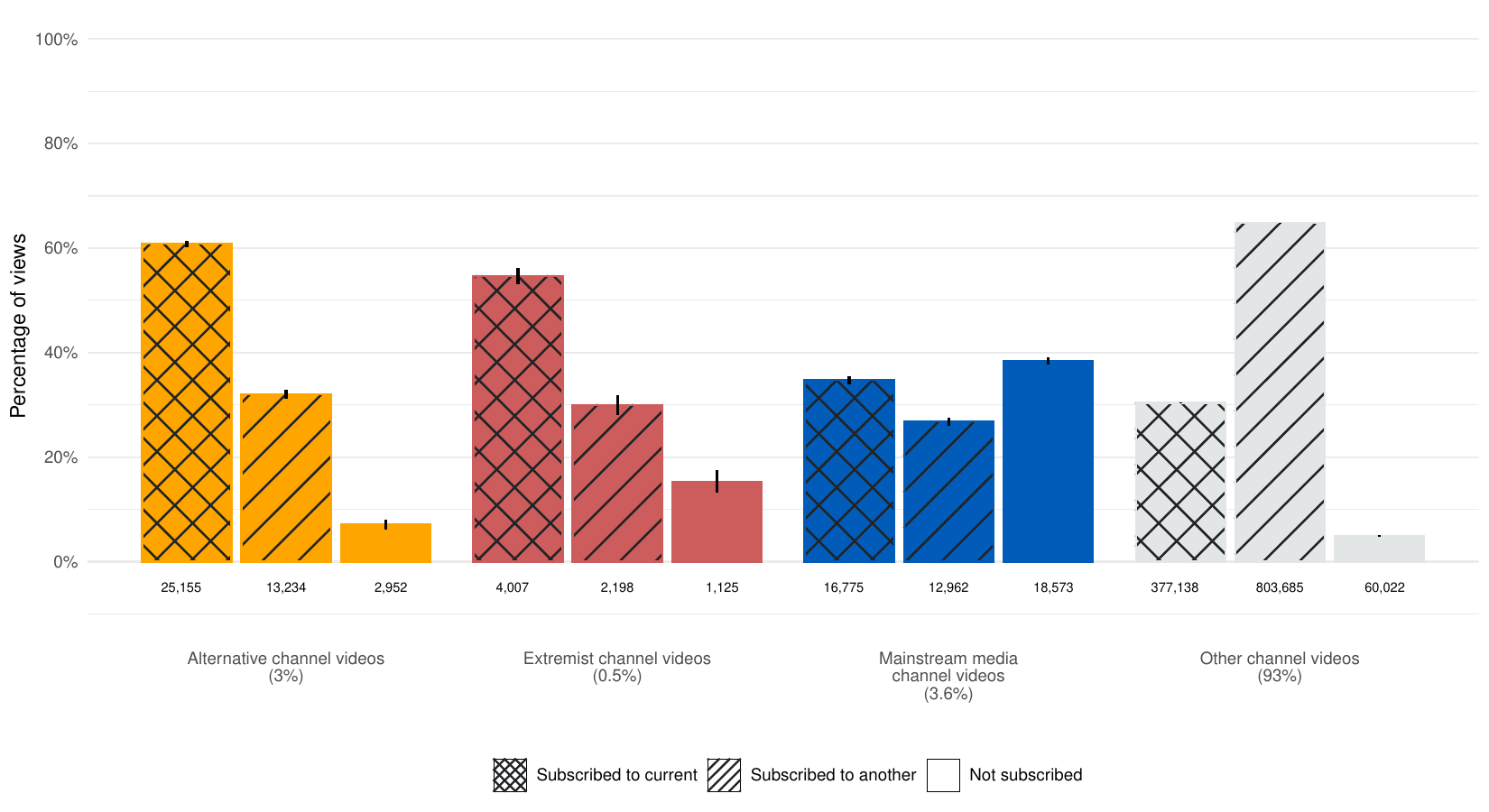}
    \end{center}
    \vskip -10pt
    \begin{footnotesize}
    \noindent Weighted percentages of views for videos from each type of channel that come from people who are subscribed to that channel (crosshatches), who subscribe to one or more different channels of the same type but not the channel currently being viewed (hatches), and who do not subscribe to any channel of that type (no hatches). Each estimate includes the corresponding 95\% confidence interval. Total view counts are displayed at bottom of each bar. Total views for videos of that type as a percentage of all views are displayed under the channel labels. 
    \end{footnotesize}
\end{figure}

\autoref{fig:subscriptions} illustrates this point in a different way by disaggregating video views according to both channel type and subscription status. We observe that 60.8\% of views for videos from alternative channels and 54.7\% of views for videos from extremist channels come from subscribers to the channel in question. If we instead define subscribers to include all people who subscribe to at least one channel of the type in question, the proportion of views from subscribers increases to 92.9\% for alternative channels and 84.7\% for extremist channels. These patterns for alternative and extremist channels are distinct from mainstream media channels, which receive 38\% of their views from people who do not subscribe to any channel in the category. 

Among the participants who viewed at least one video from an alternative or extremist channel, the time spent watching them was relatively low (and concentrated among subscribers): an overall mean of 26 minutes per week for alternative channel videos (62 minutes per week for subscribers to one or more alternative channels [6\%] versus 0.2 minutes per week for non-subscribers [9\%]) and 8 minutes for extremist channel videos (15 minutes per week for subscribers [3\%] versus 0.04 minutes per week for non-subscribers [3\%]). The comparison statistics are 12 minutes per week for mainstream media channel videos and 214 minutes per week for videos from other channels. As noted above, however, these data are highly skewed: the median time spent watching among participants who viewed at least one video from an alternative or extremist channel was 1.1 minutes for alternative channel videos and 0.6 minutes for extremist channel videos. That said, these results mirror those from \citet{hosseinmardi2020evaluating}, who observed the same partial ordering, in terms of video watch time, for anti-woke (i.e., alternative), far right (i.e., extreme), and mainstream news sources, from most to least watched.

\begin{figure}[!tb]
 \begin{center}
    \caption{Concentration of exposure to alternative and extremist channels}
    \label{fig:concentration_time}
    \includegraphics[width = .99\textwidth]{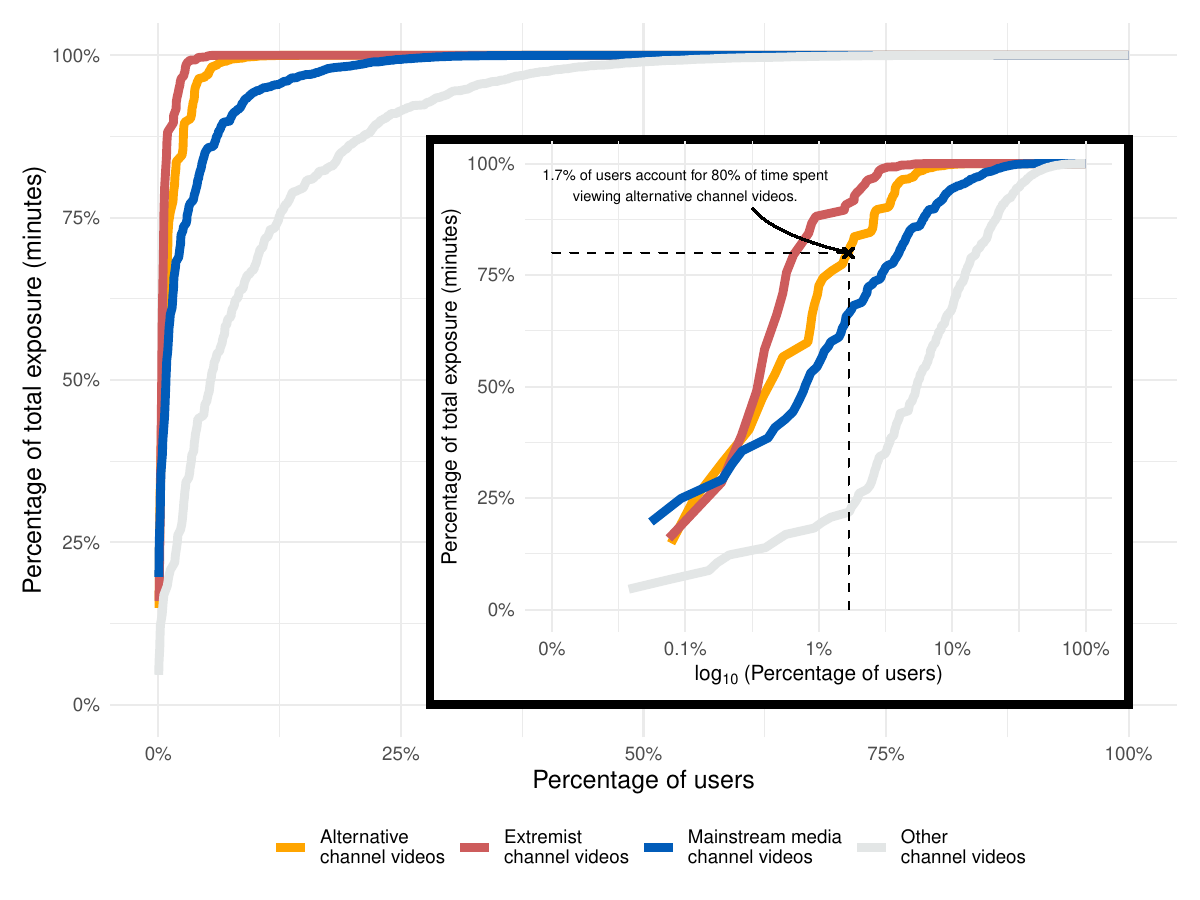}
   \end{center}
    \begin{footnotesize}
    \noindent Weighted empirical cumulative distribution function (eCDF) showing the percentage of participants responsible for a given level of total observed video viewership of alternative and extremist channels on YouTube (in minutes). Inset graph shows the same data using a log scale for the weighted eCDF.
    \end{footnotesize}    
\end{figure}

Viewership of potentially harmful videos on YouTube is heavily concentrated among a few participants, mirroring patterns observed on YouTube over the 2016--2019 time frame~\cite{hosseinmardi2020evaluating}, Twitter and untrustworthy websites \cite{grinberg2019fake,guess2020exposure}, and news content generally \cite{allen2020evaluating,yang2021online}. As \autoref{fig:concentration_time} indicates, 1.7\% of participants account for 80\% of total time spent on videos from alternative channels. This imbalance is even more severe for extremist channels, where 0.6\% of participants were responsible for 80\% of total time spent on these videos. Skew is similar when we examine view counts (\autoref{fig:concentration_visit}) rather than time spent on videos---1.9\% and 1.1\% of participants were responsible for 80\% of alternative and extremist channel viewership, respectively. We observe a similar pattern of concentration for mainstream media consumption---just 3.8\% of participants account for 80\% of the total views. (We provide a more detailed analysis of the viewership patterns of these ``superconsumers'' in the SM.)

\subsection*{Correlates of exposure}

We next evaluate demographic and attitudinal factors that are potentially correlated with time spent watching videos from alternative, extremist, and mainstream media channels. We focus specifically on hostile sexism, racial resentment, and negative feelings toward Jews---three factors that may make people vulnerable to the types of messages offered by alternative and extremist channels, which often target women, racial and ethnic minorities, and Jews \cite{lewis18,zannettou2019}. Negative attitudes towards these out groups may make people vulnerable to the types of messages offered by alternative and extremist channels. We therefore estimate the statistical models reported below on the subset of 851 respondents for whom prior scale measures of hostile sexism and racial resentment are available from the 2018 Cooperative Congressional Election Study. (Details on survey wording and measurement, including the wording for these scales, are provided in Methods below; feelings toward Jews are measured using a feeling thermometer.)

\begin{figure}[!tb]
 \begin{center}
    \caption{Predictors of video watch time}
    \label{fig:qpois_coefficients_time}
    \includegraphics[width = .99\textwidth]{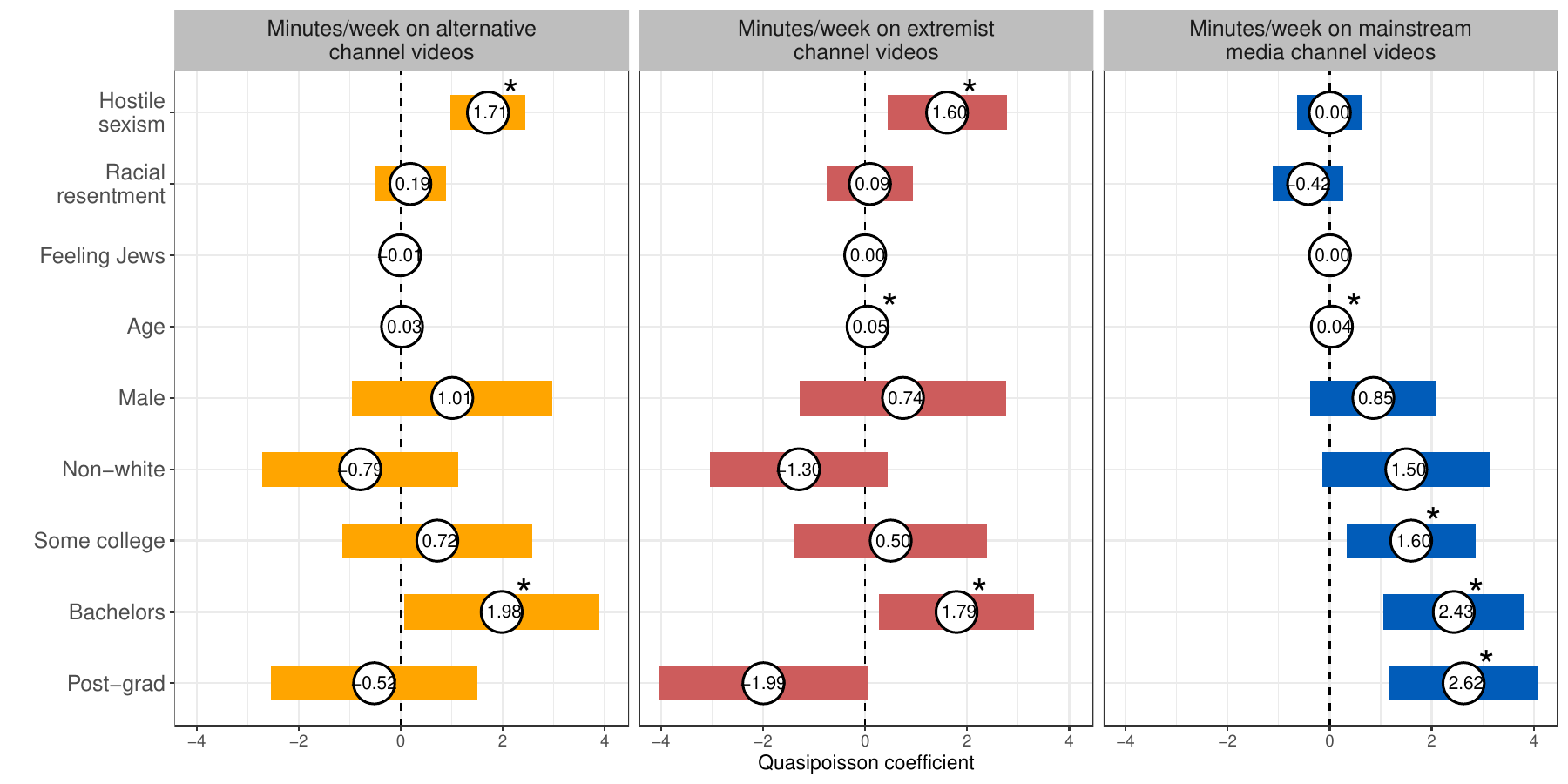}
   \end{center}
    \begin{footnotesize}
    \noindent Quasipoisson regression coefficients for correlates of the amount of time respondents spent on videos from alternative, extremist, and mainstream media channels in minutes per week. Figure includes 95\% confidence intervals calculated from robust standard errors. All results incorporate survey weights. Stars indicate coefficients that are significant at the $p<.05$ level. See \autoref{tab:fig4output} for regression table.
    \end{footnotesize}   
\end{figure}

We estimate models measuring the association between the average time per week that respondents spent on videos from alternative, extremist, or mainstream media channels and the measures listed above as well as relevant demographic characteristics: age, sex (male\slash not male), race (white\slash non-white), and indicators for different levels of education above high school (some college\slash bachelor's\slash post-grad). Results of the quasipoisson models we estimate, which account for the skew in video watch time, are shown in \autoref{fig:qpois_coefficients_time}. (See \autoref{fig:qpois_coefficients_visits} for equivalent results for the number of views of videos from alternative and extremist channels.) 

\begin{figure}[!tb]
 \begin{center}
    \caption{Hostile sexism as predictor of alternative and extremist channel viewing}
    \label{fig:qpois_predictions}
    \includegraphics[width = .99\textwidth]{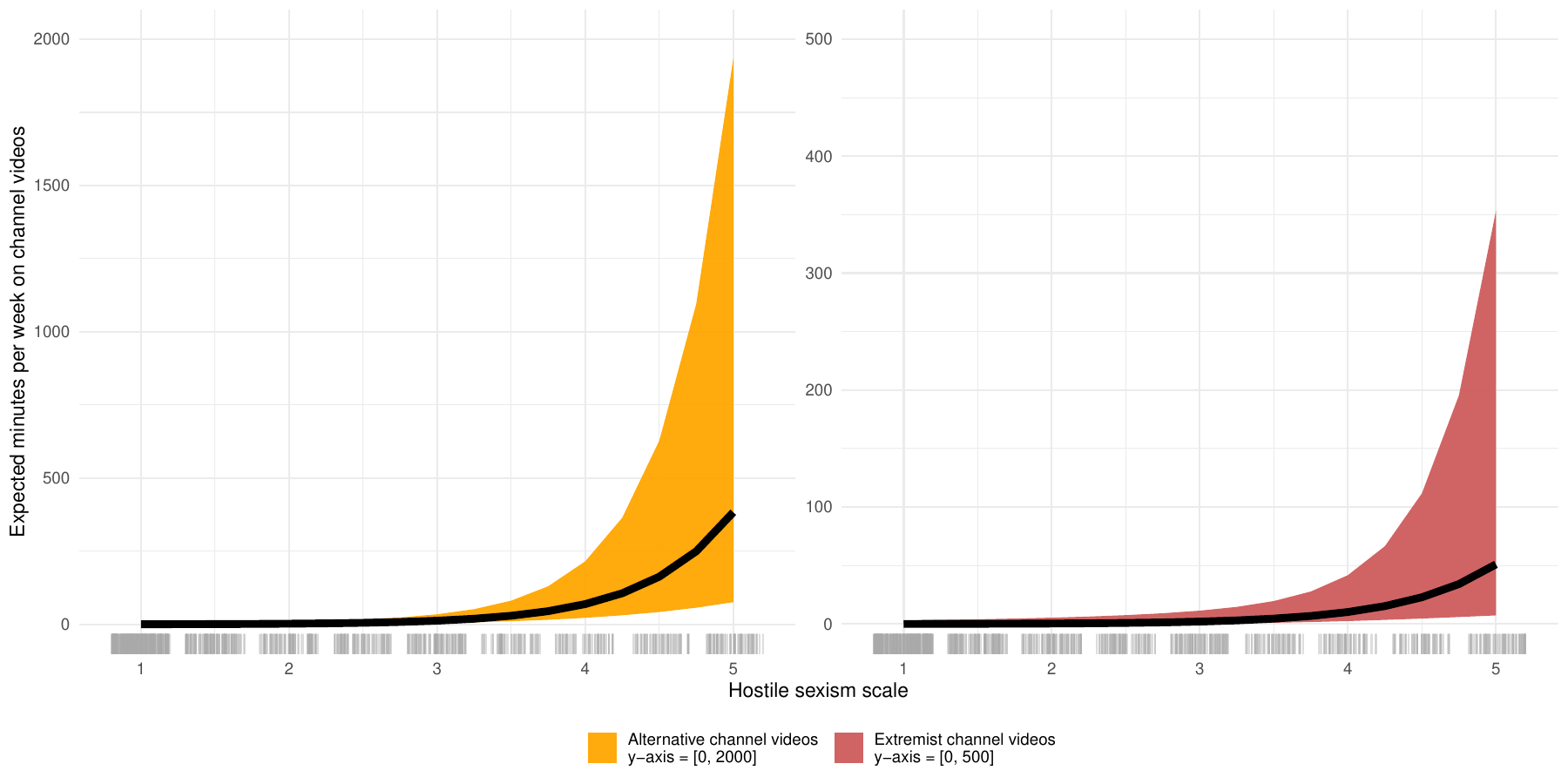}
   \end{center}
    \begin{footnotesize}
    \noindent Predictions are estimated from the models in Figure \ref{fig:qpois_coefficients_time} holding other covariates at their median (continuous variables) and modal (categorical variables) values. Colored bands represent 95\% robust confidence intervals. All results incorporate survey weights.
    \end{footnotesize}   
\end{figure}

The results indicate that prior levels of hostile sexism are significantly associated with time spent on videos from alternative channels and time spent on videos from extremist channels but not time spent watching mainstream media channels. This relationship, which is consistent with the commenter overlap observed between men's rights\slash anti-feminist channels and alt-right channels on YouTube \citep{mamie2021anti}, is not observed for prior levels of racial resentment when controlling for hostile sexism. However, both hostile sexism and racial resentment are positively associated with time spent on videos and number of views of videos from alternative and extremist channels when entered into statistical models separately (see Tables \ref{tab:hh_rr_time} and \ref{tab:hh_rr_visit}). Finally, we find no association between feelings toward Jews and viewership of any of these types of channels.  

\autoref{fig:qpois_predictions} illustrates the relationship between prior levels of hostile sexism and time spent per week watching videos from alternative or extremist channels using the model results described above. When hostile sexism is at its minimum value of 1, expected levels are 0.4 minutes per week spent watching alternative channel videos and 0.08 minutes for extremist channel videos. These predicted values increase to 383 and 51 minutes, respectively, when hostile sexism is at its maximum value of 5 (with the greatest marginal increases as hostile sexism reaches its highest levels).

\subsection*{Recommendations and YouTube ``rabbit holes''}

Critics of YouTube have emphasized the role of its algorithmic recommendations in leading people to potentially harmful content. We therefore measure which types of videos YouTube recommended to participants and how often those recommendations were followed. Next, we specifically count how often people follow recommendations to more extreme channels to which they don't subscribe in a manner that is consistent with the ``rabbit hole'' narrative. Finally, we disaggregate YouTube recommendations and following behavior based on subscription status. In general, we find that recommendations to alternative and extremist channel videos are rare and frequently shown to and followed by people who already subscribe to those channels.

\begin{figure}[!tb]
    \begin{center}
    \caption{Recommendation frequency by type of channel being watched \label{fig:waffles_recs_shown}}
    \includegraphics[width = .99\textwidth]{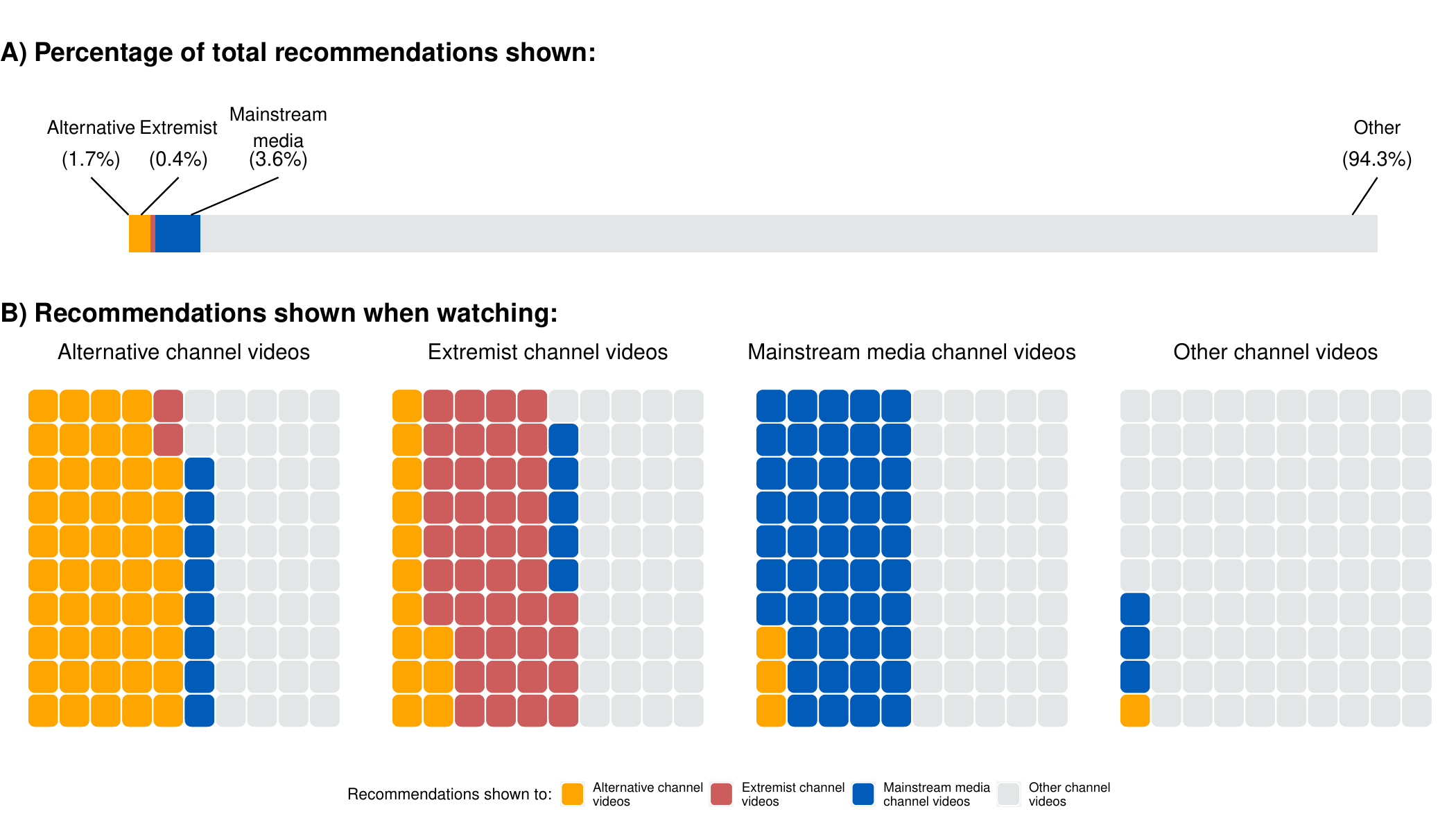}
    \end{center}
   \begin{footnotesize}
    Number of colored tiles shown are proportional to the proportion of recommendations shown for each type of video when watching videos from alternative, extremist, mainstream media, or other channels. Results are based on the full set of recommendations that we could extract from each video and incorporate survey weights. 
    \end{footnotesize}
\end{figure}

We disaggregate the recommendations shown to participants by the type of video on which the recommendation appears, which appears to play a large role in determining what YouTube recommends. As Panel A of \autoref{fig:waffles_recs_shown} shows, there are relatively few recommendations to alternative and extremist videos. As Panel B shows, recommendations to alternative and extremist channel videos are very rare when watching videos from mainstream media or other types of channels, which together make up 97\% of views in our sample. Recommendations to alternative and extremist channel videos are much more common, however, when people are already viewing videos from alternative and extremist channels, which make up 3\% and 0.5\% of views, respectively. Just under half (47.9\%) of recommendations when viewing an alternative channel video point to another alternative channel video, while 41.1\% of recommendations follow the same pattern for extremist channel videos. Substantively similar patterns of recommendations have been observed in random walk studies on YouTube~\cite{ledwich2019algorithmic,ribeiro2020auditing}.

\autoref{fig:waffles_recs_followed} in the SM provides corresponding statistics for the proportion of recommendations followed by channel type. As expected, the people who are already watching alternative and extremist channel videos are especially likely to follow recommendations to other alternative or extremist channel videos. Among people who were watching alternative channel videos, 53.7\% of recommendations followed were to alternative or extremist channel videos (compared to 50.3\% of recommendations shown). Correspondingly, 73.8\% of recommendations followed from extremist channel videos were to other extremist or alternative channel videos (versus 54.3\% of recommendations shown). The probability of following a recommendation to such a video by people not already watching an alternative or extremist channel video was negligible. (We disaggregate recommendations and follows by recommendation rank in Figures \ref{fig:recsrank}--\ref{fig:followsrank}.)

Next, we more directly test how often YouTube video recommendations create ``rabbit holes'' in which people are shown more extreme content than they would otherwise encounter. Specifically, we define four conditions that must be met to constitute a ``rabbit hole'' and report the number of views, sessions, and users that meet these criteria when sequentially applied: 

\begin{enumerate}
    \item A participant followed a recommendation to an alternative or extremist channel video: 0.17\% of all video visits among 7.3\% of participants;
    \item The recommendation that the participant followed moved them to a more extreme channel type (i.e., \{mainstream media, other\} $\rightarrow$ \{alternative\} or \{mainstream media, other, alternative\} $\rightarrow$ \{extreme\}):
    0.07\% of all video visits among 5.4\% of participants;
    \item The participant does not subscribe to the channel of the recommended video:
    0.02\% of all video visits among 4.7\% of participants;
    \item The participant does not subscribe to any channels of the same type (i.e., alternative or extremist) as the recommended video: 0.01\% of all video visits among only 3.0\% of participants.
\end{enumerate}

Based on these strict criteria, we observe very few cases of ``rabbit hole'' events. As noted above, the set of events that meet all four criteria for alternative and extremist channel videos represent only 0.01\% of all video visits and were observed among just 3.0\% of participants. The set of such sequences that specifically ended in exposure to an extremist channel video represented just 0.002\% of all visits and were only observed among 1.0\% of participants. (We provide qualitative accounts of three such sequences in the SM as well as an analysis showing no trend toward greater exposure to alternative or extremist channel videos in longer YouTube sessions.) 

\begin{figure}[!tb]
    \begin{center}
    \caption{YouTube recommendations by subscription status and channel type     \label{fig:recs-stratified}}
    \includegraphics[width = .99\textwidth]{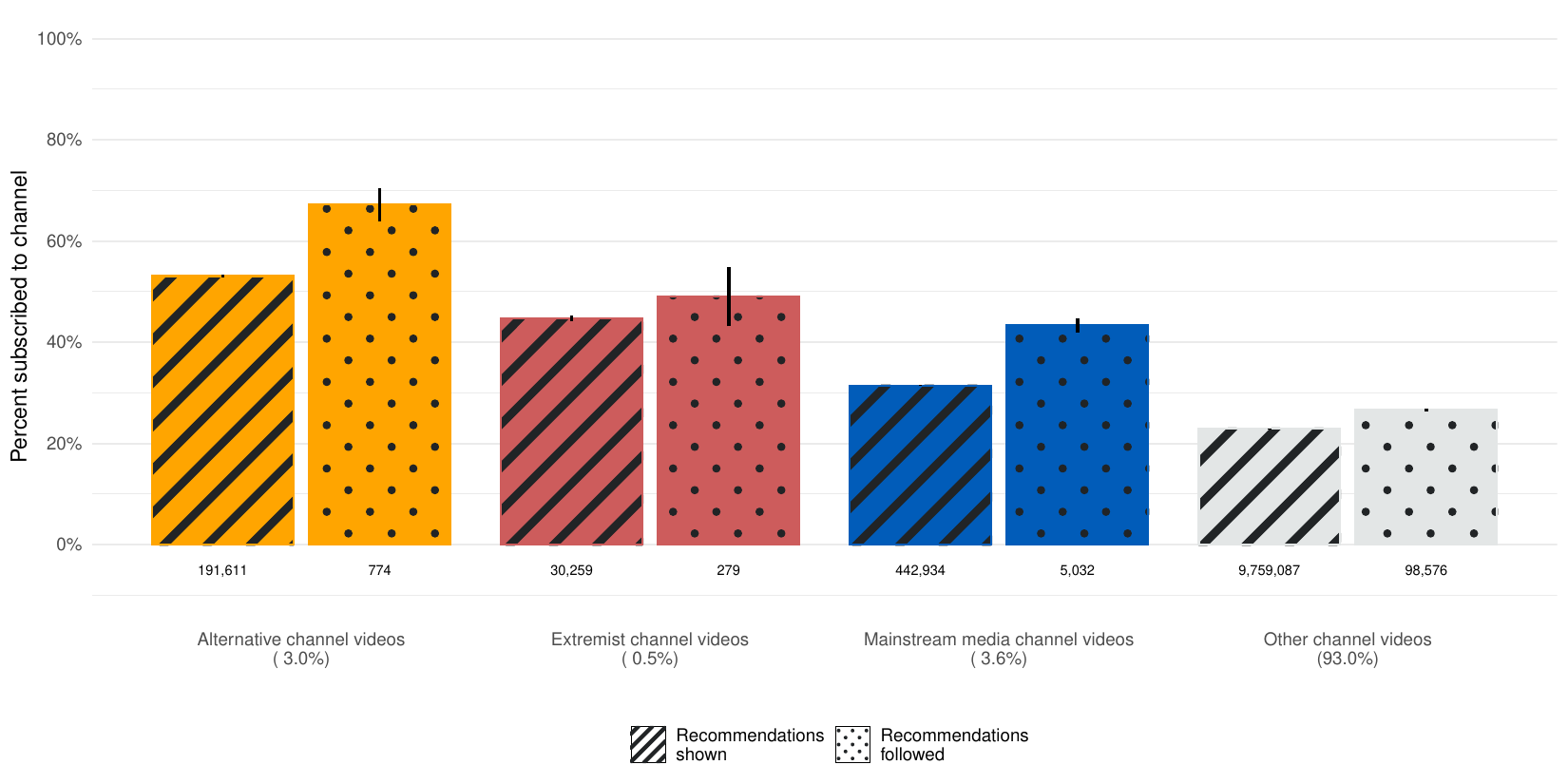}
    \end{center}
    The weighted percentage of recommendations shown and followed to people who subscribe to one or more channels of each type (including 95\% confidence intervals for both, though these are sometimes not visible due to the sample size of the recommendations shown data). The weighted percentage of views of each type of video are shown in parentheses under the labels.  
\end{figure}

We observe that recommendations to videos from alternative and extremist channels are frequently shown to channel subscribers---the same group that is most likely to follow those recommendations. As \autoref{fig:recs-stratified} demonstrates, people who subscribe to at least one alternative channel received 53.1\% of all alternative channel video recommendations and represented 67.2\% of the cases in which a participant followed a recommendation to an alternative channel video. This skew was somewhat smaller for extremist channel videos---subscribers to one or more extremist channels saw 44.7\% of recommendations to videos from extremist channels and made up 49.0\% of the cases in which respondents followed a recommendation to watch such a video. These figures are generally larger than those observed for mainstream media channels or other types of channels.

\subsection*{Internal and external referrers}

Finally, we replicate and expand an analysis conducted by \citet{hosseinmardi2020evaluating} that measures the process by which people come to watch alternative and extremist videos on YouTube. As in prior work, we denote the page that people viewed immediately prior to a video being opened (within an existing browser tab or within a new tab) as the ``referrer'' and distinguish between ``on-platform'' referrers (a YouTube channel page, the YouTube homepage, a YouTube search page, or another YouTube video) and ``off-platform'' referrers that are not part of the YouTube domain such as search engines, webmail sites, mainstream social media sites (e.g., Facebook, Twitter, Reddit), or alternative social media sites (e.g., Parler, Gab, 4chan). The complete list of external referrers in each category can be found in \autoref{tab:externalrefs}. Details on how we identify referrers are provided in Methods below.

We find that off-platform referrers are responsible for approximately half of all views of alternative and extremist channel videos, a finding that is roughly consistent with YouTube's statement that ``borderline content gets most of its views from other platforms that link to YouTube'' \citep{yt21}. Our finding is slightly higher than the 36--41\% external referrers for alternative and extreme videos observed by \citet{hosseinmardi2020evaluating}, but we include referrals from non-YouTube search engines in our total while \citet{hosseinmardi2020evaluating} do not. That said, as we show in \autoref{fig:on_platform_referrers_by_channel}, 52.4\% and 46.6\% of referrals to alternative and extremist channel videos, respectively, were off-platform sources, which is only somewhat higher than off-platform referrals for videos from mainstream media (41.7\%) or other channels (41.1\%).

With respect to on-platform referrers, we observe frequent within-category referrals by video type, with 19.6\% of referrals to alternative channel videos coming from other alternative channel videos, 21.3\% of referrals to extremist channel videos coming from other extremist channel videos, and 25.6\% of referrals to mainstream media channel videos coming from other mainstream media channel videos. This is broadly consistent with results from random walk studies on YouTube that have examined recommendations between different types of videos~\cite{ledwich2019algorithmic,ribeiro2020auditing}. Interestingly, we observe 3.8\% of referrals to extremist channel videos coming from alternative channel videos, but only 0.8\% of referrals to alternative channel videos coming from extremist channel videos, which suggests that it is rare for our participants to move from more to less extreme content in this manner. Lastly, we observe that alternative, extremist, and mainstream media channel videos all receive roughly equal referrals from videos in other channels (10.0--12.8\%) and other on-platform sources (15.1--19.1\%). Overall, these results also broadly similar to those Hosseinmardi et al. \cite{hosseinmardi2020evaluating}, who found that 36--39\% of referrals to alternative and extreme videos came from other videos, while 21--23\% of referrals came from other on-platform sources.

\begin{figure}[!tb]
    \begin{center}
    \caption{Relative frequency of referrals to YouTube videos by channel and referrer type     \label{fig:internal_external referrer}}
    \includegraphics[width = .99\textwidth]{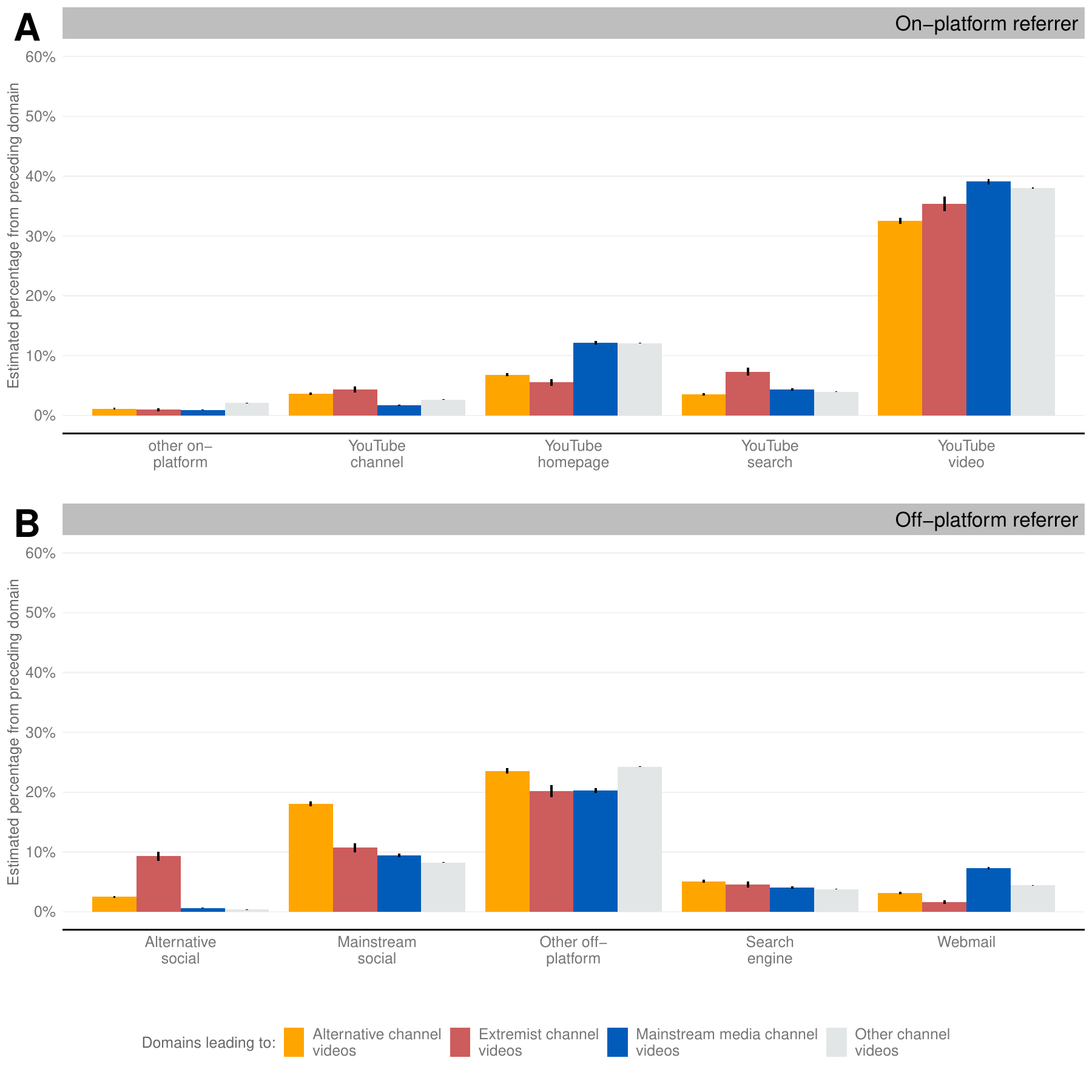}
    \end{center}
    \begin{footnotesize}
    Weighted proportion of referrals to YouTube videos of each channel type by referrer type. Other on-platform platform referrals such as YouTube playlists and personal user pages were grouped into a separate category. Similarly, off-platform domains that do not fit into any of the labeled categories in panel B are grouped together. A list of all domains included in each group can be found in the SM. 
    \end{footnotesize}   
\end{figure}

Figure~\ref{fig:internal_external referrer} reports the proportion of views to each type of YouTube channel video (alternative, extremist, mainstream media, and other) from each type of referrer. This analysis allows us to determine which types of referrers are unusually (un)common across channel types. On-platform, we note that the YouTube homepage, YouTube search, and other YouTube videos are relatively less frequent sources of referrals to alternative and extremist channel videos than videos from mainstream media channels and other channels. In contrast, channel pages are a more common referral source to alternative and extremist channel videos. Like quantitatively similar findings by \citet{hosseinmardi2020evaluating}, this highlights that participants arrive at alternative and extremist videos from a variety of referrers, not just YouTube recommendations.

Among off-platform referrers, social media platforms stand out as playing an especially important role in referring people to alternative and extremist channel videos. Participants are disproportionately more likely to reach alternative channel videos via mainstream social media sites and to reach extremist channel videos via alternative social media sites compared with videos from other types of channels. For instance, 9.3\% of extremist channel video views were preceded by a visit to an alternative social media site despite their limited reach. Platforms like Gab and 4chan may attract extremist users in part due to their lax content moderation policies. These results supplement those from \citet{hosseinmardi2020evaluating}, who found that alternative and extreme news websites generated many of the off-platform referrals to the corresponding types of videos on YouTube.

\section*{Discussion}

Using web browsing data collected in 2020, we provide behavioral measures of exposure to videos from alternative and extremist channels on YouTube. These data enable us to measure exposure to potentially harmful content on the platform and to analyze the role of YouTube's algorithms in facilitating exposure to that content after reported changes to the recommendation system in 2019.

Our data indicate that many alternative and extremist channels remain on the platform and attract a small but active audience of individuals who expressed high levels of hostile sexism and racial resentment in survey data collected in 2018. These participants frequently subscribe to the channels in question, generating more frequent recommendations. By continuing to host these channels, YouTube facilitates the growth of problematic communities (many channel views originate in referrals from alternative social media platforms where users with high levels of gender and racial resentment may congregate) and enables creators of alternative and extreme content to profit from shared YouTube advertising revenue or indirectly via affiliated stores and donation campaigns~\cite{ballard-2022-www,factcheckers-2022-poynter}.

In the data we collected in 2020, YouTube's recommendation algorithm plays a secondary role in facilitating exposure to potentially harmful content. We observe that recommendations to videos from alternative and extreme channels are far more common when people are already watching those videos or subscribed to those channels relative to videos from mainstream news and non-news channels. We also observe that people rarely follow recommendations to videos from alternative and extreme channels when they are watching videos from mainstream news and non-news channels.

While these results complicate the narrative of pervasive radicalization via ``rabbit holes'' on YouTube, our study does not imply that there never was a radicalization problem on YouTube or that the status quo is normatively unproblematic. Our data do not allow us to evaluate the previous state of the platform; YouTube's algorithms may have recommended videos from alternative and extremist channels more frequently prior to the changes made in 2019. Furthermore, given the limitations of our study (see below), our findings should be interpreted as estimating lower bounds on ``rabbit hole'' exposures in 2020 on YouTube. In addition, even very low rates of ``rabbit hole'' recommendations may be enough to expose large numbers of vulnerable people to harm, especially when extrapolated over YouTube's entire viewership and over the course of years. 

It is important to note several other limitations of the study:
\begin{itemize}
    \item Though our browser extension sample is large and diverse and we weight our results to national benchmarks, it is not fully representative and does not capture YouTube consumption among users of browsers other than Chrome and Firefox or on mobile devices. Any outside study of a platform also faces challenges in recruiting large numbers of heavy consumers of fringe content.
    \item YouTube users who were susceptible to potentially harmful content may have already suffered from its effects prior to changes to the platform's algorithms in 2019. We are therefore unable to make causal claims based on our data---participant's preexisting gender and racial resentment may have caused them to seek out congruent content on YouTube, but in some cases YouTube's algorithmic recommendations may have introduced them to such content and increased feelings of resentment even before our prior survey measures of hostile sexism and racial resentment were recorded (November/December 2018). Exposure to YouTube's algorithms before the changes in 2019 could also reduce our ability to detect new ``rabbit hole'' events during the study period in 2020 as some people who are likely to follow problematic recommendations might already be subscribed to these types of channels.
    \item Our results only cover U.S. users; they should be replicated outside the U.S. in contexts including Europe and the global South (and non-English language content). 
    \item Our results depend on channel-level classifications from scholars and subject matter experts; further research should examine whether the patterns we observe are robust to alternate measures at the channel and (if possible) video level.
    \item Our measures of views, referrals, and subscriptions contain some degree of error. In particular, as with most passive behavioral data, we cannot verify that every user paid attention to the content that appeared on their device in every instance.
\end{itemize}

Nonetheless, these results underscore the need to apply the tools of behavioral science to measure exposure to extremist content across social media platforms and to determine how these platforms may reinforce (or hinder) those patterns of behavior individually and collectively. As our findings suggest, these problems often center on the way social media platforms enable the distribution of potentially harmful content to vulnerable audiences rather than algorithmic exposure itself.

\section*{Acknowledgments} 

We are grateful to the Russell Sage Foundation, Anti-Defamation League, Carnegie Corporation of New York, the National Science Foundation, John Smith Guggenheim Memorial Foundation, and the European Research Council (ERC) for financial support; to Samantha Luks at YouGov for survey assistance; to Kasey Rhee for research assistance; to Andy Guess for helping design this project in its initial stages; and to Tanushree Mitra, Joseph B. Phillips, David Rothschild, Gianluca Stringhini, and Savvas Zannettou for comments and feedback. We also thank Virg\'{i}lio A.F. Almeida, Stephen Ansolabehere, Manoel Horta Ribeiro, Aaron Sankin, Brian Schaffner, Robert West, and Anna Zaitsev for sharing their data with us or making it publicly available. This research utilized equipment funded by NSF grant IIS-1910064. Any opinions, findings, and conclusions or recommendations expressed in this material are those of the authors and do not necessarily reflect the views of the funders. In general, all conclusions and errors are our own.

\section*{Funding} 

This study was supported by the Russell Sage Foundation, Anti-Defamation League, Carnegie Corporation of New York, and the National Science Foundation (grant IIS-1910064). This project (J.R.) received funding from the European Research Council (ERC) under the European Union’s Horizon 2020 research and innovation programme (grant agreement No. 682758). 

\section*{Data availability}

Data and code necessary to replicate the results in this study have been posted on Github (\url{https://github.com/aychen5/youtube-extremism-replication} and Dataverse (\url{https://dataverse.harvard.edu/dataset.xhtml?persistentId=doi:10.7910/DVN/UC1XM1}).

\section*{Author contributions} 

B.N., J.R., and C.W. designed the study. All the authors wrote the original manuscript. B.N., J.R., and C.W. revised the manuscript. R.E.R. collected the browser data. A.Y.C. and R.E.R. analyzed the data.

\section*{Competing interests}

The authors declare that they have no competing interests.

\bibliography{bibtex-all-alt}
\end{doublespacing}

\newpage
\begin{center}
\textbf{\LARGE{Supplementary Materials:\\
\vskip 16pt
Subscriptions and external links help drive resentful users to alternative and extremist YouTube channels}}
\end{center}

\begin{singlespacing}

\section*{Sample details and additional results}
\renewcommand{\thetable}{S\arabic{table}}
\renewcommand{\thefigure}{S\arabic{figure}}
\setcounter{table}{0}
\setcounter{figure}{0}

\subsection*{Demographic statistics by sample}\label{app:desc}

\renewcommand{\arraystretch}{0.35}
\begin{table}[!htb]

\caption{\label{tab:demog}Full and extension sample demographics}
\centering
\fontsize{9}{10}\selectfont
\begin{tabular}[t]{ccccc}
\toprule
& \multicolumn{2}{c}{\textbf{Full sample}} & 
\multicolumn{2}{c}{\textbf{Extension sample}}\\
\cmidrule(r){2-3}\cmidrule(l){4-5}
 & weighted & unweighted & weighted & unweighted\\
\midrule
\addlinespace[0em]
\multicolumn{5}{l}{\cellcolor{gray!6}{\textbf{Gender}}}\\
\hspace{1em}Female & 0.48 & 0.46 & 0.49 & 0.49\\
\hspace{1em} & (0.01) & (0.01) & (0.02) & \vphantom{22}(0.01)\\
\hspace{1em}Male & 0.52 & 0.54 & 0.51 & 0.51\\
\hspace{1em} & (0.01) & (0.01) & (0.02) & \vphantom{21}(0.01)\\
\addlinespace[0em]
\multicolumn{5}{l}{\cellcolor{gray!6}{\textbf{Race}}}\\
\hspace{1em}White & 0.68 & 0.76 & 0.69 & 0.75\\
\hspace{1em} & (0.01) & (0.01) & (0.02) & \vphantom{20}(0.01)\\
\hspace{1em}Black & 0.12 & 0.08 & 0.14 & 0.08\\
\hspace{1em} & (0.01) & (0.00) & (0.02) & \vphantom{2}(0.01)\\
\hspace{1em}Hispanic & 0.10 & 0.07 & 0.10 & 0.07\\
\hspace{1em} & (0.01) & (0.00) & (0.02) & \vphantom{1}(0.01)\\
\hspace{1em}Asian & 0.04 & 0.04 & 0.04 & 0.04\\
\hspace{1em} & (0.01) & (0.00) & (0.01) & \vphantom{1}(0.01)\\
\addlinespace[0em]
\multicolumn{5}{l}{\cellcolor{gray!6}{\textbf{2016 presidential vote}}}\\
\hspace{1em}Donald Trump & 0.33 & 0.40 & 0.19 & 0.20\\
\hspace{1em} & (0.01) & (0.01) & (0.02) & \vphantom{19}(0.01)\\
\hspace{1em}Hillary Clinton & 0.28 & 0.31 & 0.40 & 0.49\\
\hspace{1em} & (0.01) & (0.01) & (0.02) & \vphantom{18}(0.01)\\
\addlinespace[0em]
\multicolumn{5}{l}{\cellcolor{gray!6}{\textbf{Employment status}}}\\
\hspace{1em}Employed & 0.46 & 0.49 & 0.48 & 0.51\\
\hspace{1em} & (0.01) & (0.01) & (0.02) & \vphantom{17}(0.01)\\
\hspace{1em}Unemployed & 0.12 & 0.10 & 0.12 & 0.10\\
\hspace{1em} & (0.01) & (0.00) & (0.02) & (0.01)\\
\addlinespace[0em]
\multicolumn{5}{l}{\cellcolor{gray!6}{\textbf{Education}}}\\
\hspace{1em}High school graduate & 0.35 & 0.19 & 0.26 & 0.14\\
\hspace{1em} & (0.01) & (0.01) & (0.02) & \vphantom{16}(0.01)\\
\hspace{1em}Some college & 0.35 & 0.37 & 0.37 & 0.35\\
\hspace{1em} & (0.01) & (0.01) & (0.02) & \vphantom{15}(0.01)\\
\hspace{1em}4-year & 0.19 & 0.26 & 0.24 & 0.28\\
\hspace{1em} & (0.01) & (0.01) & (0.02) & \vphantom{14}(0.01)\\
\hspace{1em}Post-grad & 0.11 & 0.18 & 0.13 & 0.23\\
\hspace{1em} & (0.01) & (0.01) & (0.01) & \vphantom{3}(0.01)\\
\addlinespace[0em]
\multicolumn{5}{l}{\cellcolor{gray!6}{\textbf{Religion}}}\\
\hspace{1em}Atheist/Agnostic & 0.37 & 0.35 & 0.47 & 0.46\\
\hspace{1em} & (0.01) & (0.01) & (0.02) & \vphantom{13}(0.01)\\
\hspace{1em}Protestant & 0.32 & 0.34 & 0.26 & 0.27\\
\hspace{1em} & (0.01) & (0.01) & (0.02) & \vphantom{12}(0.01)\\
\hspace{1em}Roman Catholic & 0.18 & 0.18 & 0.15 & 0.14\\
\hspace{1em} & (0.01) & (0.01) & (0.02) & \vphantom{11}(0.01)\\
\addlinespace[0em]
\multicolumn{5}{l}{\cellcolor{gray!6}{\textbf{Marital status}}}\\
\hspace{1em}Divorced & 0.11 & 0.12 & 0.10 & 0.12\\
\hspace{1em} & (0.01) & (0.01) & (0.01) & \vphantom{2}(0.01)\\
\hspace{1em}Married & 0.43 & 0.53 & 0.39 & 0.48\\
\hspace{1em} & (0.01) & (0.01) & (0.02) & \vphantom{10}(0.01)\\
\hspace{1em}Never married & 0.35 & 0.26 & 0.39 & 0.30\\
\hspace{1em} & (0.01) & (0.01) & (0.02) & \vphantom{9}(0.01)\\
\addlinespace[0em]
\addlinespace[0em]
\multicolumn{5}{l}{\cellcolor{gray!6}{\textbf{Party identification}}}\\
\hspace{1em}Democrat & 0.37 & 0.35 & 0.51 & 0.54\\
\hspace{1em} & (0.01) & (0.01) & (0.02) & \vphantom{5}(0.01)\\
\hspace{1em}Independent & 0.32 & 0.32 & 0.29 & 0.28\\
\hspace{1em} & (0.01) & (0.01) & (0.02) & \vphantom{4}(0.01)\\
\hspace{1em}Republican & 0.31 & 0.33 & 0.20 & 0.18\\
\hspace{1em} & (0.01) & (0.01) & (0.02) & \vphantom{3}(0.01)\\
\addlinespace[0em]
\multicolumn{5}{l}{\cellcolor{gray!6}{\textbf{Age}}}\\
\hspace{1em}18-34 & 0.27 & 0.16 & 0.33 & 0.21\\
\hspace{1em} & (0.01) & (0.01) & (0.02) & \vphantom{2}(0.01)\\
\hspace{1em}35-54 & 0.33 & 0.34 & 0.31 & 0.37\\
\hspace{1em} & (0.01) & (0.01) & (0.02) & \vphantom{1}(0.01)\\
\hspace{1em}55-64 & 0.18 & 0.23 & 0.18 & 0.24\\
\hspace{1em} & (0.01) & (0.01) & (0.01) & (0.01)\\
\hspace{1em}65+ & 0.21 & 0.27 & 0.18 & 0.19\\
\hspace{1em} & (0.01) & (0.01) & (0.02) & (0.01)\\
\addlinespace[0em]
\multicolumn{5}{l}{\cellcolor{gray!6}{\textbf{Sample size}}}\\
\hspace{1em}N &  4000 & 4000 & 1236 & 1236\\

\bottomrule
\multicolumn{5}{l}{\rule{0pt}{1em}Weighted estimates use YouGov survey weights. Standard errors are in parentheses.}\\
\end{tabular}
\end{table}

\clearpage
\newpage
\subsection*{Enrollment and consumption over time}

\begin{figure}[!htb]
\begin{center}
    \caption{Total participants with browser activity data over time \label{fig:attrition}}
    \includegraphics[width=.99\textwidth]{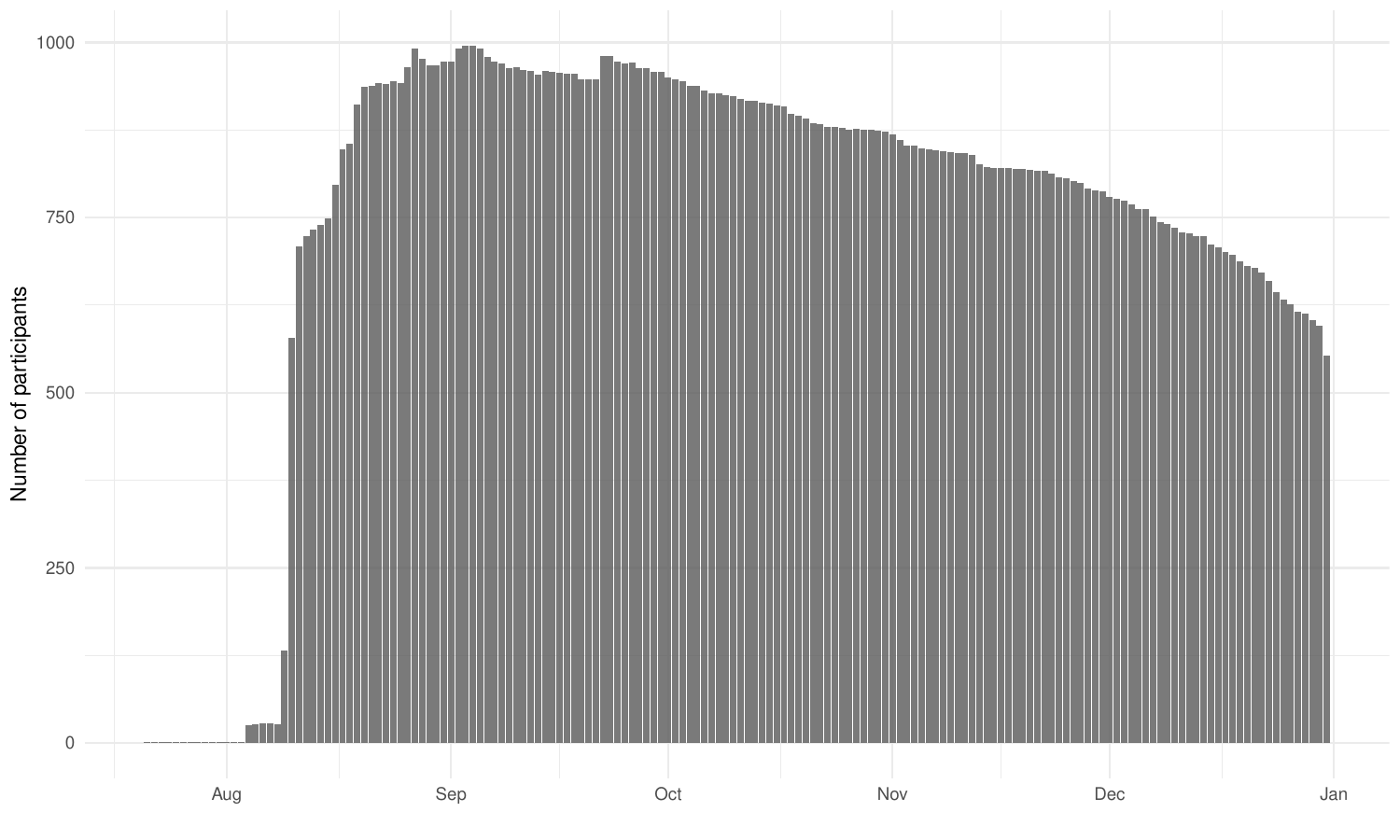}
    \end{center}
    \vskip -10pt
    \footnotesize{Day-level totals of the number of study participants with browser activity data. All results incorporate survey weights.}
\end{figure}

\begin{figure}[!htb]
\begin{center}
    \caption{Consumption levels over time by channel type \label{fig:consumption_over_time_type}}
    \includegraphics[width=.99\textwidth]{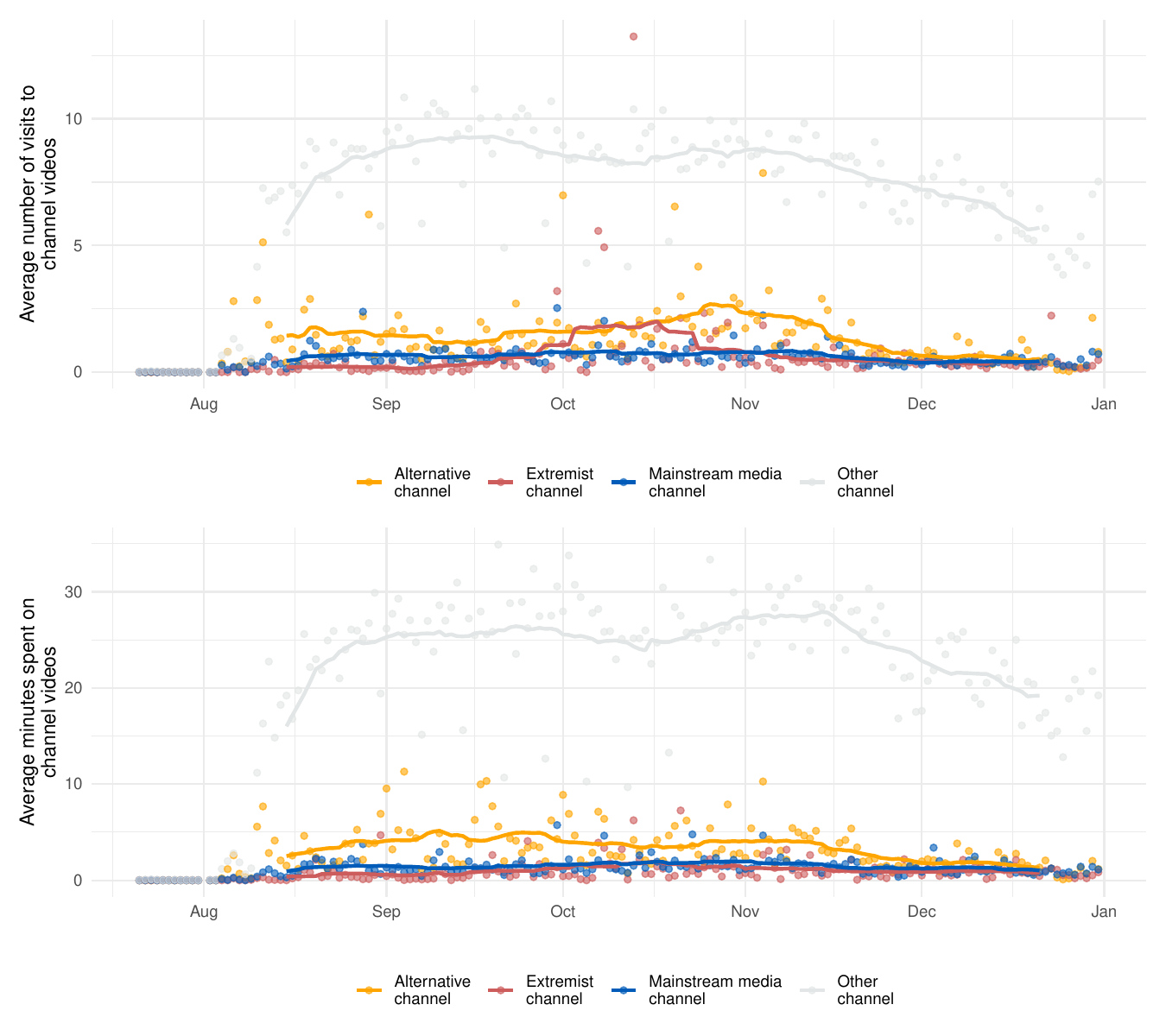}
    \end{center}
    \vskip -10pt
    \footnotesize{Each point represents the weighted mean number of views (top panel) or minutes spent (bottom panel) on videos from each channel type per day. Trend lines are three-week moving averages. All results incorporate survey weights.} 
\end{figure}

\newpage
\begin{figure}[!htb]
\begin{center}
    \caption{YouTube video diets of individuals who viewed any alternative channel video \label{fig:consumption_any_alt}}
    \includegraphics[width=.99\textwidth]{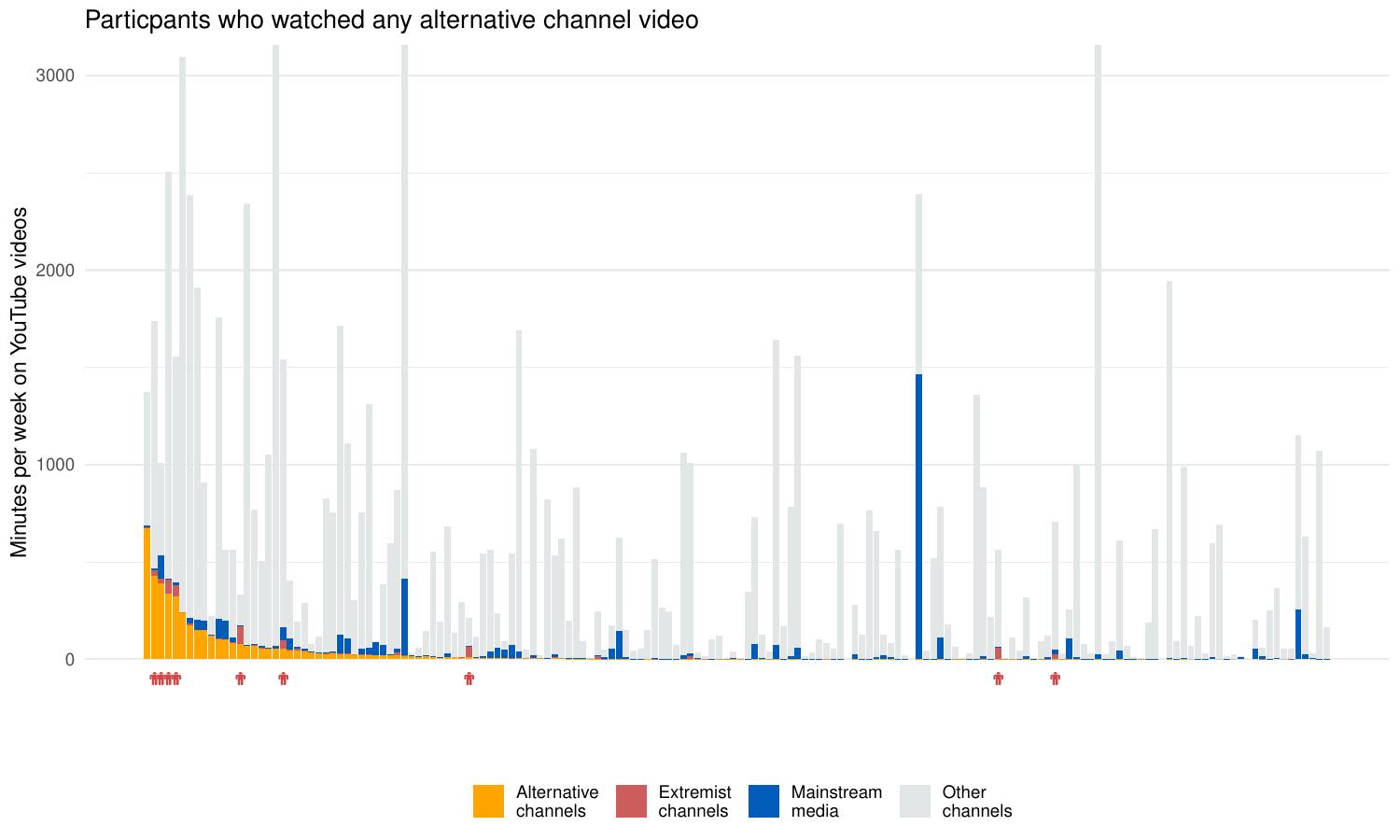}
    \end{center}
    \vskip -10pt
    \footnotesize{All results incorporate survey weights.} 
\end{figure}

\begin{figure}[!htb]
\begin{center}
    \caption{YouTube video diets of individuals who viewed any extremist channel video \label{fig:consumption_any_ext}}
    \includegraphics[width=.99\textwidth]{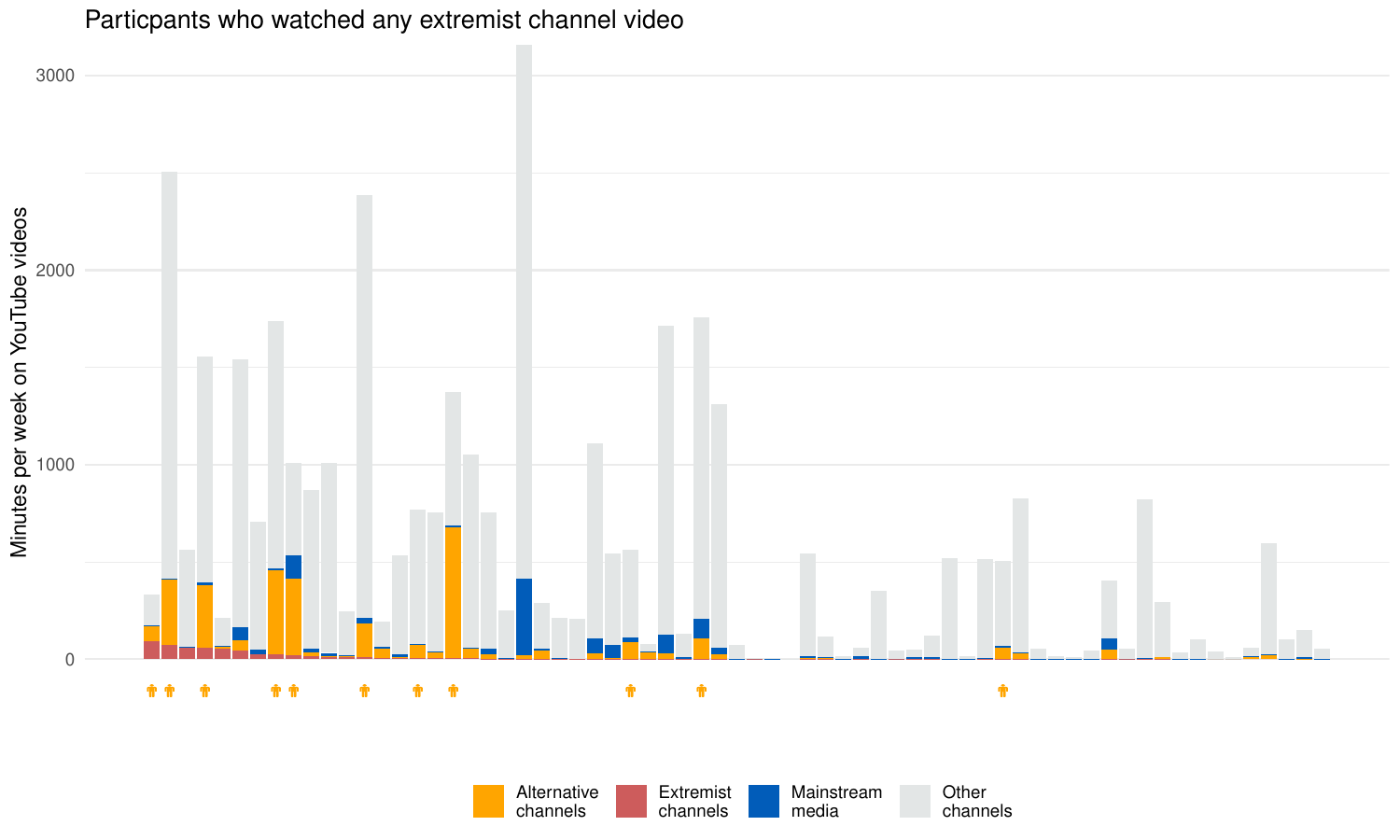}
    \end{center}
    \vskip -10pt
    \footnotesize{All results incorporate survey weights.} 
\end{figure}

\clearpage
\newpage
\subsection*{Alternative and extremist superconsumers}

\begin{figure}[!b]
    \begin{center}
         \caption{YouTube video diets of alternative and extremist superconsumers}
    \label{fig:alternative_extremist_superconsumers}
    \includegraphics[width=0.99\textwidth]{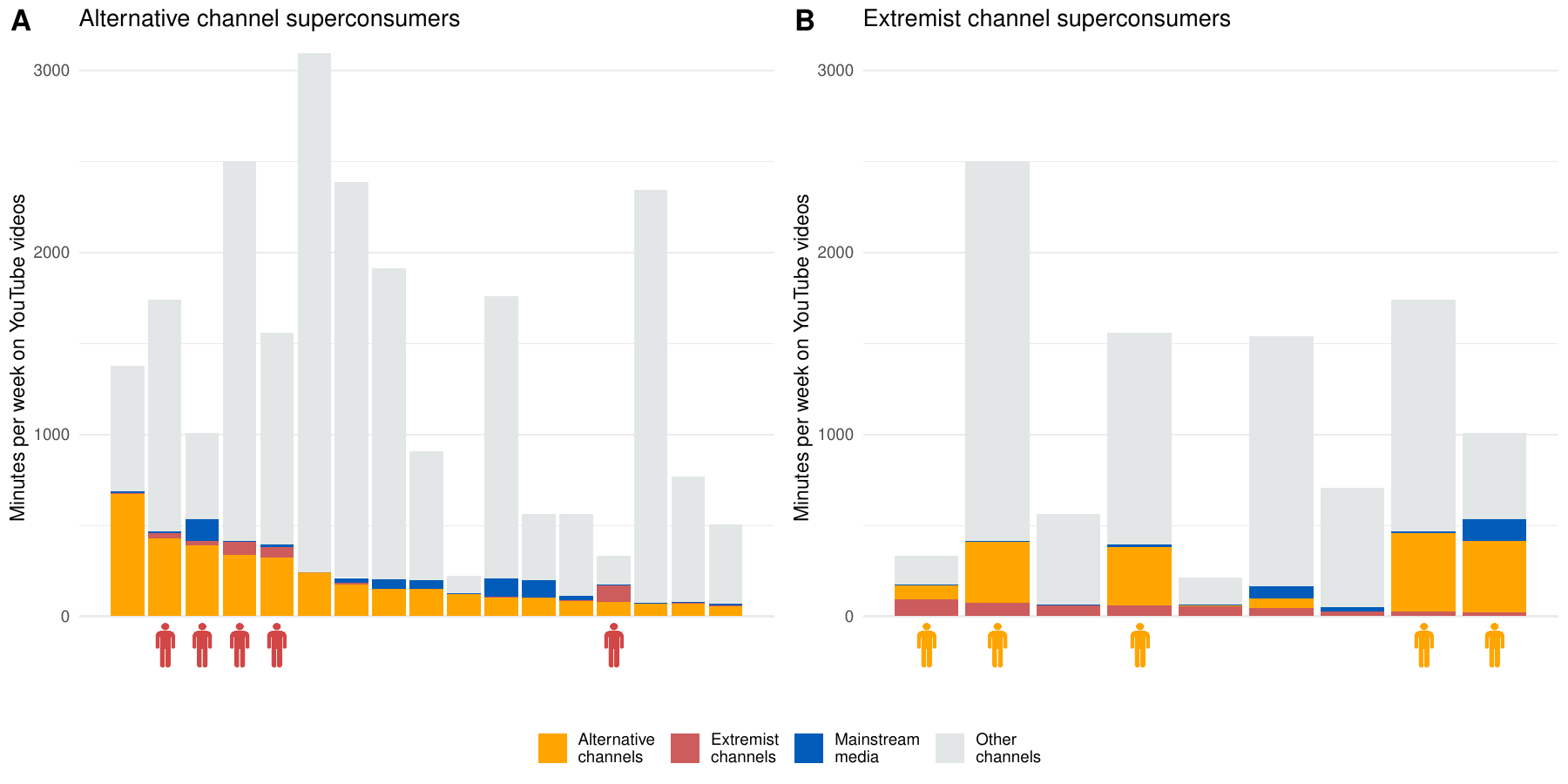}
    \end{center}
    \vskip -10pt
\begin{footnotesize}
\noindent Total YouTube behavior of alternative (panel A) and extremist (panel B) superconsumers measured in minutes per week of video watch time. Each bar represents one individual and the height of the bar represents total view time of YouTube videos by channel type. The alternative superconsumers are ordered left to right by time spent on videos from alternative channels (orange portions of bars); the extremist superconsumers in the right panel are ordered left to right by time spent on videos from extremist channels (red portions of the bars). Red icons under bars in the left panel represent individuals who are also extremist superconsumers; orange icons under bars in the right panel represent individuals who are also alternative content superconsumers. All results incorporate survey weights.
\end{footnotesize}
\end{figure}

\autoref{fig:alternative_extremist_superconsumers} presents watch time totals for the people responsible for 80\% of the viewership of videos from alternative and extremist channels in our sample. We note two facts about superconsumers. First, they often watch a great deal of YouTube. Alternative channel superconsumers spend a median of 29 hours each week watching YouTube, while the median time that extremist channel superconsumers spend watching is 16 hours per week. By comparison, the median time per week across all participants is 0.2 hours. Second, there is substantial overlap between the two sets of superconsumers, who represent just 2\% of all participants. Figures \ref{fig:consumption_any_alt} and \ref{fig:consumption_any_ext} show the YouTube video diets by channel type for individuals who viewed any alternative or extremist channel video during the study.

\clearpage
\newpage
\subsection*{Additional data on recommendations and referrers}\label{app:recsfollowed}

\begin{figure}[!ht]
    \begin{center}
    \caption{Recommendation follows by video channel type \label{fig:waffles_recs_followed}}
    \includegraphics[width = .99\textwidth]{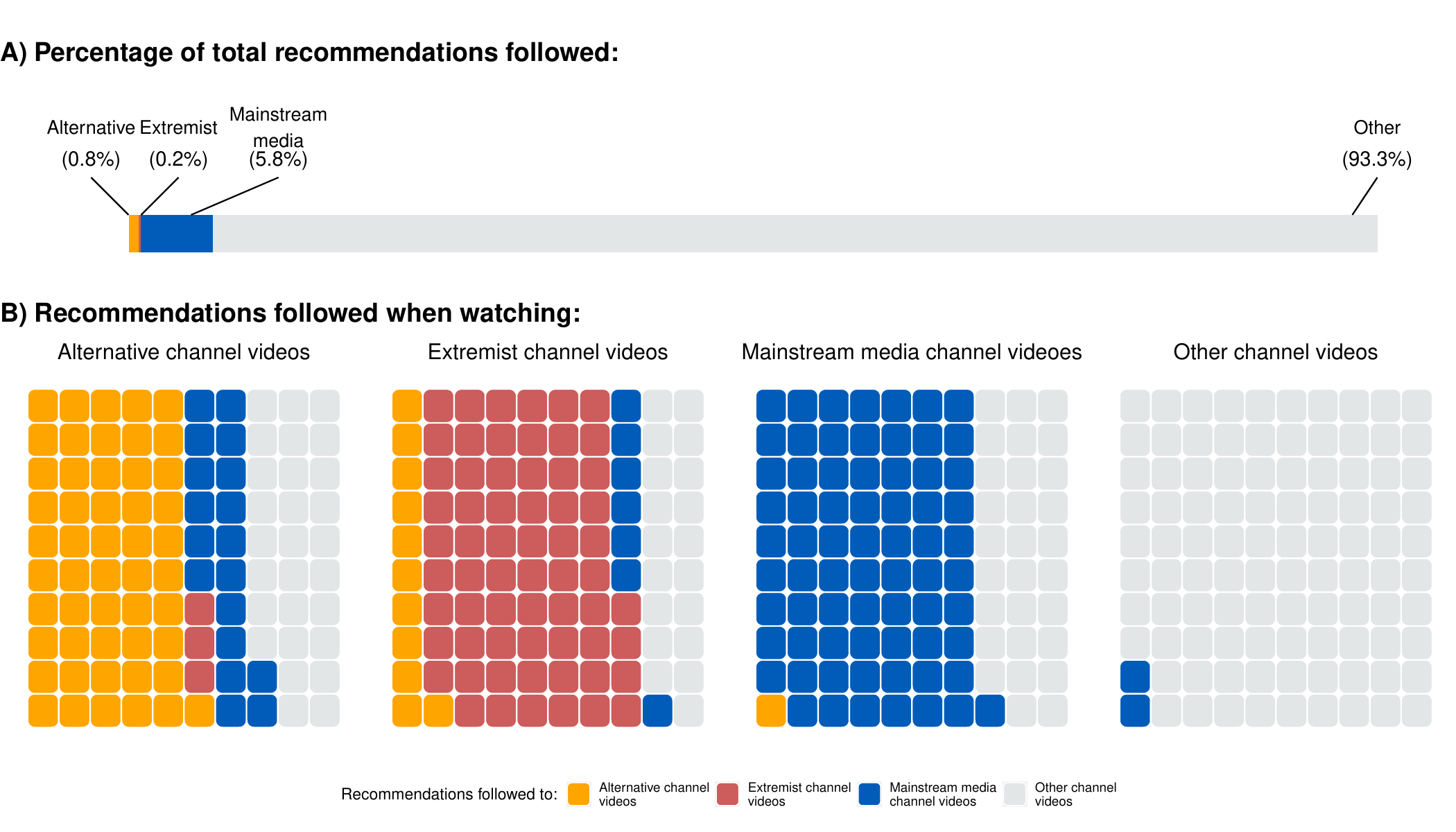}
    \end{center}
    Number of colored tiles shown are proportional to the proportion of recommendations followed to each type of video when watching videos from alternative, extremist, mainstream media, or other channels. Results are based on the full set of recommendations that we could extract from each video and incorporate survey weights. 
\end{figure}

\begin{figure}[!ht]
    \begin{center}
    \caption{Pages viewed immediately prior to YouTube videos by channel type     \label{fig:on_platform_referrers_by_channel}}
    \includegraphics[width = .99\textwidth]{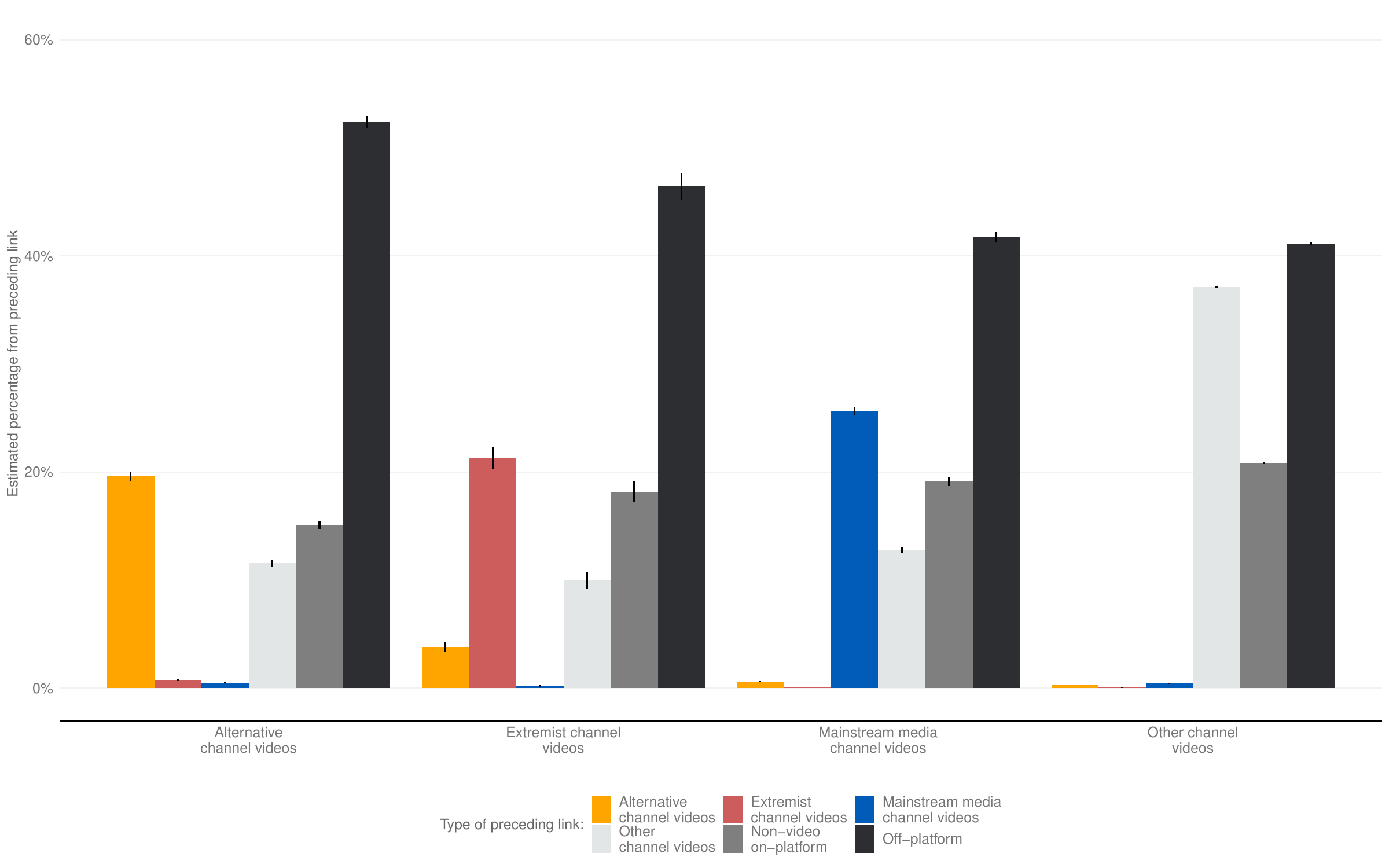}
    \end{center}
    \begin{footnotesize}
    Weighted proportion of each type of URL recorded immediately before viewing a YouTube video of a given channel type. Observations where the preceding link was not a YouTube video are shown in the ``non-video, on-platform'' and ``off-platform'' bars. (``Non-video, on-platform'' referrers combines YouTube channel pages, YouTube homepage, and YouTube search.) 
    \end{footnotesize}   
\end{figure}

\newpage
\clearpage
\subsection*{Additional regressions}\label{app:regressions}

The Poisson GLM for rates takes the form:

$\begin{aligned}
log(\lambda_i) &= log(t_i) + \sum_{j=1}^{p}\beta_jx_{ij}
\end{aligned}$

Let $\lambda_{i}$ be either the expected number of minutes or the expected number of views of alternative, extremist, or mainstream media channel videos. $t_{i}$ is the total number of weeks we have activity data for user $i$. $j$ indexes the predictors (racial resentment, hostile sexism, feelings toward Jews, age, gender, education, and race). Due to overdispersion in the data, we relax the mean-variance equivalence assumption ($Var[y|x] = \phi E[y|x]$) of Poisson models in which $\phi$ (dispersion) is restricted to 1 and estimate $\phi$ directly from the data through quasi-MLE.

\begin{figure}[!htb]
\begin{center}
    \caption{Zero-inflated models on correlates of time on YouTube video by channel type \label{fig:zip_coefficients_time}}    
    \includegraphics[width=.9\textwidth]{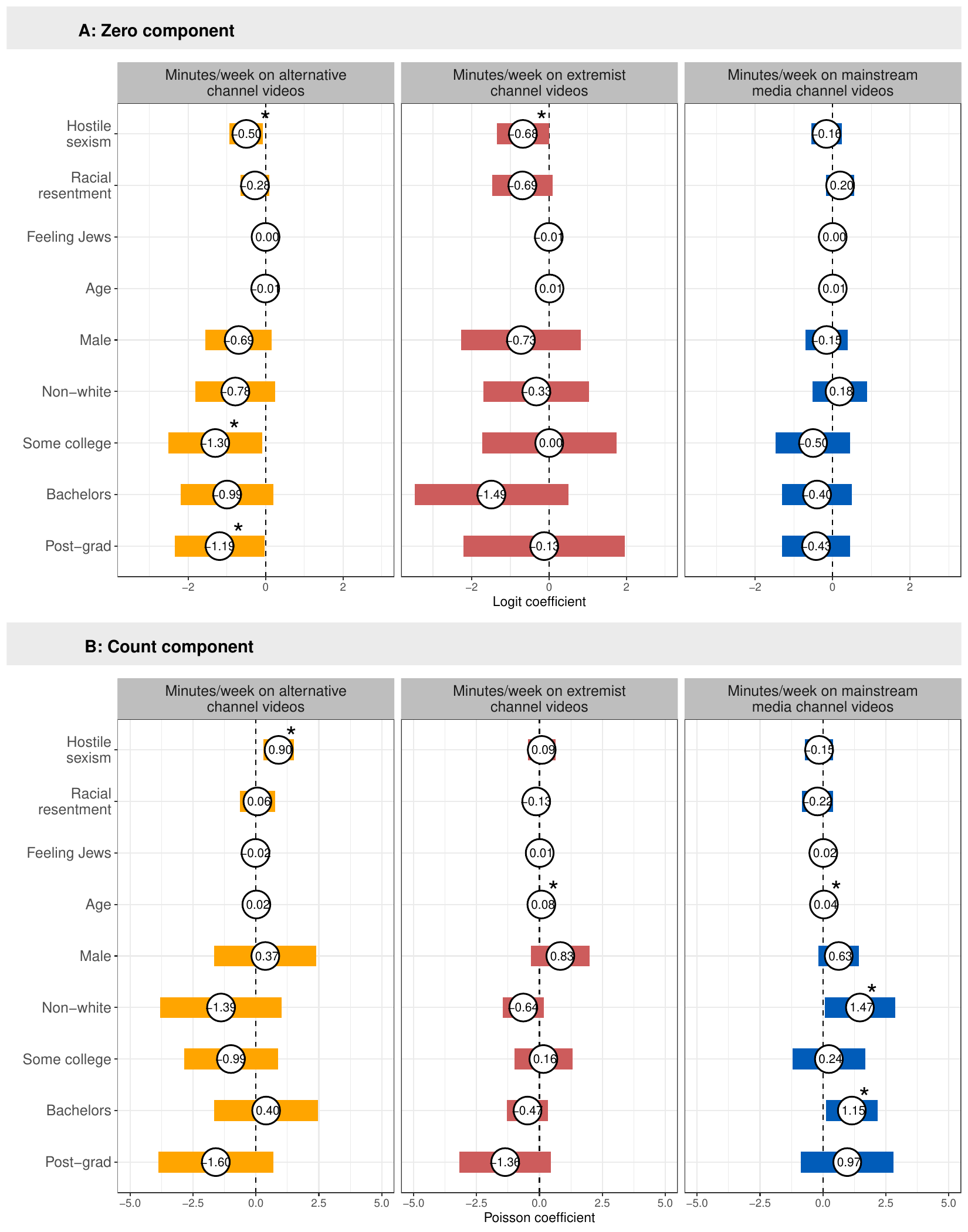}
    \end{center}
    Zero-inflated Poisson coefficients for correlates of the time per week spent on videos from alternative, extremist, and mainstream media channels. Figure includes 95\% confidence intervals calculated from robust standard errors. All results incorporate survey weights. Stars indicate coefficients that are significant at the $p<.05$ level. See \autoref{tab:figa5table} for the regression table.
\end{figure}

Figure \ref{fig:qpois_coefficients_time} in the main text and \autoref{tab:fig4output} below report quasipoisson estimates using this estimation approach for time spent on videos from alternative and extremist channels. Figure \ref{fig:zip_coefficients_time} and \autoref{tab:figa5table} report corresponding results from zero-inflated Poisson models in which the zero component is modelled with a Binomial regression and a secondary process generating the counts including zeros is governed by a Poisson model. 

\begin{table}[!htbp] 
\caption{Correlates of time on YouTube videos by channel type\label{tab:fig4output}} 
\fontsize{9}{10}\selectfont
\begin{center}
\begin{tabular}{@{\extracolsep{5pt}}lcccccc} 
\toprule
 & \multicolumn{3}{c}{\textit{Dependent variable: Time elapsed}} \\
\cmidrule(r){2-4}
& Alternative & Extremist & Mainstream \\ 
& channel videos & channel videos & channel videos \\ 
& (1) & (2) & (3) \\ 
\midrule
Hostile sexism    & $1.71^{***}$  & $1.60^{**}$    & $0.00$       \\
                  & $(0.37)$      & $(0.60)$       & $(0.32)$     \\
Racial resentment & $0.19$        & $0.09$         & $-0.42$      \\
                  & $(0.35)$      & $(0.43)$       & $(0.36)$     \\
Feeling Jews      & $-0.01$       & $-0.00$        & $0.00$       \\
                  & $(0.01)$      & $(0.01)$       & $(0.02)$     \\
Age               & $0.03$        & $0.05^{**}$    & $0.04^{***}$ \\
                  & $(0.02)$      & $(0.02)$       & $(0.01)$     \\
Male              & $1.01$        & $0.74$         & $0.85$       \\
                  & $(1.00)$      & $(1.03)$       & $(0.63)$     \\
Non-white         & $-0.79$       & $-1.30$        & $1.50$       \\
                  & $(0.98)$      & $(0.89)$       & $(0.84)$     \\
Some college      & $0.72$        & $0.50$         & $1.60^{*}$   \\
                  & $(0.95)$      & $(0.97)$       & $(0.64)$     \\
Bachelor's degree & $1.98^{*}$    & $1.79^{*}$     & $2.43^{***}$ \\
                  & $(0.98)$      & $(0.77)$       & $(0.71)$     \\
Post-grad         & $-0.52$       & $-1.99$        & $2.62^{***}$ \\
                  & $(1.03)$      & $(1.04)$       & $(0.74)$     \\
Intercept         & $-8.06^{***}$ & $-10.73^{***}$ & $-3.12$      \\
                  & $(2.04)$      & $(2.56)$       & $(2.07)$     \\
\midrule
N  & $851$         & $851$          & $851$        \\
\bottomrule
\multicolumn{4}{l}{\scriptsize{$^{***}p<0.001$; $^{**}p<0.01$; $^{*}p<0.05$}}
\end{tabular}
\end{center}
\footnotesize{Quasipoisson coefficients for correlates of time per week spent on videos from alternative, extremist, and mainstream media channels. Robust standard errors are in parentheses.}
\end{table}

\newpage
\begin{table}
\caption{Zero-inflated Poisson models for correlates of time on YouTube video by channel type}
\label{tab:figa5table}
\begin{center}\fontsize{9}{10}\selectfont
\begin{tabular}{l c c c}
\toprule
 & \multicolumn{3}{c}{\textit{Dependent variable: Time elapsed}} \\
\cmidrule(r){2-4}\\
& Alternative & Extremist & Mainstream \\ 
& channel videos & channel videos & channel videos \\ 
& (1) & (2) & (3) \\
\midrule      
Zero component & & & \\
\midrule      
Hostile sexism    & $-0.50^{*}$  & $-0.68^{*}$  & $-0.16$     \\
                  & $(0.22)$     & $(0.34)$     & $(0.20)$    \\
Racial resentment & $-0.28$      & $-0.69$      & $0.20$      \\
                  & $(0.19)$     & $(0.40)$     & $(0.18)$    \\
Feeling Jews      & $-0.00$      & $-0.01$      & $0.00$      \\
                  & $(0.01)$     & $(0.01)$     & $(0.01)$    \\
Age               & $-0.01$      & $0.01$       & $0.01$      \\
                  & $(0.01)$     & $(0.02)$     & $(0.01)$    \\
Male              & $-0.69$      & $-0.73$      & $-0.15$     \\
                  & $(0.44)$     & $(0.79)$     & $(0.28)$    \\
Non-white         & $-0.78$      & $-0.33$      & $0.18$      \\
                  & $(0.53)$     & $(0.70)$     & $(0.36)$    \\
Some college      & $-1.30^{*}$  & $0.00$       & $-0.50$     \\
                  & $(0.62)$     & $(0.89)$     & $(0.49)$    \\
Bachelor's degree & $-0.99$      & $-1.49$      & $-0.40$     \\
                  & $(0.61)$     & $(1.01)$     & $(0.46)$    \\
Post-grad         & $-1.19^{*}$  & $-0.13$      & $-0.43$     \\
                  & $(0.59)$     & $(1.06)$     & $(0.45)$    \\
Intercept         & $6.88^{***}$ & $9.34^{***}$ & $1.00$      \\
                  & $(1.11)$     & $(1.79)$     & $(1.04)$    \\
\midrule
 Count component& & & \\
\midrule
Hostile sexism    & $0.90^{**}$  & $0.09$       & $-0.15$     \\
                  & $(0.31)$     & $(0.28)$     & $(0.28)$    \\
Racial resentment & $0.06$       & $-0.13$      & $-0.22$     \\
                  & $(0.36)$     & $(0.21)$     & $(0.31)$    \\
Feeling Jews      & $-0.02$      & $0.01$       & $0.02$      \\
                  & $(0.01)$     & $(0.02)$     & $(0.02)$    \\
Age               & $0.02$       & $0.08^{***}$ & $0.04^{**}$ \\
                  & $(0.02)$     & $(0.02)$     & $(0.01)$    \\
Male              & $0.37$       & $0.83$       & $0.63$      \\
                  & $(1.03)$     & $(0.59)$     & $(0.41)$    \\
Non-white         & $-1.39$      & $-0.64$      & $1.47^{*}$  \\
                  & $(1.23)$     & $(0.41)$     & $(0.71)$    \\
Some college      & $-0.99$      & $0.16$       & $0.24$      \\
                  & $(0.95)$     & $(0.59)$     & $(0.74)$    \\
Bachelor's degree & $0.40$       & $-0.47$      & $1.15^{*}$  \\
                  & $(1.06)$     & $(0.42)$     & $(0.52)$    \\
Post-grad         & $-1.60$      & $-1.36$      & $0.97$      \\
                  & $(1.16)$     & $(0.93)$     & $(0.94)$    \\
Intercept         & $0.27$       & $-5.06^{*}$  & $-4.00^{*}$ \\
                  & $(1.69)$     & $(2.03)$     & $(1.66)$    \\
\midrule
N        & $851$        & $851$        & $851$       \\
\bottomrule
\end{tabular}
\end{center}
\footnotesize{Zero-inflated Poisson coefficients for correlates of the time per week spent on videos from alternative, extremist, and mainstream media channels. Robust standard errors are in parentheses.\scriptsize{$^{***}p<0.001$; $^{**}p<0.01$; $^{*}p<0.05$}}
\end{table}

\begin{figure}[!htbp]
\begin{center}
    \caption{Correlates of YouTube video views by channel type \label{fig:qpois_coefficients_visits}}    
    \includegraphics[width=.99\textwidth]{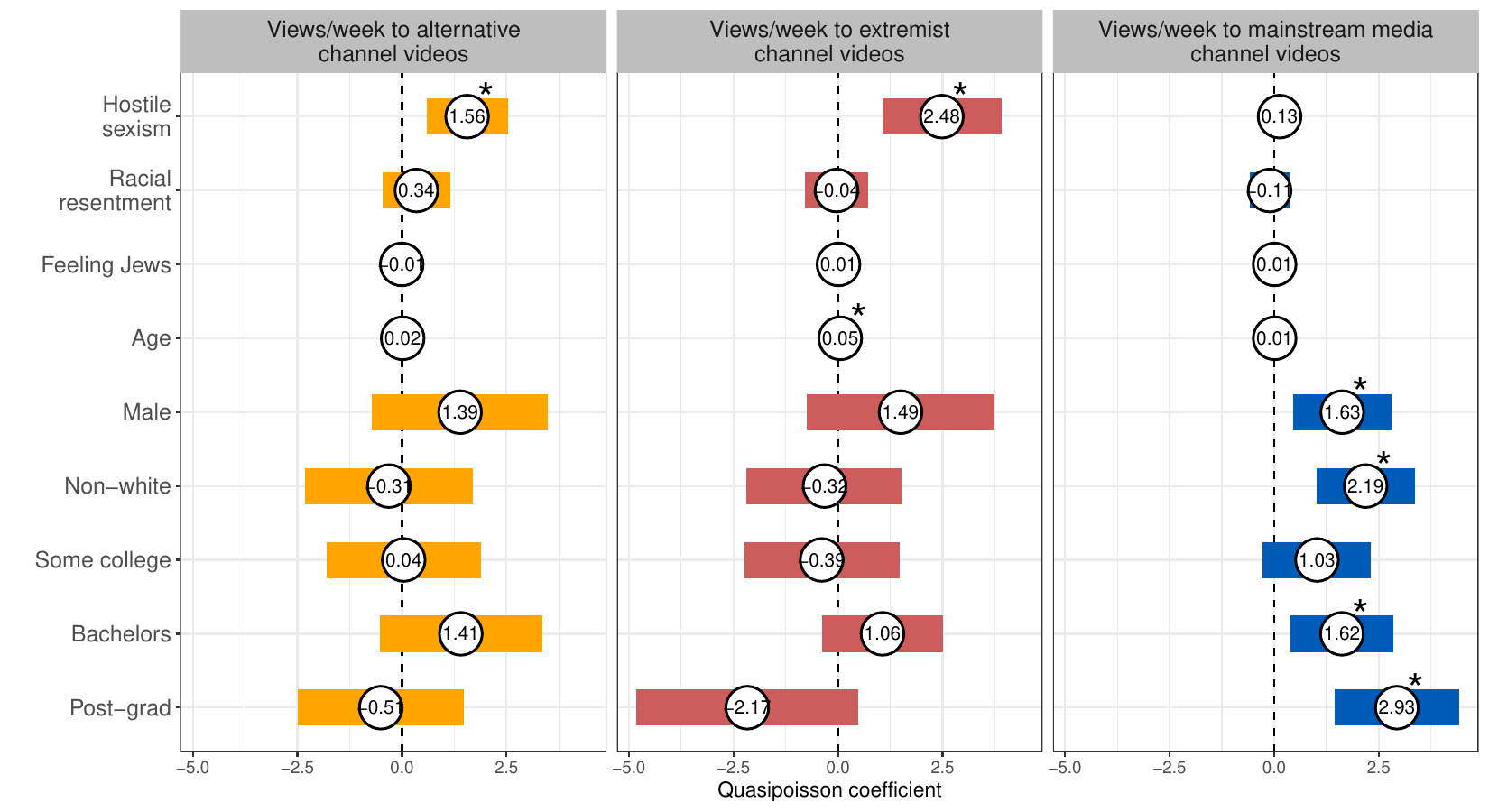}
    \end{center}
     \noindent Quasipoisson regression coefficients for correlates of the number of respondent views per week of videos from alternative, extremist, and mainstream media channels. Figure includes 95\% confidence intervals calculated from robust standard errors. All results incorporate survey weights. Stars indicate coefficients that are significant at the $p<.05$ level. See \autoref{tab:figa6output} for the regression table.
\end{figure}

\newpage
\clearpage
Figure \ref{fig:qpois_coefficients_visits} and \autoref{tab:figa6output} instead report quasipoisson estimates for the number of views of videos from alternative, extremist, and mainstream media channels (rather than time spent).

\begin{table}[!htbp] 
\caption{Correlates of YouTube video views by channel type\label{tab:figa6output}} 
\fontsize{9}{10}\selectfont
\begin{center}
\begin{tabular}{@{\extracolsep{5pt}}lcccccc} 
\toprule
 & \multicolumn{3}{c}{\textit{Dependent variable: Views}} \\
\cmidrule(r){2-4}
& Alternative & Extremist & Mainstream \\ 
& (1) & (2) & (3) \\ 
\midrule  
Hostile sexism    & $1.56^{**}$   & $2.48^{***}$   & $0.13$        \\
                  & $(0.50)$      & $(0.73)$       & $(0.16)$      \\
Racial resentment & $0.34$        & $-0.04$        & $-0.11$       \\
                  & $(0.41)$      & $(0.38)$       & $(0.25)$      \\
Feeling Jews      & $-0.01$       & $0.01$         & $0.01$        \\
                  & $(0.01)$      & $(0.01)$       & $(0.01)$      \\
Age               & $0.02$        & $0.05^{**}$    & $0.01$        \\
                  & $(0.02)$      & $(0.02)$       & $(0.01)$      \\
Male              & $1.39$        & $1.49$         & $1.63^{**}$   \\
                  & $(1.07)$      & $(1.15)$       & $(0.60)$      \\
Non-white         & $-0.31$       & $-0.32$        & $2.19^{***}$  \\
                  & $(1.03)$      & $(0.95)$       & $(0.60)$      \\
Some college      & $0.04$        & $-0.39$        & $1.03$        \\
                  & $(0.94)$      & $(0.95)$       & $(0.66)$      \\
Bachelor's degree & $1.41$        & $1.06$         & $1.62^{**}$   \\
                  & $(0.99)$      & $(0.73)$       & $(0.63)$      \\
Post-grad         & $-0.51$       & $-2.17$        & $2.93^{***}$  \\
                  & $(1.01)$      & $(1.35)$       & $(0.76)$      \\
Intercept         & $-8.75^{***}$ & $-16.15^{***}$ & $-4.64^{***}$ \\
                  & $(2.62)$      & $(4.46)$       & $(1.38)$      \\
\midrule
N        & $851$         & $851$          & $851$         \\
\bottomrule
\multicolumn{4}{l}{\scriptsize{$^{***}p<0.001$; $^{**}p<0.01$; $^{*}p<0.05$}}
\end{tabular}
\end{center}
\footnotesize{Quasipoisson coefficients for correlates of views per week spent on videos from alternative, extremist, and mainstream media channels. Robust standard errors are in parentheses.}
\end{table}

\begin{figure}[!htb]
\begin{center}
    \caption{Correlates of YouTube video exposure by channel type (with party controls)\label{fig:qpois_coefficients_pid}}    
    \includegraphics[width=.75\textwidth]{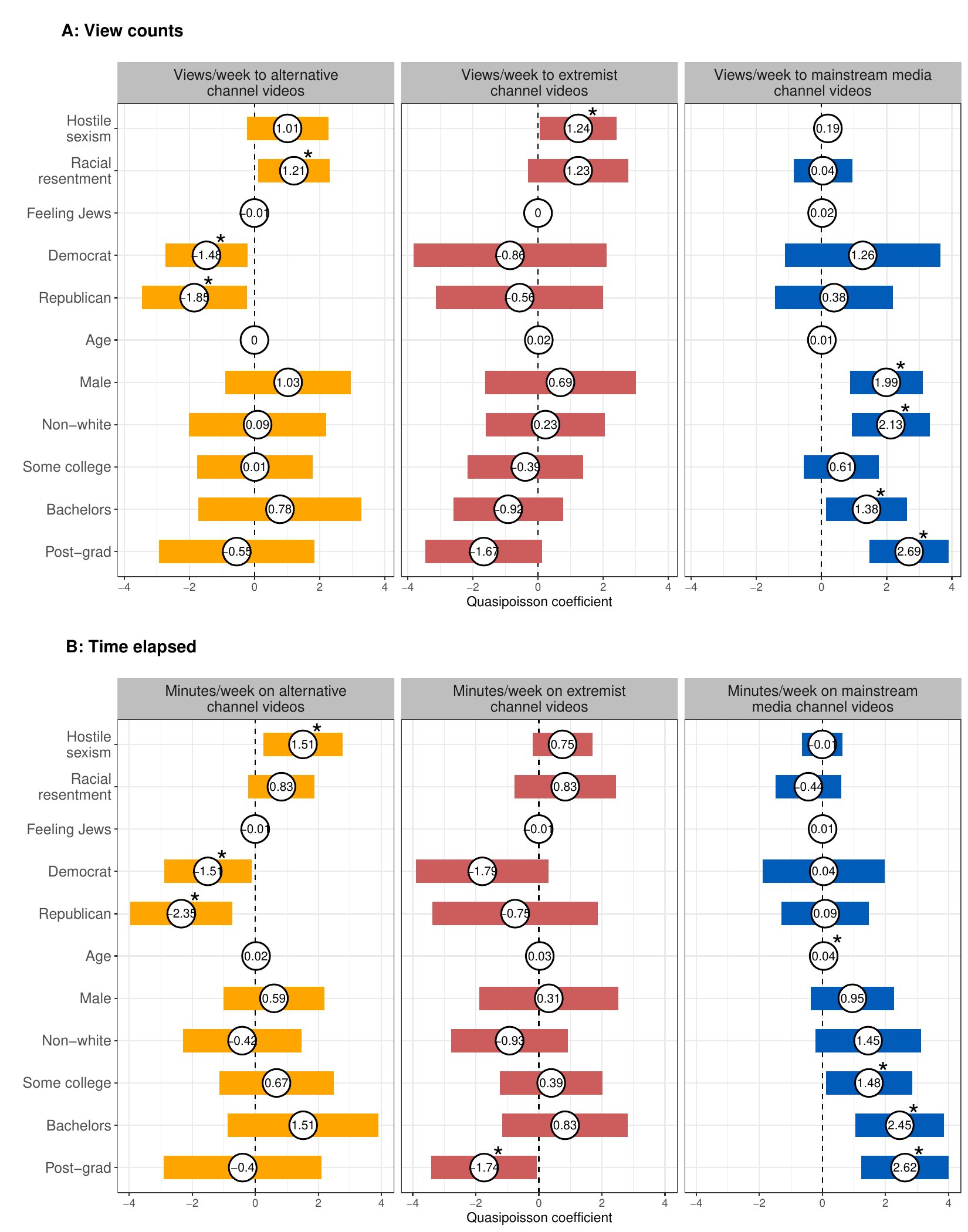}
    \end{center}
    Quasipoisson regression coefficients for correlates of the number of respondent video views per week (panel A) and time spent (panel B) per week on videos from alternative, extremist, and mainstream media channels. Figure includes 95\% confidence intervals calculated from robust standard errors. All results incorporate survey weights. See \autoref{tab:figa7output} for the regression table.
\end{figure}

\newpage
\clearpage
Due to concerns about post-treatment bias, we omit controls for party identification from the models reported in the main text. However, \autoref{fig:qpois_coefficients_pid} (\autoref{tab:figa7output}) reports quasipoisson results mirroring those in Figure \ref{fig:qpois_coefficients_time} (\autoref{tab:fig4output}) and Figure \ref{fig:qpois_coefficients_visits} (\autoref{tab:figa6output})  but which additionally control for Democratic and Republican self-identification (including leaners). 

\begin{table}[!htbp] 
\caption{Correlates of YouTube video exposure by channel type (with party controls) \label{tab:figa7output}} 
\fontsize{9}{10}\selectfont
\begin{center}
\begin{tabular}{@{\extracolsep{5pt}}lcccccc} 
\toprule
 & \multicolumn{3}{c}{\textit{Dependent variable: Views}} & \multicolumn{3}{c}{\textit{Dependent variable: Time elapsed}} \\
\cmidrule(r){2-4}\cmidrule(l){5-7}
& Alternative & Extremist & Mainstream & Alternative & Extremist & Mainstream \\ 
& (1) & (2) & (3) & (4) & (5) & (6)\\ 
\midrule  
Hostile sexism    & $1.01$      & $1.24^{*}$   & $0.19$       & $1.51^{*}$   & $0.75$       & $-0.01$       \\
                  & $(0.64)$    & $(0.60)$     & $(0.15)$     & $(0.64)$     & $(0.49)$     & $(0.33)$      \\
Racial resentment & $1.21^{*}$  & $1.23$       & $0.04$       & $0.83$       & $0.83$       & $-0.44$       \\
                  & $(0.56)$    & $(0.79)$     & $(0.46)$     & $(0.54)$     & $(0.82)$     & $(0.53)$      \\
Feeling Jews      & $-0.01$     & $-0.00$      & $0.02$       & $-0.01$      & $-0.01$      & $0.01$        \\
                  & $(0.01)$    & $(0.01)$     & $(0.01)$     & $(0.01)$     & $(0.01)$     & $(0.02)$      \\
Democrat          & $-1.48^{*}$ & $-0.86$      & $1.26$       & $-1.51^{*}$  & $-1.79$      & $0.04$        \\
                  & $(0.64)$    & $(1.51)$     & $(1.22)$     & $(0.71)$     & $(1.07)$     & $(0.99)$      \\
Republican        & $-1.85^{*}$ & $-0.56$      & $0.38$       & $-2.35^{**}$ & $-0.75$      & $0.09$        \\
                  & $(0.82)$    & $(1.31)$     & $(0.92)$     & $(0.83)$     & $(1.34)$     & $(0.71)$      \\
Age               & $-0.00$     & $0.02$       & $0.01$       & $0.02$       & $0.03$       & $0.04^{**}$   \\
                  & $(0.03)$    & $(0.03)$     & $(0.01)$     & $(0.03)$     & $(0.02)$     & $(0.01)$      \\
Male              & $1.03$      & $0.69$       & $1.99^{***}$ & $0.59$       & $0.31$       & $0.95$        \\
                  & $(0.98)$    & $(1.18)$     & $(0.57)$     & $(0.82)$     & $(1.12)$     & $(0.68)$      \\
Non-white         & $0.09$      & $0.23$       & $2.13^{***}$ & $-0.42$      & $-0.93$      & $1.45$        \\
                  & $(1.07)$    & $(0.93)$     & $(0.61)$     & $(0.96)$     & $(0.94)$     & $(0.85)$      \\
Some college      & $0.01$      & $-0.39$      & $0.61$       & $0.67$       & $0.39$       & $1.48^{*}$    \\
                  & $(0.91)$    & $(0.91)$     & $(0.59)$     & $(0.92)$     & $(0.83)$     & $(0.70)$      \\
Bachelor's degree & $0.78$      & $-0.92$      & $1.38^{*}$   & $1.51$       & $0.83$       & $2.45^{***}$  \\
                  & $(1.27)$    & $(0.86)$     & $(0.63)$     & $(1.22)$     & $(1.02)$     & $(0.72)$      \\
Post-grad         & $-0.55$     & $-1.67$      & $2.69^{***}$ & $-0.40$      & $-1.74^{*}$  & $2.62^{***}$  \\
                  & $(1.22)$    & $(0.91)$     & $(0.62)$     & $(1.28)$     & $(0.85)$     & $(0.71)$      \\
Intercept         & $-7.73^{*}$ & $-12.39^{*}$ & $-6.53^{*}$  & $-4.12$      & $-3.62$      & $0.92$        \\
                  & $(3.03)$    & $(5.18)$     & $(2.78)$     & $(3.23)$     & $(3.70)$     & $(2.40)$      \\
\midrule
N       & $847$       & $847$        & $847$        & $847$        & $847$        & $847$         \\
\bottomrule
\multicolumn{7}{l}{\scriptsize{$^{***}p<0.001$; $^{**}p<0.01$; $^{*}p<0.05$}}
\end{tabular}
\end{center}
\footnotesize{Quasipoisson models for correlates of views and time per week spent on videos from alternative, extremist, and mainstream media channels. Robust standard errors are in parentheses. All results incorporate survey weights.}
\end{table}

\newpage
\clearpage
Tables \ref{tab:hh_rr_time} and \ref{tab:hh_rr_visit} report quasipoisson estimates in which racial resentment and hostile sexism are entered into separate models rather than jointly as presented above. 

\begin{table}[!htbp]
  \caption{Correlates of time spent on YouTube videos by channel type (separating hostile sexism and racial resentment) \label{tab:hh_rr_time}} 
\fontsize{9}{10}\selectfont
\begin{center}
\begin{tabular}{@{\extracolsep{5pt}}lcccccc} 
\toprule
 & \multicolumn{6}{c}{\textit{Dependent variable: Time elapsed}} \\ 
\cline{2-7}
& \multicolumn{2}{c}{Alternative} & \multicolumn{2}{c}{Extremist} & \multicolumn{2}{c}{Mainstream} \\ 
& \multicolumn{2}{c}{channel videos} & \multicolumn{2}{c}{channel videos} & \multicolumn{2}{c}{channel videos} \\ 
& (1) & (2) & (3) & (4) & (5) & (6)\\ 
\midrule
Hostile sexism    & $1.80^{***}$  &               & $1.64^{***}$ &              & $-0.35$       &               \\
                  & $(0.23)$      &               & $(0.49)$     &              & $(0.27)$      &               \\
Racial resentment &               & $1.01^{***}$  &              & $0.90^{**}$  &               & $-0.42$       \\
                  &               & $(0.29)$      &              & $(0.30)$     &               & $(0.25)$      \\
Feeling Jews      & $-0.01$       & $-0.02^{*}$   & $-0.00$      & $-0.00$      & $0.00$        & $0.00$        \\
                  & $(0.01)$      & $(0.01)$      & $(0.01)$     & $(0.01)$     & $(0.02)$      & $(0.02)$      \\
Age               & $0.03$        & $0.02$        & $0.05^{**}$  & $0.04^{*}$   & $0.04^{**}$   & $0.04^{**}$   \\
                  & $(0.02)$      & $(0.02)$      & $(0.01)$     & $(0.02)$     & $(0.01)$      & $(0.01)$      \\
Male              & $0.98$        & $1.50$        & $0.74$       & $1.23$       & $0.88$        & $0.86$        \\
                  & $(1.01)$      & $(0.97)$      & $(1.03)$     & $(1.10)$     & $(0.60)$      & $(0.63)$      \\
Non-white         & $-0.82$       & $-1.08$       & $-1.28$      & $-1.64$      & $1.47$        & $1.50$        \\
                  & $(0.99)$      & $(0.89)$      & $(0.88)$     & $(0.88)$     & $(0.81)$      & $(0.84)$      \\
Some college      & $0.69$        & $0.87$        & $0.48$       & $0.68$       & $1.57^{*}$    & $1.60^{*}$    \\
                  & $(0.98)$      & $(0.89)$      & $(0.95)$     & $(0.96)$     & $(0.68)$      & $(0.64)$      \\
Bachelor's degree & $1.97^{*}$    & $1.86$        & $1.76^{*}$   & $1.71$       & $2.45^{***}$  & $2.43^{***}$  \\
                  & $(0.98)$      & $(1.01)$      & $(0.83)$     & $(0.90)$     & $(0.71)$      & $(0.72)$      \\
Post-grad         & $-0.61$       & $-0.52$       & $-2.10^{*}$  & $-1.89^{*}$  & $2.74^{***}$  & $2.62^{***}$  \\
                  & $(1.01)$      & $(0.99)$      & $(0.95)$     & $(0.81)$     & $(0.69)$      & $(0.73)$      \\
Intercept         & $-3.62$       & $0.00$        & $-6.41^{*}$  & $-3.39^{*}$  & $0.96$        & $0.98$        \\
                  & $(2.21)$      & $(1.47)$      & $(3.09)$     & $(1.34)$     & $(1.98)$      & $(2.10)$      \\
\midrule
N         & $851$         & $851$         & $851$        & $851$        & $851$         & $851$         \\
\bottomrule
\multicolumn{7}{l}{\scriptsize{$^{***}p<0.001$; $^{**}p<0.01$; $^{*}p<0.05$}}
\end{tabular}
\end{center}
\footnotesize{Quasipoisson coefficients for correlates of time per week spent on videos from alternative, extremist, and mainstream media channels. Robust standard errors are in parentheses. All results incorporate survey weights.}
\end{table}

\begin{table}[!htbp]
  \caption{Correlates of visits to YouTube videos by channel type (separating hostile sexism and racial resentment) \label{tab:hh_rr_visit}} 
\fontsize{9}{10}\selectfont
\begin{center}
\begin{tabular}{@{\extracolsep{5pt}}lcccccc} 
\toprule
 & \multicolumn{6}{c}{\textit{Dependent variable: Views}} \\ 
\cline{2-7}
& \multicolumn{2}{c}{Alternative} & \multicolumn{2}{c}{Extremist} & \multicolumn{2}{c}{Mainstream} \\ 
& \multicolumn{2}{c}{channel videos} & \multicolumn{2}{c}{channel videos} & \multicolumn{2}{c}{channel videos} \\ 
& (1) & (2) & (3) & (4) & (5) & (6)\\ 
\midrule
Hostile sexism    & $1.74^{***}$ &               & $2.47^{***}$  &               & $0.04$        &              \\
                  & $(0.29)$     &               & $(0.69)$      &               & $(0.16)$      &              \\
Racial resentment &              & $1.11^{***}$  &               & $0.90^{**}$   &               & $-0.00$      \\
                  &              & $(0.32)$      &               & $(0.32)$      &               & $(0.19)$     \\
Feeling Jews      & $-0.01$      & $-0.01$       & $0.01$        & $0.01$        & $0.01$        & $0.01$       \\
                  & $(0.01)$     & $(0.01)$      & $(0.01)$      & $(0.02)$      & $(0.01)$      & $(0.01)$     \\
Age               & $0.02$       & $0.01$        & $0.05^{**}$   & $0.04^{*}$    & $0.01$        & $0.01$       \\
                  & $(0.02)$     & $(0.02)$      & $(0.02)$      & $(0.02)$      & $(0.01)$      & $(0.01)$     \\
Male              & $1.32$       & $1.77$        & $1.49$        & $1.84$        & $1.62^{**}$   & $1.64^{**}$  \\
                  & $(1.07)$     & $(0.97)$      & $(1.15)$      & $(1.16)$      & $(0.59)$      & $(0.58)$     \\
Non-white         & $-0.36$      & $-0.57$       & $-0.33$       & $-1.09$       & $2.17^{***}$  & $2.19^{***}$ \\
                  & $(1.05)$     & $(0.95)$      & $(0.92)$      & $(1.01)$      & $(0.60)$      & $(0.59)$     \\
Some college      & $-0.04$      & $0.22$        & $-0.38$       & $-0.25$       & $1.03$        & $1.04$       \\
                  & $(1.02)$     & $(0.86)$      & $(0.95)$      & $(1.05)$      & $(0.66)$      & $(0.66)$     \\
Bachelor's degree & $1.37$       & $1.31$        & $1.08$        & $1.04$        & $1.63^{**}$   & $1.61^{**}$  \\
                  & $(0.98)$     & $(1.01)$      & $(0.88)$      & $(1.05)$      & $(0.63)$      & $(0.63)$     \\
Post-grad         & $-0.62$      & $-0.38$       & $-2.10$       & $-1.54$       & $2.96^{***}$  & $2.95^{***}$ \\
                  & $(0.97)$     & $(0.86)$      & $(1.18)$      & $(0.91)$      & $(0.71)$      & $(0.75)$     \\
Intercept         & $-8.23^{**}$ & $-5.14^{***}$ & $-16.26^{**}$ & $-9.07^{***}$ & $-4.60^{***}$ & $-4.47^{**}$ \\
                  & $(2.66)$     & $(1.40)$      & $(4.99)$      & $(1.96)$      & $(1.37)$      & $(1.44)$     \\
\midrule
N        & $851$        & $851$        & $851$        & $851$       & $851$        & $851$        \\
\bottomrule
\multicolumn{7}{l}{\scriptsize{$^{***}p<0.001$; $^{**}p<0.01$; $^{*}p<0.05$}}
\end{tabular}
\end{center}
\footnotesize{Quasipoisson coefficients for correlates of visits per week on videos from alternative, extremist, and mainstream media channels. Robust standard errors are in parentheses. All results incorporate survey weights.}
\end{table}

\newpage
\clearpage

\begin{figure}[!htb]
\begin{center}
    \caption{Correlates of exposure to YouTube videos by channel type (alternate racial attitude measure) \label{fig:qpois_coefficients_FIRE}}    
    \includegraphics[width=.75\textwidth]{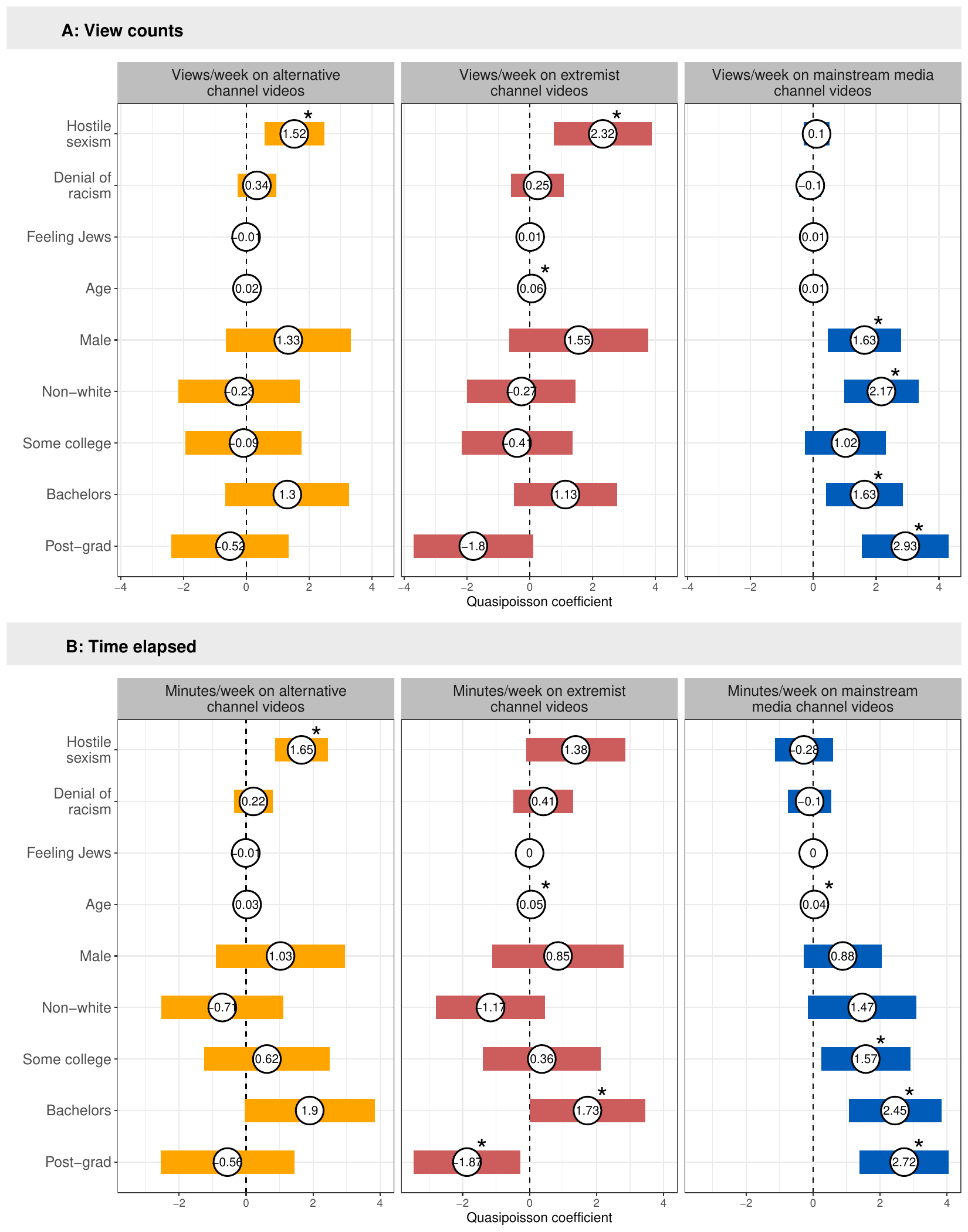}
    \end{center}
    Quasipoisson regression coefficients for correlates of the number of respondent views and time spent per week on videos from alternative, extremist, and mainstream media channels. Figure includes 95\% confidence intervals calculated from robust standard errors. All results incorporate survey weights. See \autoref{tab:figa8output} for regression table.
\end{figure}

Finally, we provide results in Figure \ref{fig:qpois_coefficients_FIRE} and \autoref{tab:figa8output} below in which we use survey respondents' prior responses to two questions measuring denial of institutional racism \citep{desante2020less} in the 2018 Cooperative Congressional Survey as an alternate measure of racial attitudes. Our findings are similar to those reported above using prior levels of racial resentment instead.

\begin{table}[!htbp] 
\caption{Correlates of exposure to YouTube videos by channel type (with alternative racial resentment) \label{tab:figa8output}} 
\fontsize{9}{10}\selectfont
\begin{center}
\begin{tabular}{@{\extracolsep{5pt}}lcccccc} 
\toprule
 & \multicolumn{3}{c}{\textit{Dependent variable: Views}} & \multicolumn{3}{c}{\textit{Dependent variable: Time elapsed}} \\
\cmidrule(r){2-4}\cmidrule(l){5-7}
& Alternative & Extremist & Mainstream & Alternative & Extremist & Mainstream \\ 
& (1) & (2) & (3) & (4) & (5) & (6)\\ 
\midrule
Hostile sexism    & $1.52^{**}$  & $2.32^{**}$   & $0.10$        & $1.65^{***}$  & $1.38$       & $-0.28$       \\
                  & $(0.49)$     & $(0.79)$      & $(0.21)$      & $(0.40)$      & $(0.75)$     & $(0.44)$      \\
Denial of racism  & $0.34$       & $0.25$        & $-0.10$       & $0.22$        & $0.41$       & $-0.10$       \\
                  & $(0.31)$     & $(0.43)$      & $(0.18)$      & $(0.29)$      & $(0.45)$     & $(0.33)$      \\
Feeling Jews      & $-0.01$      & $0.01$        & $0.01$        & $-0.01$       & $-0.00$      & $0.00$        \\
                  & $(0.01)$     & $(0.01)$      & $(0.01)$      & $(0.01)$      & $(0.01)$     & $(0.02)$      \\
Age               & $0.02$       & $0.06^{**}$   & $0.01$        & $0.03$        & $0.05^{***}$ & $0.04^{**}$   \\
                  & $(0.02)$     & $(0.02)$      & $(0.01)$      & $(0.02)$      & $(0.01)$     & $(0.01)$      \\
Male              & $1.33$       & $1.55$        & $1.63^{**}$   & $1.03$        & $0.85$       & $0.88$        \\
                  & $(1.01)$     & $(1.13)$      & $(0.59)$      & $(0.98)$      & $(1.00)$     & $(0.59)$      \\
Non-white         & $-0.23$      & $-0.27$       & $2.17^{***}$  & $-0.71$       & $-1.17$      & $1.47$        \\
                  & $(0.99)$     & $(0.88)$      & $(0.60)$      & $(0.93)$      & $(0.83)$     & $(0.82)$      \\
Some college      & $-0.09$      & $-0.41$       & $1.02$        & $0.62$        & $0.36$       & $1.57^{*}$    \\
                  & $(0.94)$     & $(0.90)$      & $(0.66)$      & $(0.96)$      & $(0.89)$     & $(0.68)$      \\
Bachelor's degree & $1.30$       & $1.13$        & $1.63^{**}$   & $1.90$        & $1.73^{*}$   & $2.45^{***}$  \\
                  & $(1.01)$     & $(0.84)$      & $(0.63)$      & $(0.99)$      & $(0.88)$     & $(0.71)$      \\
Post-grad         & $-0.52$      & $-1.80$       & $2.93^{***}$  & $-0.56$       & $-1.87^{*}$  & $2.72^{***}$  \\
                  & $(0.95)$     & $(0.97)$      & $(0.70)$      & $(1.02)$      & $(0.81)$     & $(0.68)$      \\
Intercept         & $-8.46^{**}$ & $-16.72^{**}$ & $-4.58^{***}$ & $-3.76$       & $-6.94^{*}$  & $0.98$        \\
                  & $(2.73)$     & $(5.33)$      & $(1.38)$      & $(2.27)$      & $(3.22)$     & $(2.04)$      \\
\midrule
N         & $851$        & $851$         & $851$         & $851$         & $851$        & $851$         \\
\bottomrule
\multicolumn{7}{l}{\scriptsize{$^{***}p<0.001$; $^{**}p<0.01$; $^{*}p<0.05$}}
\end{tabular}
\end{center}
\footnotesize{Quasipoisson coefficients for correlates of of views and time per week spent on videos from alternative, extremist, and mainstream media channels. Robust standard errors are in parentheses.}
\end{table}

\clearpage
\newpage
\subsection*{Browser extension validation}\label{app:pulse}

Browser activity statistics are reported throughout the paper. Below, we evaluate the validity of browser activity by comparing it to browser history data. %
    The browser extension also recorded participants' browser history (URLs with timestamps that are recorded each time a participant loads a new web page). For comparability, we limit browser history data to the period for which both browser history and activity data are available.

\begin{figure}[!htb]
\begin{center}
    \caption{Correlation between browser history and activity \label{fig:corr_history_activity}}    
    \includegraphics[width=.7\textwidth]{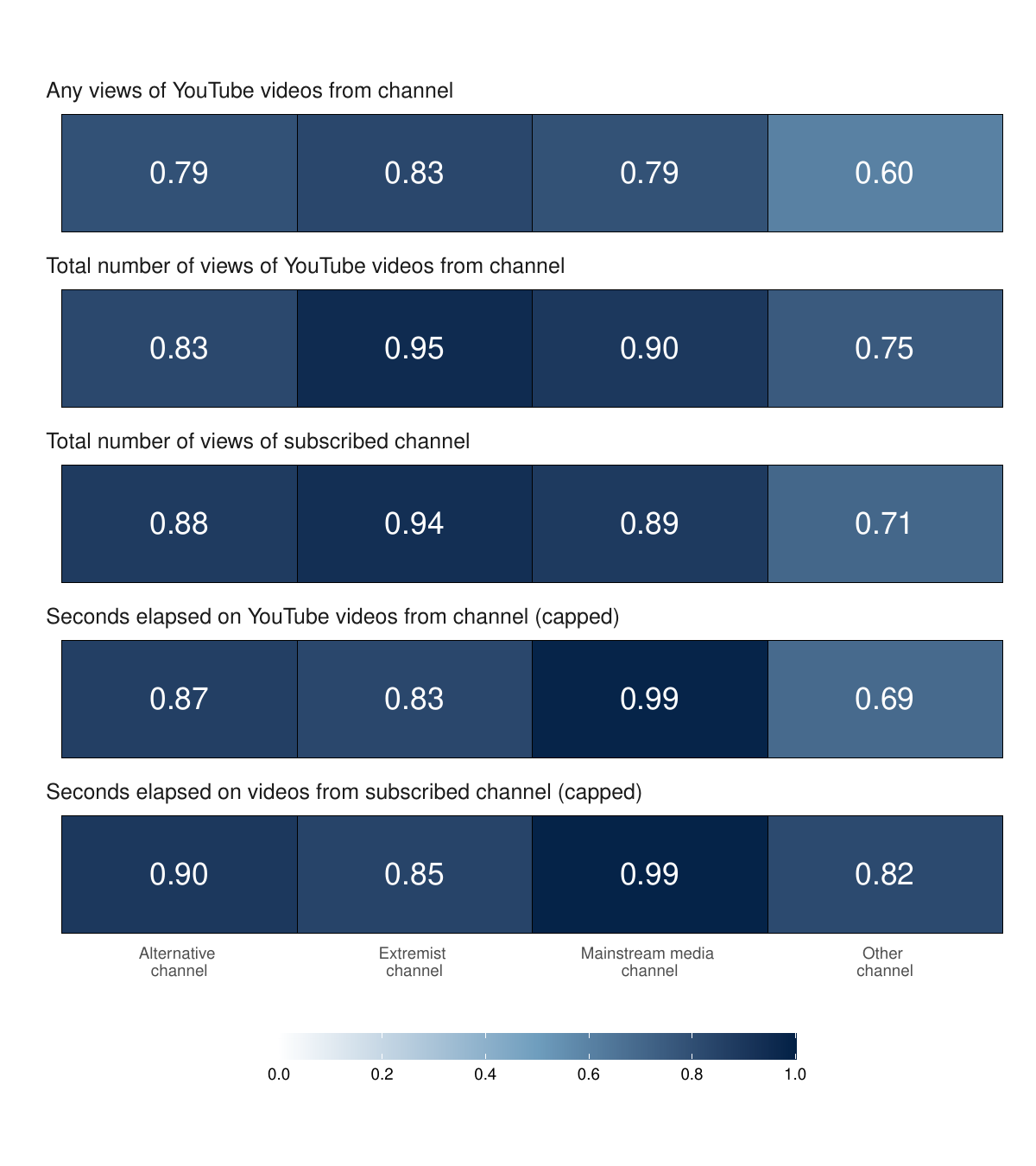}
    \end{center}
    All results incorporate survey weights.
\end{figure}

\autoref{fig:corr_history_activity} shows the Pearson correlation coefficients between browser history and activity data across five variables for alternative, extremist, mainstream media, and other YouTube channels: a binary measure of viewing any video from that type of channel, the total number of views of videos from that type of channel, the total number of views of videos from subscribed channels of that type, the number of seconds elapsed on all YouTube videos from channels of that type, and the number of seconds elapsed on all YouTube videos from channels of that type. Correlations range from $r=0.60$ to $0.99$ and are consistently high for alternative channel videos ($0.79 \leq r \leq 0.90$) and extremist channel videos ($0.83 \leq r \leq 0.95$).

\clearpage
\newpage
\subsection*{Differential browsing behavior after install}\label{app:diff_install_behave}

As shown in \autoref{tab:its} below, we find no discernible change in the proportion of time that participants spent on alternative or extremist channels after installing the extension. We performed this analysis to verify that participants did not modify their web browsing behavior after installation, an important consideration in validating our measurement approach. Leveraging browser history data, which captures three months of web activity prior to the installation of the extension, we test if the proportion of time participants spend on alternative and extremist channels changes after installation in levels or slopes. Using OLS with robust standard errors clustered by participant, we estimate the two-way fixed effects model in \autoref{equ:ITS}  where $\alpha_i$ is a participant-level fixed effect (for each $i$ = 1, ..., 1098), $\gamma_t$ is a day-level fixed effect (for $t$ = Apr. 22, 2020, ..., Dec. 31, 2020), and $\text{\textit{Installed}}_{i,t}$ is a binary variable testing whether the mean proportion of time participants spend on alternative and extremist channels changes after installation. We also estimate the model in \autoref{equ:ITS2} which adds the term $\text{\textit{Days after install}}_{i,t}$ to test for a linear time trend in alternative and extremist channel viewership after installation. The dependent variable in both models is the proportion of seconds spent on either alternative or extremist channel videos per day.

\begin{equation}\label{equ:ITS}
    Y_{i,t} = \alpha_i + \gamma_t + \beta_1 \, \text{\textit{Installed}}_{i,t} + \epsilon_{i,t}
\end{equation}

\begin{equation}\label{equ:ITS2}    
    Y_{i,t} = \alpha_i + \gamma_t + \beta_1 \, \text{\textit{Installed}}_{i,t}  + \beta_2 \, \text{\textit{Days after install}}_{i,t} + \epsilon_{i,t}
\end{equation}

\begin{table}[!htbp]  
  \caption{Predictors of proportion of time spent on alternative/extremist videos by day \label{tab:its}} 
\begin{center}
\begin{tabular}{@{\extracolsep{5pt}}lcc} 
\toprule
 & (1) & (2)\\
\midrule
 Installed & 0.00660 & 0.00627 \\ 
  & (0.00778)  & (0.00789) \\ 
  && \\ 
 Days after install & & -0.00013 \\ 
  && (0.00210) \\ 
  && \\ 
\midrule 
Day fixed effects & $\checkmark$ & $\checkmark$\\
User fixed effects & $\checkmark$ & $\checkmark$\\
\midrule  
N & 63,216 & 63,216\\ 
\bottomrule
\end{tabular} 
\end{center}
\footnotesize{OLS model results with robust standard errors clustered by participant in parentheses. Estimates include survey weights. $^{*}$p$<$0.1; $^{**}$p$<$0.05; $^{***}$p$<$0.01.}
\end{table}

\clearpage
\newpage
\subsection*{Attitudes toward YouTube}\label{app:youtube_attitudes}

\begin{figure}[!htb]
\begin{center}
    \caption{Differences in perceptions of YouTube between full sample and extension sample \label{fig:youtube_attitudes}}    
    \includegraphics[width=.99\textwidth]{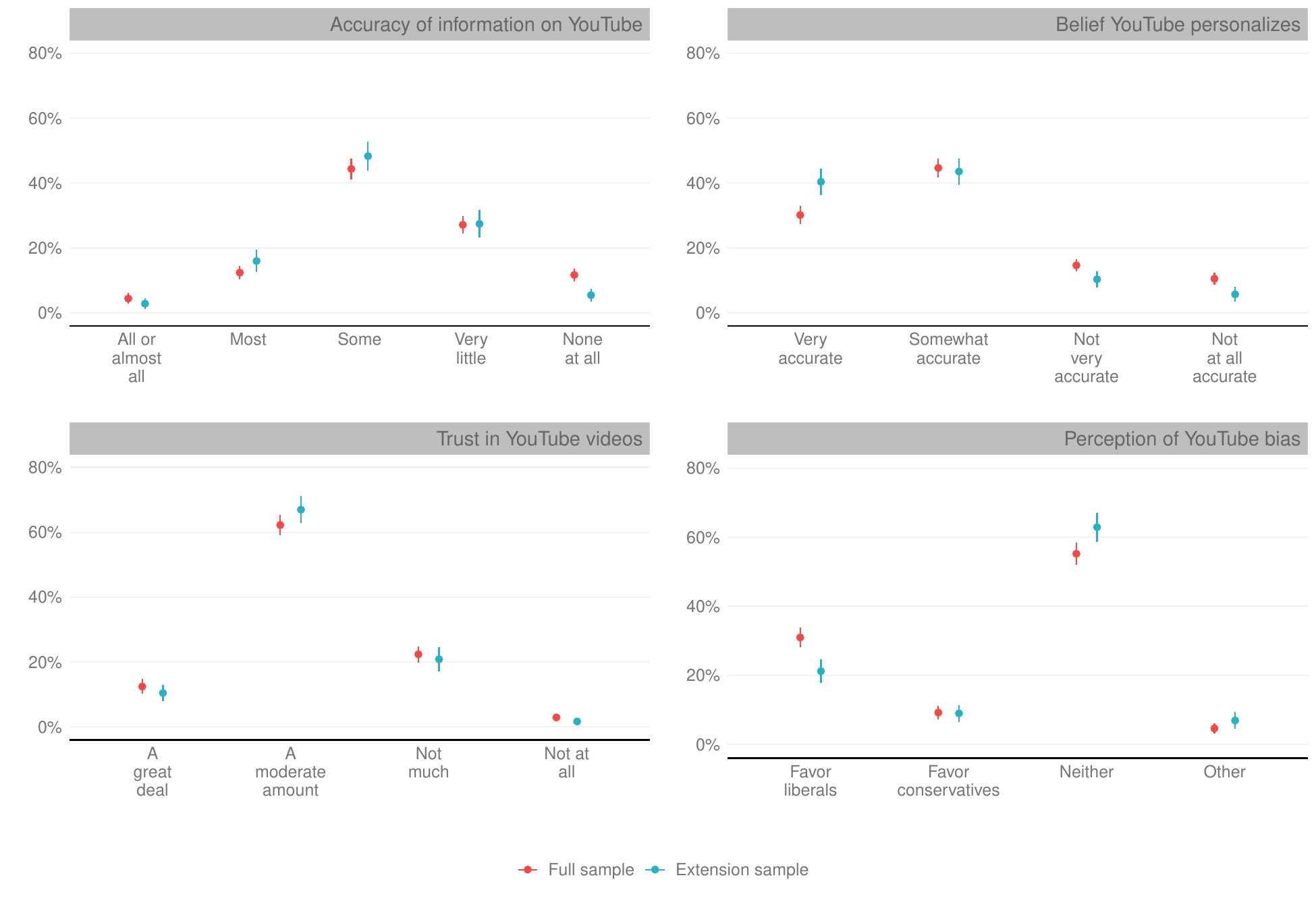}
    \end{center}
    All results incorporate survey weights.
\end{figure}

\clearpage
\newpage
\section*{Session trajectories}

We provide three examples of participant viewing paths that led to extremist channel videos in a manner consistent with the rabbit hole narrative below:
\begin{itemize}
    \item A participant conducted a search for an alternative channel's name (Dinesh D'Souza), viewed a video from that channel, and then followed a recommendation to an extremist channel video (PragerU).
    \item In another session, a participant visited the YouTube homepage, viewed a video from an ``other'' channel (English Heritage), then viewed a video from the alternative channel Carpe Donktum titled ``Stop The Steal.us,'' and then followed a recommendation to a video from the extremist channel Styxhexenhammer666 titled ``MSM Hopes You'll Just Accept the Election Despite Outstanding Evidence of Fraud.'' Following that, the participant viewed a video from an ``other'' channel that is now private titled ``Target Smart Early Voting Data Gives President Trump the Eventual Victory After Recounts.''
    \item A participant viewed an other channel video (WIRED; ``Every Race In Middle-Earth Explained'') and then followed a recommendation to an extremist channel video (Survive the Jive, ``Ancient History of Ireland, Newgrange, Celts, Vikings'').
\end{itemize}

To test for ``rabbit hole''-style patterns of exposure, we also consider whether YouTube users are more likely to encounter potentially harmful content in longer sessions \cite{hosseinmardi2020evaluating}. We construct sessions by separating a sorted timeline of respondents' YouTube activity at each point at which they (1) dwell on a non-video URL (e.g., the YouTube homepage) for greater than 10 minutes, (2) spend longer than the duration of the video in question plus 30 minutes before interacting with the page, or (3) spend longer than four hours on a video. We call the number of YouTube videos between these breakpoints a session and define each session by its length (number of videos viewed). 

First, we note several descriptive findings about YouTube sessions. They are relatively numerous---the median number of sessions for a participant is 19.4 during the study period---and frequently short. In total, 18.6\% of sessions on YouTube do not include a video view, 15\% are singletons in which respondents view just one video, and 42.1\% include 2--10 videos. Just 24.3\% of sessions have length 11 or longer. However, due to skewness in the distribution of YouTube consumption by session length, 77\% of videos are watched in these sessions of length eleven or greater.
 
\begin{figure}[!htb]
    \begin{center}
    \caption{Percentage of views to each channel type by video number within session \label{fig:within_sessions}}    
    \includegraphics[width=.99\textwidth]{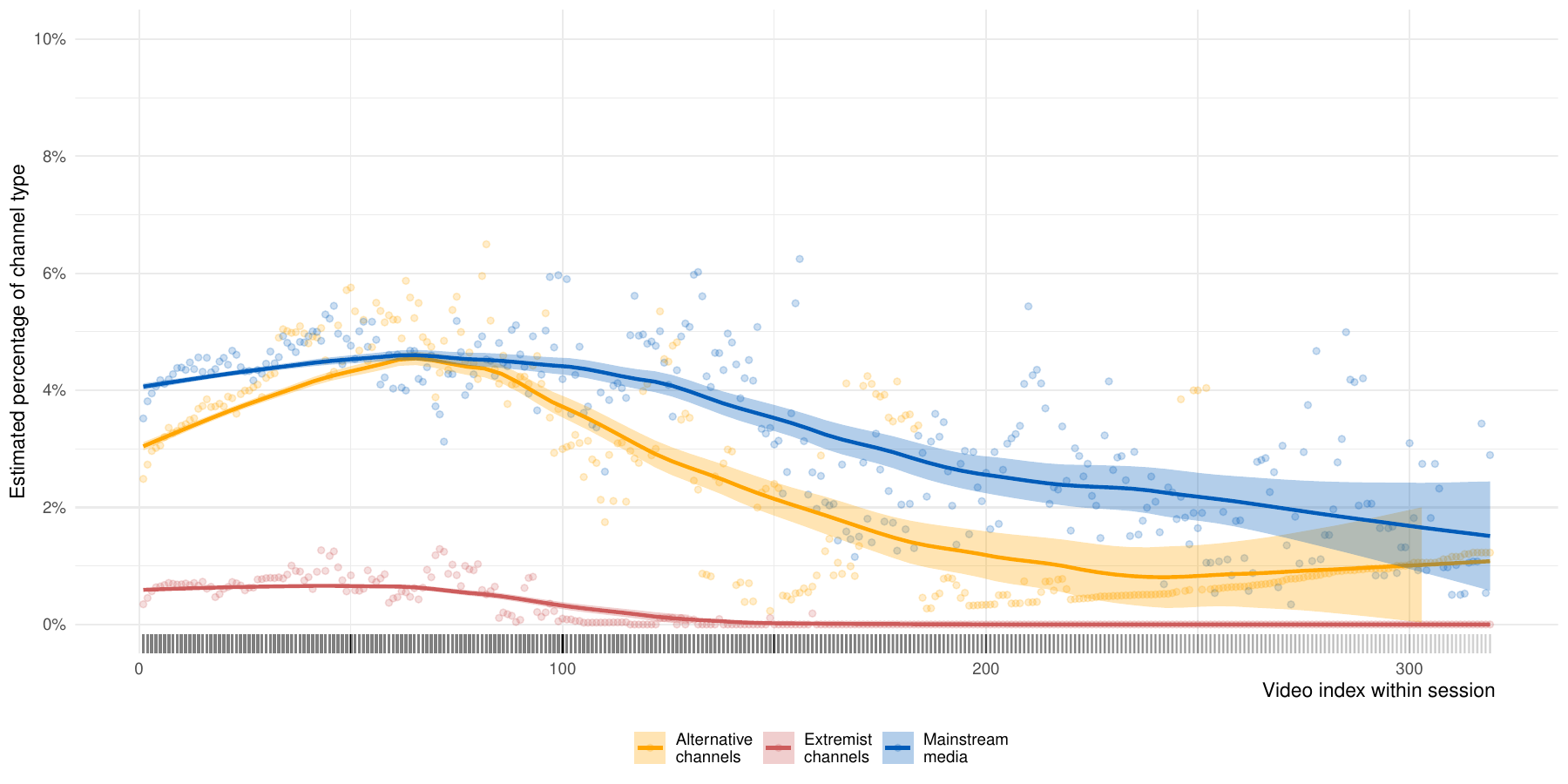}
    \end{center}
    Each point represents the average percentage of videos from a channel type at a given session length. Lines are loess curves fit with a linear function and a 0.5 span. All results incorporate survey weights.
\end{figure}

\autoref{fig:within_sessions} considers how the probability of viewing an alternative or extremist channel video varies by the point in a session over sessions of length 1--319 (the range of lengths that capture 99\% of the sessions in our data). Each point in the graph represents the estimated probability of viewing a particular type of video at a particular session length. We find no clear evidence that the probability of viewing an alternative or extremist channel video increases as sessions lengthen; the probabilities are generally stable. The equivalent probability for mainstream media channel videos, which we provide for comparison, is also relatively stable.

\begin{figure}[!htb]
    \begin{center}
    \caption{Percentage of views to each channel type by total session length \label{fig:extension_sessions_length_99}}
    \includegraphics[width=.99\textwidth]{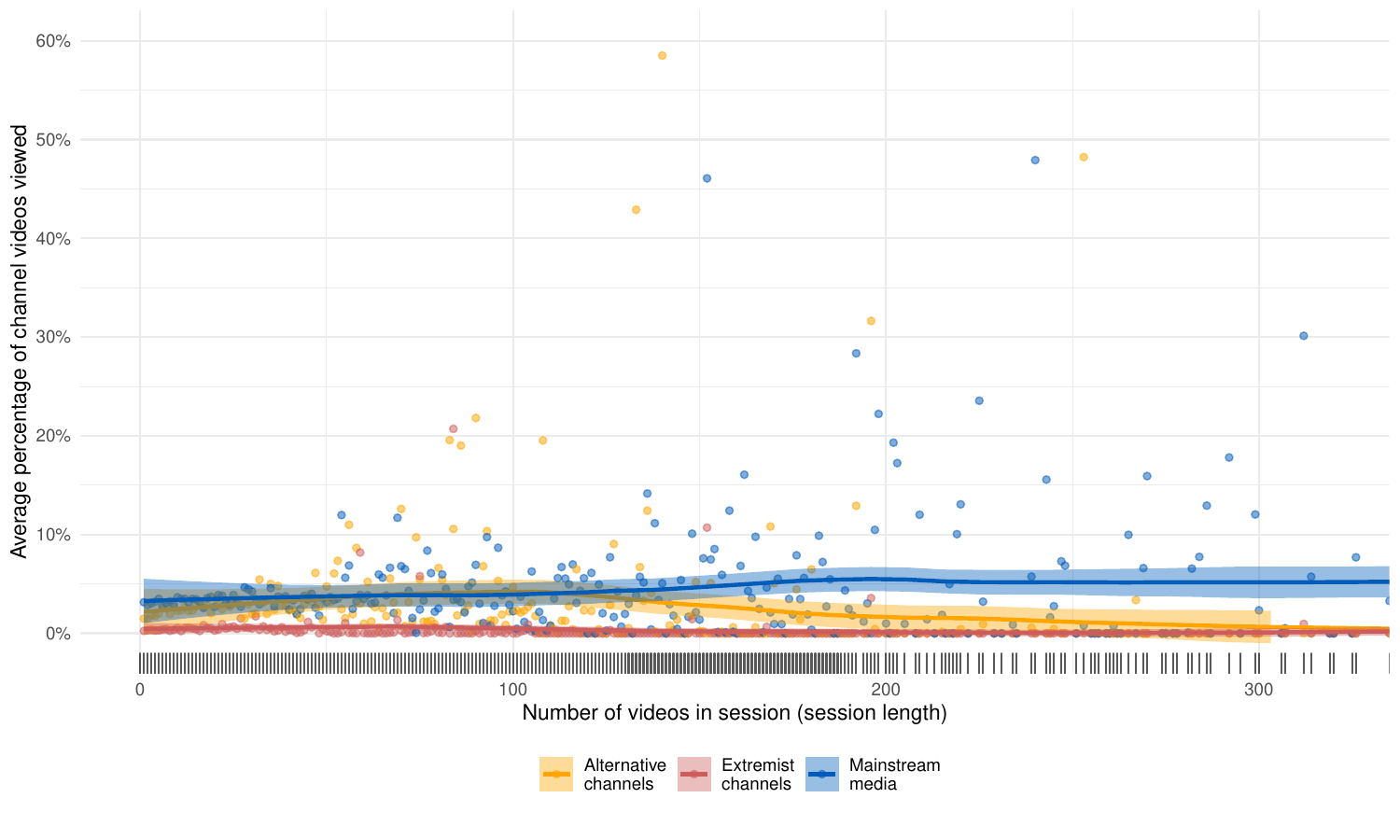}
    \end{center}
    Each point represents the average percentage of videos from a channel type of all videos viewed in sessions of a fixed session length. Lines are loess curves fit with a linear function and a 0.5 span. All results incorporate survey weights.
\end{figure}

\begin{figure}[!hb]
 \begin{center}
    \caption{Concentration of exposure to alternative and extremist channels (view counts)}
    \label{fig:concentration_visit}
    \includegraphics[width = .99\textwidth]{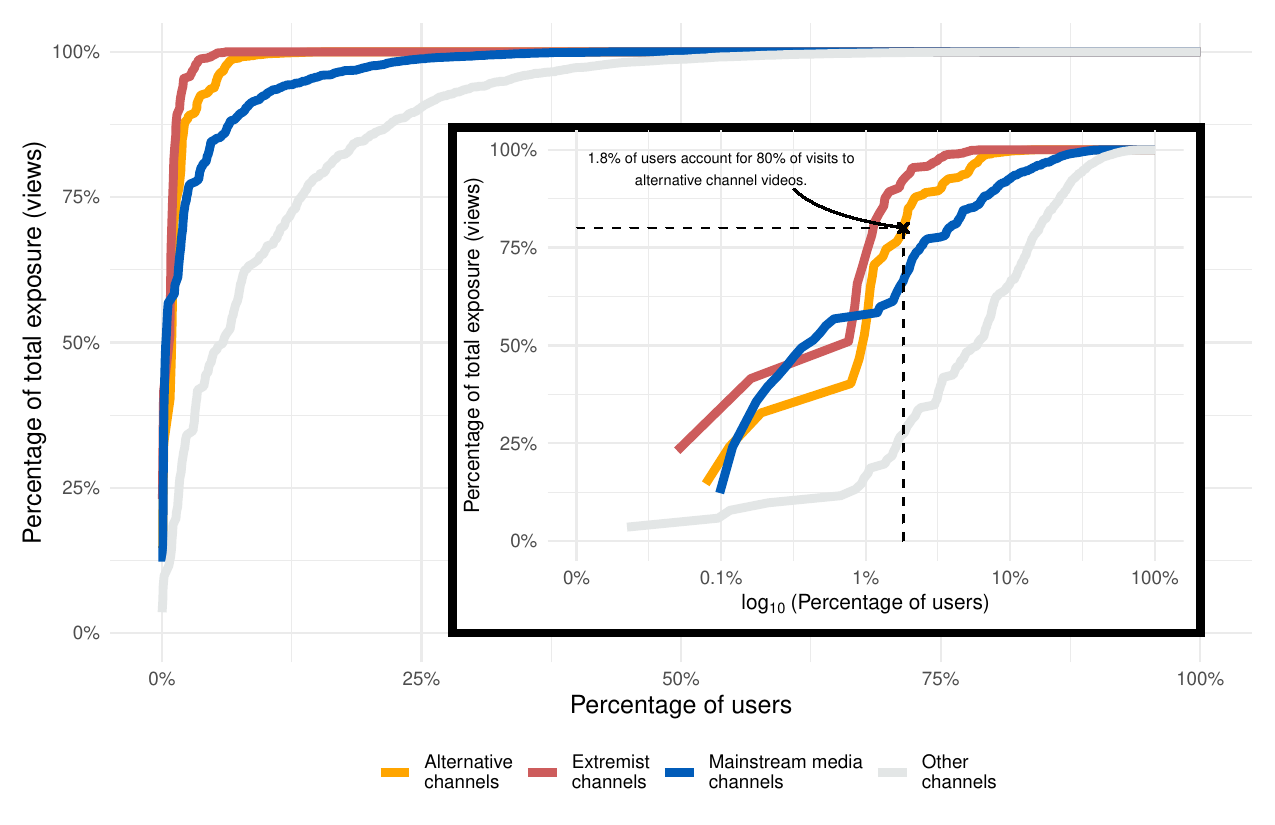}
   \end{center}
    \begin{footnotesize}
    \noindent Weighted empirical cumulative distribution function showing the percentage of participants responsible for a given level of total observed video viewership of alternative and extremist channels on YouTube (by view count). All results incorporate survey weights.
    \end{footnotesize}    
\end{figure}

\begin{figure}[!hb]
 \begin{center}
    \caption{Recommendations seen by rank conditional on video channel type}
    \label{fig:recsrank}
    \includegraphics[width = .99\textwidth]{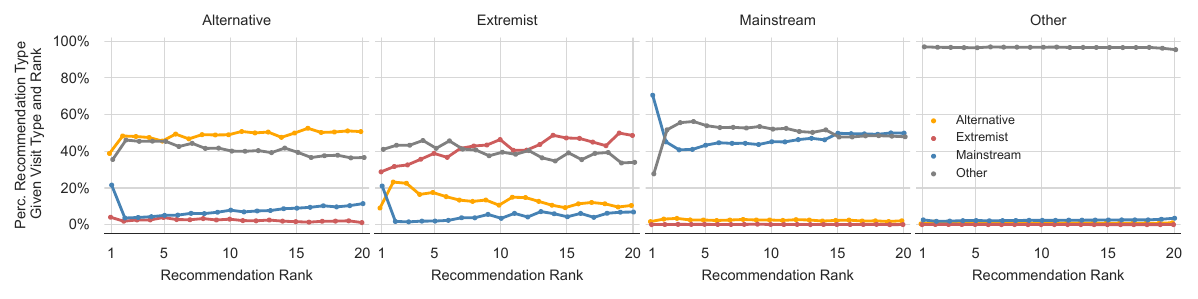}
   \end{center}
    \begin{footnotesize}
    \noindent Video type recommended by rank when visiting a video of the channel type named at the top of the panel. The results incorporate survey weights. 
    \end{footnotesize}    
\end{figure}

\begin{figure}[!hb]
 \begin{center}
    \caption{Recommendations followed by rank conditional on video channel type}
    \label{fig:followsrank}
    \includegraphics[width = .99\textwidth]{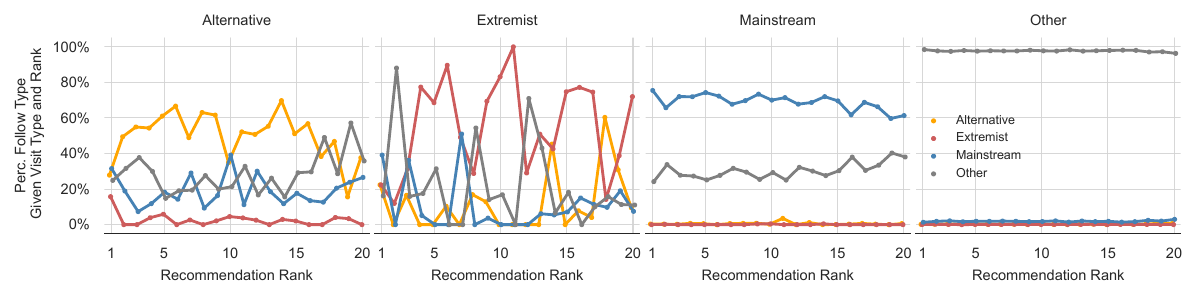}
   \end{center}
    \begin{footnotesize}
    \noindent Video type recommendation follows by rank when visiting a video of the channel type named at the top of the panel. The results incorporate survey weights.  
    \end{footnotesize}    
\end{figure}

\autoref{fig:extension_sessions_length_99} instead examines whether the \emph{total} proportion of videos watched from alternative and extremist channels by session is greater in longer sessions. A point represents the percentage of videos of a particular type that were watched in sessions of a given total length. We find no evidence that longer sessions have higher proportions of alternative or extremist channel videos.

\newpage
\clearpage
\subsection*{External referrers}

\begin{table}[!htbp]
\caption{\label{tab:externalrefs}External referrers to alternative and extremist channel videos}
\centering
\begin{tabular}[t]{llrr}
\toprule
Referrer type & Preceding domain & \% to extremist channel & \%  to alternative channel\\
\midrule
Alternative social & 4chan.org & 0.000 & 0.000\\
 & banned.video & 0.000 & 0.008\\
 & parler.com & 0.159 & 0.556\\
 & gab.com & 0.384 & 0.476\\
 & boards.4chan.org & 1.524 & 1.400\\
 & boards.4channel.org & 5.007 & 2.116\\
 & twitchy.com & 14.154 & 0.479\\
\addlinespace
Mainstream social & bumble.com & 0.000 & 0.035\\
 & discord.com & 0.000 & 0.050\\
 & pinterest.com & 0.000 & 0.017\\
 & tumblr.com & 0.000 & 0.155\\
 & twitch.tv & 0.000 & 1.040\\
 & tinder.com & 0.030 & 0.245\\
 & apps.facebook.com & 0.160 & 0.345\\
 & instagram.com & 0.238 & 1.045\\
 & messenger.com & 0.506 & 1.166\\
 & linkedin.com & 0.515 & 0.069\\
 & reddit.com & 0.760 & 3.555\\
 & old.reddit.com & 2.861 & 3.497\\
 & facebook.com & 6.394 & 8.527\\
 & twitter.com & 12.095 & 14.975\\
\addlinespace
Search engine social & search.yahoo.com & 0.000 & 0.050\\
 & yahoo.com & 0.085 & 0.076\\
 & duckduckgo.com & 0.402 & 0.823\\
 & bing.com & 0.948 & 0.700\\
 & google.com & 8.237 & 8.033\\
\addlinespace
Webmail & mail.com & 0.000 & 0.064\\
 & outlook.office.com & 0.000 & 0.010\\
 & outlook.office365.com & 0.000 & 0.014\\
 & mail.aol.com & 0.088 & 0.217\\
 & outlook.live.com & 0.125 & 0.285\\
 & mail.yahoo.com & 0.863 & 1.444\\
 & mail.google.com & 2.289 & 3.926\\
\bottomrule
\end{tabular}
All results incorporate survey weights.
\end{table}

\newpage
\clearpage
\subsection*{Exposure concentration by views}

\newpage
\clearpage
\subsection*{Recommendations seen and followed by rank}

\newpage
\clearpage
\section*{Channel labeling criteria}

In this appendix, we aggregate the methods used by the authors of prior work to identify and label specific YouTube channels. 

\subsection*{Ribeiro et al. \cite{ribeiro2020auditing}}

Ribeiro et al. \cite{ribeiro2020auditing} used the following process to identify a set of channels:
\begin{quote}
(1) We choose a set of seed channels. Seeds were extracted from the I.D.W. unofficial website [7], Anti Defamation League's report on the Alt-lite/the Alt-right [3] and Data \& Society's report on YouTube Radicalization [24]. We pick popular channels that are representative of the community we are interested in. Each seed was independently annotated two times and discarded in case there was any disagreement.

(2) We choose a set of keywords related to the sub-communities. For each keyword, we use YouTube's search functionality and consider the first 200 results in English. We then add channels that broadly relate in topic to the community in question. For example, for the Alt-right, keywords included both terms associated with their narratives, such as The Jewish Question and White Genocide, as well as the names or nicknames of famous Alt-righters, such as weev and Christopher Cantwell.

(3) We iteratively search the related and featured channels collected in steps (1) and (2), adding relevant channels (as defined in 2). Note that these are two ways channel can link to each other. Featured channels may be chosen by YouTube content creators: if your friend has a channel and you want to support it, you can put it on your "Featured Channels" tab. Related channels are created by YouTube's recommender system.

(4) We repeat step (3), iteratively collecting another hop of featured/recommended channels from those obtained in (3). The annotation process done here followed the same instructions as the one explained in detail for data collection step (c). Steps (2)--(4), were done by a co-author with more than 50 hours of watch-time of the communities of interest. Notice that, in steps (2)--(4), we are not labeling the channels, but creating a pool of channels to be further inspected and labeled in subsequent steps. The complete list of seeds obtained from (1) and of keywords used in (2) may be found in Appendix A. A clear distinction between featured and recommended channels may be found in Appendix B.
\end{quote}

\noindent Ribeiro et al. used the following process to label and validate channels.
\begin{quote}
(c) Channel labeling was done in multiple steps. All channels are either seeds (Type 1) or obtained through YouTube's recommendation/search engine (Types 2 and 3). Notice that Type 1 channels were assigned labels at the time of their collection. For the others, we had 2 of the authors annotate them carefully. They both had significant experience with the communities being studied, and were given the following instructions:

\begin{quote}
Carefully inspect each one of the channels in this table, taking a look at the most popular videos, and watching, altogether, at least 5 minutes of content from that channel. Then you should decide if the channel belongs to the Alt-right, the Alt-lite, the Intellectual Dark Web (I.D.W.), or whether you think it doesn't fit any of the communities. To get a grasp on who belongs to the I.D.W., read [42], and check out the website with some of the alleged members of the group [7]. Yet, we ask you to consider the label holistically, including channels that have content from these creators and with a similar spirit to also belong in this category. To distinguish between the Alt-right and the Alt-lite, read [3] and [28]. It is important to stress the difference between civic nationalism and racial nationalism in that case. Please consider the Alt-right label only to the most extreme content. You are encouraged to search on the internet for the name of the content creator to help you make your decision.
\end{quote}

The annotation process lasted for 3 weeks. In case they disagreed, they had to discuss the cases individually until a conclusion was reached. Interanotator agreement was of 75.57%
\end{quote}

\subsection*{Ledwich and Zaitsev \cite{ledwich2019algorithmic}}

Ledwich and Zaitsev \cite{ledwich2019algorithmic} explain how they labeled YouTube channels:
\begin{quote}
The tagging process allowed each channel to be characterized by a maximum of four different tags to create meaningful and fair categories for the content. In addition to labeling created by the two authors, we recruited an additional volunteer labeler, who was well versed in the YouTube political sphere, and whom we trusted to label channels by their existing content accurately. When two or more labelers defined a channel by the same label, that label was assigned to the channel. When the labelers disagreed and ended in a draw situation, the tag was not assigned. The majority was needed for a tag to be applied.

\ldots

To assign a label, we investigated which topics the channels discussed and from which perspective...The only way to conduct this labeling was to watch the content on the channels until the labelers found enough evidence for assigning specific labels. For some channels, this was relatively straightforward: the channels had introductory videos that stated their political perspectives \ldots In other cases, the labelers could not assign a label based on introduction or description but had to watch several videos on the channel to determine the political leanings. On average, every labeler watched over 60 hours of YouTube videos to define the political leanings without miscategorizing the channel and thus misrepresenting the views of the content creators.
\end{quote}

\noindent In their study, they label the following types of channels using the quoted criteria.
\begin{itemize}
    \item \textbf{Anti-SJW}: ``Channel has to have a significant focus on criticizing ``Social Justice'' (see next category) with a positive view of the marketplace of ideas and discussing controversial topics. To tag a channel, this should be a common focus in their content.'' Raters had 74\% agreement on channels of this type.
    \item \textbf{MRA}: ``Focus on advocating for rights for men. See men as the oppressed sex and will focus on examples where men are currently oppressed. Incels, who identify as victims of sex inequality, would also be included in this category.'' Raters had 97\% agreement on channels of this type.
    \item \textbf{White Identitarian}: ``Identifies-with/is-proud-of the superiority of ``whites'' and Western Civilization.'' Raters had 94\% agreement on channels of this type.
\end{itemize}

\subsection*{Lewis \cite{lewis18}}

Lewis  \cite{lewis18} describes the following process for identifying and validating channels:
\begin{quote}
To understand the AIN in-depth, I analyzed both the content of YouTube influencers (that is, what they are saying) as well as their collaborations (who they are broadcasting with). The latter presented a significant research challenge, as YouTube does not provide metadata about guest appearances. To get around this, I manually collected data from each influencer's video titles, and at times, video content, to determine each of the guests they hosted in their content between January 1, 2017 and April 1, 2018. I found new influencers through a snowball approach: for each guest on an influencer's channel, I would visit their own channel (if one existed) to see who they, in turn, hosted.

Overall, I collected data for approximately 65 influencers across 81 channels \ldots I watched content from each of these channels and performed an in-depth content analysis on the transcripts for two of them. Overall, I watched hundreds of hours of content from these 65 content creators.

At the time of data collection, this group of influencers was as close as I could get to a snapshot of the Alternative Influence Network. However, the boundaries of this network are loose and constantly changing. Since the time of my data collection, newly popular influencers have begun to collaborate with others in the network, and some of those I tracked in April have since deleted their channels or removed their content. The data also does not represent the full extent of networking and collaboration that occurs between influencers. Many of them, for example, comment on each other's videos; they reference each other's ideas in their content; and they interact on platforms like Twitter and Instagram in addition to YouTube. In other words, the data I collected is illustrative, not comprehensive.
\end{quote}

\subsection*{Charles \cite{charles2020main}}

Charles \cite{charles2020main} describes the following process for identifying and labeling channels:
\begin{quote}
The first step was to identify a network of channels containing white supremacist content on YouTube, and then to analyze a representative sample of the themes, rhetoric, messaging, presentation in the videos uploaded to those channels. In the first stage, I gathered channels via user interface snowball sampling, using the `related channel' feature on each channel---as well as any cross-channel appearances by content creators. Channels were tagged and categorized, then ranked by subscriber count within those categories.

\ldots

The first stage of this study used a modified style of snowball sampling, called user interface snowball sampling (UISS), to build a repository of YouTube channels for stage two's analysis \ldots Rather than using recommendations from gatekeepers, this study uses the `related channels' bar to find similar channels, as well as channels whose content creators appear in the videos of that channel. As more channels were found, I stopped periodically to analyze each channel for white supremacist themes (see Table 1). In order to be considered for analysis, the channel had to include at least one of the themes from Table 1.

The initial categorization was performed using six sampled videos: the two most viewed, the two most recently uploaded, and two randomly selected from the hundred most recent uploads (using a random number generator). This approach aimed to represent the nature of the content on that channel, determining whether it contains any of the white supremacist themes described in the literature. Channel samples that did not contain any of these themes were excluded from analysis and their related channels were not snowballed. The process was repeated until the point of data saturation (Schensul \& LeCompte, 2010). This was apparent by generation four when already-sampled channels began to dominate the related channels sections and when the few, new channels were so low in subscribers that they would not make the final cut in stage two.

\ldots

[T]he study started with avowed white nationalist Richard Spencer's YouTube channel, AltRight.com and proceeded from there, using YouTube's related channel feature and cross-channel appearances to approximate the size and composition of white supremacist communities on YouTube. 
\end{quote}

\noindent Charles used the following themes to identify channels (drawn from Table 1 in \cite{charles2020main}): Neo-Nazi, Nationalism, Genocide, Christian Identity/Racist Asatru, Opposition to Interracial Marriage, White Pro-nationalism, Islamophobia, Anti-Feminism, Non-white Criminality, Anti-Immigrant, White Supremacy, Anti-Semitism, Conspiracies, Apocalypticism.

\subsection*{Aaron Sankin \cite{sankin-2019-gizmodo}}

Journalist Aaron Sankin \cite{sankin-2019-gizmodo} describes the following process for curating and validating a list of extremist channels:
\begin{quote}
[W]e used lists of organizations promoting hate from the Southern Poverty Law Center, Hope Not Hate, the Canadian Anti-Hate Network, and the Counter Extremism Project, in addition to channels recommended on the white supremacist forum Stormfront, to create a compendium of 226 extremist YouTube channels earlier this year.

While less than scientific (and suffering from a definite selection bias), this list of channels provided a hazy window to watch what YouTube’s promises to counteract hate looked like in practice. And since June 5th, just 31 channels from our list of more than 200 have been terminated for hate speech. (Eight others were either banned before this date or went offline for unspecified reasons.)

Before publishing this story, we shared our list with Google, which told us almost 60 percent of the channels on it have had at least one video removed, with more than 3,000 individual videos removed from them in total. The company also emphasized it was still ramping up enforcement. These numbers, however, suggest YouTube is aware of many of the hate speech issues concerning the remaining 187 channels---and has allowed them to stay active.
\end{quote}

\newpage
\clearpage
\section*{Ethics and consent language}

\subsection*{Survey informed consent}
This research project is being conducted by Andrew Guess from Princeton University, Brendan Nyhan from Dartmouth College, and Christo Wilson from Northeastern University. It is a study to learn more about public opinion on issues in the news. Your participation is voluntary. Participation involves completion of a short survey as well as the option to participate in additional components of the study that would collect confidential data on your online behavior. This would entail confidential tracking data of your online website visits which you have already agreed to as part of your YouGov Pulse participation, and could include up to 1 year of data already collected prior to this survey. You may choose to not answer any or all questions and to not participate in any portion of the study that you choose. The researchers will not store information that could identify you with your survey responses. Identifying information will not be used in any presentation or publication written about this project. You must be age 18 or older to participate. Questions about this project may be directed to Brendan Nyhan, Professor of Government, at Brendan.J.Nyhan@dartmouth.edu.\\
 
\noindent If you agree to participate in this survey, click ``I agree'' below.\\
-I agree to participate\\
-I do not agree to participate\\

\subsection*{Browser extension informed consent (invitation)}
This extension implements a user study being conducted by researchers at Northeastern University, Dartmouth, Princeton, and University of Exeter. If you choose to participate, this browser extension will confidentially collect four types of data from your browser.\\

\noindent 1.	Metadata for web browsing (e.g. URL visited with time of visit), exposure to embedded URLs on websites (e.g. YouTube videos), and interactions with websites (e.g. clicks and video viewing time). This data is collected until the study is completed.\\

\noindent 2.	Copies of the HTML seen on specific sites: Google Search, Google News, YouTube, Facebook Newsfeed, and Twitter Feed. We remove all identifying information before it leaves the browser. This confidential data is collected until the study is completed.\\

\noindent 3.	Browsing history, Google and YouTube account histories (e.g. searches, comments, clicks), and online advertising preferences (Google, Bluekai, Facebook). This data is initially collected for the year prior to the installation of our browser extension, and we then check these sources once every two weeks to collect updates until the study is completed.\\

\noindent 4.	Snapshots of selected URLs from your browser. For each URL, the extension saves a copy of the HTML that renders, effectively capturing what you would have seen had you visited that website yourself. Once per week we conduct searches on Google Search, Google News, Youtube, and Twitter, and collect the current frontpage of Google News, YouTube, and Twitter. These web page visits will occur in the background and will not affect the normal functioning of your browser.\\

\noindent Additionally, if you choose to participate, you will be asked to take a survey in which we ask you several questions about your demographics, web usage, and media preferences. These data, as well as those mentioned above, will be used to analyze the correlations between your online behavior and your interest profiles.\\

\noindent After the study is complete on December 31, 2020, the extension will uninstall itself. All data collected will be kept strictly confidential and used for research purposes only. We will not share your responses with anyone who is not involved in this research.\\

\noindent Minimizing risks: None of the raw data collected through our browser extension during this study will be publicly released, and the survey data will not be given or sold to a third party without the panelist's consent. All raw data will be stored on a secure server at Northeastern University, and access to that server will be limited to members of the research group. Only aggregated data will be released, which minimizes the possibility of reidentification. All data that is collected from our survey and from participants' browsers will be stripped of personally identifiable information to the best of our ability. \\

\noindent The decision to participate in this research project is voluntary. You do not have to participate, and there is no penalty if you choose not to participate in this research or if you choose to stop participating. You may choose to stop participating at any time, and you may request that we delete all data collected from your browser.\\

\subsection*{Browser extension informed consent (installation page)}

\noindent Welcome to the study!\\

\noindent This extension implements a user study being conducted by researchers at Northeastern University, Dartmouth, Princeton, and the University of Exeter. If you choose to participate, this browser extension will confidentially collect four types of data from your browser.\\

\noindent 1. Metadata for web browsing (e.g. URL visited with time of visit), exposure to embedded URLs on websites (e.g. YouTube videos), and interactions with websites (e.g. clicks and video viewing time). This data is collected until the study is completed.\\

\noindent 2. Copies of the HTML seen on specific sites: Google Search, Google News, YouTube, Facebook Newsfeed, and Twitter Feed. We remove all identifying information before it leaves the browser. This confidential data is collected until the study is completed.\\

\noindent 3. Browsing history, Google and YouTube account histories (e.g. searches, comments, clicks), and online advertising preferences (Google, Bluekai, Facebook). This data is initially collected for the year prior to the installation of our browser extension, and we then check these sources once every two weeks to collect updates until the study is completed.\\

\noindent 4. Snapshots of selected URLs from your browser. For each URL, the extension saves a copy of the HTML that renders, effectively capturing what you would have seen had you visited that website yourself. Once per week we conduct searches on Google Search, Google News, YouTube, and Twitter, and collect the current frontpage of Google News, YouTube, and Twitter. These web page visits will occur in the background and will not affect the normal functioning of your browser. There is a theoretical risk of ``profile pollution'' – that this extension will impact your online profiles, i.e., ``pollute'' them with actions that you did not take. To mitigate this risk, the extension will only visit content that is benign and will only execute searches for general terms. Our previous work has found that historical information of this kind has minimal impact on online services.\\

\noindent Additionally, if you choose to participate, you will be asked to take a survey in which we ask you several questions about your demographics, web usage, and media preferences. These data, as well as those mentioned above, will be used to analyze the correlations between your online behavior and your interest profiles.\\

\noindent After the study is complete on December 31, 2020, the extension will uninstall itself. All data collected will be kept strictly confidential and used for research purposes only. We will not share your responses with anyone who is not involved in this research.\\

\noindent You must be at least 18 years old to take part in this study. The decision to participate in this research project is voluntary. You do not have to participant and you can refuse to participate. Even if you begin our experiment, you can stop at any time. You may request that we delete all data collected from your web browser at any time.\\

\noindent We have minimized the risks. We are collecting basic demographic information, information about your internet habits, and copies of web pages that you visit. To the greatest extent possible, information that identifies you will be removed from all collected web data.\\

\noindent Your role in this study is confidential. However, because of the nature of electronic systems, it is possible, though unlikely, that respondents could be identified by some electronic record associated with the response. Neither the researchers nor anyone involved with this study will be collecting those data. Any reports or publications based on this research will use only aggregate data and will not identify you or any individual as being affiliated with this project.\\

\clearpage
\newpage
\section*{Survey codebook}
\setcounter{table}{0}
\setcounter{figure}{0}

\end{singlespacing}



\section*{Supplementary material}

# Survey codebook

```
================================================================================
Project Code:     XXXXX
Project Name:     XXXXXXXXXXXXX
Prepared for:     XXXXXXX
Interviews: 4000
Field Period:  July 21, 2020 – September 22, 2020
Project Manager: Sam Luks – 650.462.8009
================================================================================


================================================================================
                              Variable List
================================================================================
caseid                          Case ID
weight                          Weight
samplegroup                     Sample group
consent                         consent
q1                              Ideology
yt_freq                         How frequently you use YouTube
pid3                            3 point party ID
pid7                            7 point Party ID
q2                              Interested in politics
q3                              Trump job approvement
feeling_DemParty                Feeling thermometer -- Democratic Party
feeling_DemParty_dk_flag        feeling_DemParty – don't know flag
feeling_Trump                   Feeling thermometer -- Donald Trump
feeling_Trump_dk_flag           feeling_Trump – don't know flag
feeling_Biden                   Feeling thermometer -- Joe Biden
feeling_Biden_dk_flag           feeling_Biden – don't know flag
feeling_NewsMedia               Feeling thermometer -- The news media
feeling_NewsMedia_dk_flag       feeling_NewsMedia – don't know flag
feeling_Jews                    Feeling thermometer -- Jews
feeling_Jews_dk_flag            feeling_Jews – don't know flag
feeling_Israel                  Feeling thermometer -- Israel
feeling_Israel_dk_flag          feeling_Israel – don't know flag
feeling_Muslims                 Feeling thermometer -- Muslims
feeling_Muslims_dk_flag         feeling_Muslims – don't know flag
feeling_Norway                  Feeling thermometer -- Norway
feeling_Norway_dk_flag          feeling_Norway – don't know flag
feeling_LGBT                    Feeling thermometer -- People who identify as
                                lesbian, gay, bisexual, or transgender
feeling_LGBT_dk_flag            feeling_LGBT – don't know flag
feeling_Christians              Feeling thermometer -- Christians
feeling_Christians_dk_flag      feeling_Christians – don't know flag
feeling_Blacks                  Feeling thermometer -- Blacks
feeling_Blacks_dk_flag          feeling_Blacks – don't know flag
feeling_White                   Feeling thermometer -- Whites
feeling_White_dk_flag           feeling_White – don't know flag
feeling_Hispanics               Feeling thermometer -- Hispanics
feeling_Hispanics_dk_flag       feeling_Hispanics – don't know flag
feeling_Asians                  Feeling thermometer -- Asians
feeling_Asians_dk_flag          feeling_Asians – don't know flag
```

| Variable | Description |
|---|---|
| feeling_Feminists | Feeling thermometer -- Feminists |
| feeling_Feminists_dk_flag | feeling_Feminists – don't know flag |
| q4_1 | Whether agree with the statement (racial resentment) -- Irish, Italians, Jewish and many other minorities overcame prejudice and worked their way up. Blacks should do the same without any special favors. |
| q4_2 | Whether agree with the statement (racial resentment) -- Generations of slavery and discrimination have created conditions that make it difficult for blacks to work their way out of the lower class. |
| q4_3 | Whether agree with the statement (racial resentment) -- Over the past few years, blacks have gotten less than they deserve. |
| q4_4 | Whether agree with the statement (racial resentment) -- It's really a matter of some people not trying hard enough, if blacks would only try harder they could be just as well off as whites. |
| q4_5 | Whether agree with the statement (racial resentment) -- White people in the U.S. have certain advantages because of the color of their skin. |
| q4_6 | Whether agree with the statement (racial resentment) -- Racial problems in the U.S. are rare, isolated situations. |
| q4_7 | Whether agree with the statement (feminists) -- When women lose to men in a fair competition, they typically complain about being discriminated against. |
| q4_8 | Whether agree with the statement (feminists) -- Feminists are making entirely reasonable demands of men. |
| q5 | How often play video games |
| social_isolation_1 | How often the statement is descriptive -- How often do you feel that you lack companionship? |
| social_isolation_2 | How often the statement is descriptive -- How often do you feel left out? |
| social_isolation_3 | How often the statement is descriptive -- How often do you feel isolated from others? |
| q6_2 | Whether the statement is True -- Given enough provocation, I may hit a person |
| q6_4 | Whether the statement is True -- I often find myself disagreeing with people |
| q6_5 | Whether the statement is True -- I can't help getting into arguments when people disagree with me |
| q6_7 | Whether the statement is True -- I have trouble controlling my temper |
| q6_9 | Whether the statement is True -- I flare up |

| | |
|---|---|
| | quickly but get over it quickly |
| q6_10 | Whether the statement is True -- At times I feel I have gotten a raw deal out of life |
| q6_1 | Whether the statement is True -- There are people who have pushed me so far that we have come to blows |
| q6_3 | Whether the statement is True -- I have threatened people I know |
| q6_6 | Whether the statement is True -- My friends say I'm somewhat argumentative |
| q6_8 | Whether the statement is True -- Sometimes I fly off the handle for no good reason |
| q6_11 | Whether the statement is True -- Other people always seem to get the breaks |
| q6_12 | Whether the statement is True -- I wonder why sometimes I feel so bitter about things |
| q7_1 | Whether agree with the statement (conspiracy predispositions) -- Much of our lives are being controlled by plots hatched in secret places. |
| q7_2 | Whether agree with the statement (conspiracy predispositions) -- Even though we live in a democracy, a few people will always run things anyway. |
| q7_3 | Whether agree with the statement (conspiracy predispositions) -- The people who really "run" the country are not known to the voter. |
| q7_4 | Whether agree with the statement (conspiracy predispositions) -- Big events like wars, recessions, and the outcomes of elections are controlled by small groups of people who are working in secret against the rest of us. |
| q8 | How much trust you have in the mass media |
| q9 | How accurate is the news posted online |
| q10a_1 | Fox News |
| q10a_2 | The New York Times |
| q10a_3 | CNN |
| q10a_4 | The Washington Post |
| q10a_5 | MSNBC |
| q10a_6 | Breitbart |
| q10a_7 | InfoWars |
| q11_1 | How much you trust you have in this news source -- Fox News |
| q11_2 | How much you trust you have in this news source -- The New York Times |
| q11_3 | How much you trust you have in this news source -- CNN |
| q11_4 | How much you trust you have in this news source -- The Washington Post |
| q11_5 | How much you trust you have in this news source -- MSNBC |
| q11_6 | How much you trust you have in this news source |

| | | |
|---|---|---|
| | | -- Breitbart |
| q11_7 | How much you trust you have in this news source -- InfoWars | |
| q12_1 | How much trust you have in information from the following source -- Organizations you follow on YouTube or social media platforms (Twitter, Facebook, Instagram, Snapchat, etc.) | |
| q12_2 | How much trust you have in information from the following source -- Celebrities you follow on YouTube or social media platforms (Twitter, Facebook, Instagram, Snapchat, etc.) | |
| q12_3 | How much trust you have in information from the following source -- People you follow but do not personally know on YouTube or social media platforms (Twitter, Facebook, Instagram, Snapchat, etc.) | |
| q12_4 | How much trust you have in information from the following source -- People you follow and personally know on YouTube or social media platforms (Twitter, Facebook, Instagram, Snapchat, etc.) | |
| q12_5 | How much trust you have in information from the following source -- People you personally know and talk to offline | |
| q12_6 | How much trust you have in information from the following source -- The mass media (such as newspapers, TV and radio) | |
| q13 | How frequently you use Google | |
| q14 | How much of the information you find using google is accurate | |
| q15 | Google personalizes the search results | |
| q17 | How much of the information you find using YouTube is accurate | |
| q18 | YouTube personalizes the videos | |
| q28 | How satisfied you are with the search result quality on Google | |
| q29 | How much trust you have in information on Google | |
| q30 | What Google search results favor – Liberals or conservatives | |
| q31 | How satisfied you are with the video quality on YouTube | |
| q32 | How much trust you have in YouTube videos | |
| q33 | What YouTube videos favor – Liberals or conservatives | |
| q34_1 | How concerned you feel about the following -- Getting coronavirus yourself | |
| q34_2 | How concerned you feel about the following -- Family members getting coronavirus | |
| q35 | How much of a threat is the coronavirus for US people | |
| q36_2 | Believe the following or not -- Avoiding larger | |

| | |
|---|---|
| | gatherings of people can help prevent the spread of the coronavirus |
| q36_3 | Believe the following or not -- Masks are an effective way to prevent the spread of the coronavirus |
| q36_4 | Believe the following or not -- Coronavirus can be spread by people who do not show symptoms |
| q36_6 | Believe the following or not -- The medication hydroxychloroquine is proven to cure or prevent COVID-19, the illness caused by the novel coronavirus |
| q36_8 | Believe the following or not -- The Chinese government created the coronavirus that causes COVID-19 as a bioweapon |
| q36_10 | Believe the following or not -- A group funded by Bill Gates patented the coronavirus that causes COVID-19 |
| q36_1 | Believe the following or not -- Frequent handwashing is a way to protect against the coronavirus |
| q36_5 | Believe the following or not -- A new loss of taste or smell is a symptom of the coronavirus |
| q36_7 | Believe the following or not -- The coronavirus is being spread by 5G cell phone technology |
| q36_9 | Believe the following or not -- The media is exaggerating the threat from the coronavirus to damage President Trump |
| q37_1 | Agree with the following or not -- Getting vaccines is a good way to protect children from disease |
| q37_2 | Agree with the following or not -- Generally I do what my doctor recommends about vaccines |
| q37_3 | Agree with the following or not -- New vaccines are recommended only if they are safe |
| q37_4 | Agree with the following or not -- Children do not need vaccines for diseases that are not common anymore |
| q37_5 | Agree with the following or not -- I am concerned about serious side effects of vaccines |
| q37_6 | Agree with the following or not -- Some vaccines cause autism in healthy children |
| q37_7 | Agree with the following or not -- Vaccinations are one of the most significant achievements in improving public health |
| q38 | How often you don't take surveys seriously |
| q39 | Did you make an effort to look up information |
| platform_1 | Desktop or laptop computer |
| platform_2 | Tablet |
| platform_3 | Smartphone |
| browsers_1 | Chrome |
| browsers_2 | Firefox |

```
browsers_3                     Safari
browsers_4                     Microsoft Edge
browsers_5                     Internet Explorer
browsers_6                     None of the above
browser_top                    browser_top
elig_extension                 elig_extension
extension_install              Agree to install extension
birthyr                        Birth Year
gender                         Gender
race                           Race
educ                           Education
marstat                        Marital Status
employ                         Employment Status
faminc_new                     Family income
presvote16post                 2016 President Vote Post Election
inputstate                     State of Residence
votereg                        Voter Registration Status
ideo5                          Ideology (1)
newsint                        Political Interest
religpew                       Religion
pew_churatd                    Church attendance (Pew version)
pew_bornagain                  Born Again (Pew version)
pew_religimp                   Importance of religion (Pew version)
pew_prayer                     Frequency of Prayer (Pew version)
starttime                      Questionnaire Start Time
endtime                        Questionnaire End Time
```

                              Verbatims
================================================================================
```
session_visa                   ID to link to extension installation
pid3_t                         3 point party ID - other
q30_open                       What Google search results favor - Other
q33_open                       What YouTube videos favor - Other
q40                            Comments on the survey
whynot                         Reason for not installing extension
```

                         Variable map and codebook
================================================================================
Name:           caseid
Description:    Case ID

                Numeric Variable - no categories

                answered        : 4000
================================================================================
Name:           weight
Description:    Weight

                Numeric Variable - no categories

                answered        : 4000

```
================================================================================
Name:              samplegroup
Description:       Sample group

           Count    Code   Label
           -----    ----   -----
            2000     1     CCES 2018 recontact
            1000     2     CCES 2018 with high racial resentment recontact
            1000     3     High YouTube users

================================================================================
Name:              consent
Description:       consent

           Count    Code   Label
           -----    ----   -----
            4000     1     I agree to participate
               0     2     I do not agree to participate

================================================================================
Name:              q1
Description:       Ideology

           Count    Code   Label
           -----    ----   -----
             638     1     Very liberal
             528     2     Somewhat liberal
             224     3     Slightly liberal
             909     4     Moderate; middle of the road
             243     5     Slightly conservative
             551     6     Somewhat conservative
             905     7     Very conservative
               2    98     skipped

================================================================================
Name:              yt_freq
Description:       How frequently you use YouTube

           Count    Code   Label
           -----    ----   -----
             440     1     Almost constantly
            1300     2     Several times a day
             417     3     About once a day
             614     4     A few times a week
             235     5     About once a week
             368     6     A few times a month
              89     7     Once a month
             345     8     Less often than once a month
             192     9     Never

================================================================================
```

```
Name:           pid3
Description:    3 point party ID

           Count   Code   Label
           -----   ----   -----
            1307    1     Democrat
            1235    2     Republican
            1170    3     Independent
             222    4     Other
              66    5     Not sure

=============================================================================
Name:           pid7
Description:    7 point Party ID

           Count   Code   Label
           -----   ----   -----
             932    1     Strong Democrat
             375    2     Not very strong Democrat
             355    3     Lean Democrat
             572    4     Independent
             483    5     Lean Republican
             327    6     Not very strong Republican
             908    7     Strong Republican
              48    8     Not sure
               0    9     Don't know

=============================================================================
Name:           q2
Description:    Interested in politics

           Count   Code   Label
           -----   ----   -----
            1704    1     Extremely interested
            1164    2     Very interested
             711    3     Somewhat interested
             276    4     Not very interested
             145    5     Not at all interested

=============================================================================
Name:           q3
Description:    Trump job approvement

           Count   Code   Label
           -----   ----   -----
            1458    1     Strongly approve
             540    2     Somewhat approve
             255    3     Somewhat disapprove
            1746    4     Strongly disapprove
               1    8     skipped
```

```
================================================================================
Name:              feeling_DemParty
Description:       Feeling thermometer -- Democratic Party

                   Numeric Variable - no categories

                   answered         : 4000
                   don't know       : 55
================================================================================
Name:              feeling_DemParty_dk_flag
Description:       feeling_DemParty - don't know flag

                   Numeric Variable - no categories

                   answered         : 4000
================================================================================
Name:              feeling_Trump
Description:       Feeling thermometer -- Donald Trump

                   Numeric Variable - no categories

                   answered         : 4000
                   don't know       : 34
================================================================================
Name:              feeling_Trump_dk_flag
Description:       feeling_Trump - don't know flag

                   Numeric Variable - no categories

                   answered         : 4000
================================================================================
Name:              feeling_Biden
Description:       Feeling thermometer -- Joe Biden

                   Numeric Variable - no categories

                   answered         : 4000
                   don't know       : 63
================================================================================
Name:              feeling_Biden_dk_flag
Description:       feeling_Biden - don't know flag

                   Numeric Variable - no categories

                   answered         : 4000
================================================================================
Name:              feeling_NewsMedia
Description:       Feeling thermometer -- The news media

                   Numeric Variable - no categories
```

```
                    answered      : 4000
                    don't know    : 55
================================================================================
Name:               feeling_NewsMedia_dk_flag
Description:        feeling_NewsMedia - don't know flag

                    Numeric Variable - no categories

                    answered      : 4000
================================================================================
Name:               feeling_Jews
Description:        Feeling thermometer -- Jews

                    Numeric Variable - no categories

                    answered      : 4000
                    don't know    : 167
================================================================================
Name:               feeling_Jews_dk_flag
Description:        feeling_Jews - don't know flag

                    Numeric Variable - no categories

                    answered      : 4000
================================================================================
Name:               feeling_Israel
Description:        Feeling thermometer -- Israel

                    Numeric Variable - no categories

                    answered      : 4000
                    don't know    : 268
================================================================================
Name:               feeling_Israel_dk_flag
Description:        feeling_Israel - don't know flag

                    Numeric Variable - no categories

                    answered      : 4000
================================================================================
Name:               feeling_Muslims
Description:        Feeling thermometer -- Muslims

                    Numeric Variable - no categories

                    answered      : 4000
                    don't know    : 163
================================================================================
Name:               feeling_Muslims_dk_flag
Description:        feeling_Muslims - don't know flag
```

```
                Numeric Variable - no categories

                answered        : 4000
===============================================================================
Name:           feeling_Norway
Description:    Feeling thermometer -- Norway

                Numeric Variable - no categories

                answered        : 5
                not asked       : 3995
===============================================================================
Name:           feeling_Norway_dk_flag
Description:    feeling_Norway - don't know flag

                Numeric Variable - no categories

                answered        : 4000
===============================================================================
Name:           feeling_LGBT
Description:    Feeling thermometer -- People who identify as lesbian, gay,
                bisexual, or transgender

                Numeric Variable - no categories

                answered        : 5
                not asked       : 3995
===============================================================================
Name:           feeling_LGBT_dk_flag
Description:    feeling_LGBT - don't know flag

                Numeric Variable - no categories

                answered        : 4000
===============================================================================
Name:           feeling_Christians
Description:    Feeling thermometer -- Christians

                Numeric Variable - no categories

                answered        : 4000
                don't know      : 85
===============================================================================
Name:           feeling_Christians_dk_flag
Description:    feeling_Christians - don't know flag

                Numeric Variable - no categories

                answered        : 4000
===============================================================================
Name:           feeling_Blacks
```

```
Description:    Feeling thermometer -- Blacks

                Numeric Variable - no categories

                answered          : 4000
                don't know        : 92
===============================================================================
Name:           feeling_Blacks_dk_flag
Description:    feeling_Blacks - don't know flag

                Numeric Variable - no categories

                answered          : 4000
===============================================================================
Name:           feeling_White
Description:    Feeling thermometer -- Whites

                Numeric Variable - no categories

                answered          : 4000
                don't know        : 76
===============================================================================
Name:           feeling_White_dk_flag
Description:    feeling_White - don't know flag

                Numeric Variable - no categories

                answered          : 4000
===============================================================================
Name:           feeling_Hispanics
Description:    Feeling thermometer -- Hispanics

                Numeric Variable - no categories

                answered          : 5
                not asked         : 3995
===============================================================================
Name:           feeling_Hispanics_dk_flag
Description:    feeling_Hispanics - don't know flag

                Numeric Variable - no categories

                answered          : 4000
===============================================================================
Name:           feeling_Asians
Description:    Feeling thermometer -- Asians

                Numeric Variable - no categories

                answered          : 5
                not asked         : 3995
```

```
===============================================================================
Name:              feeling_Asians_dk_flag
Description:       feeling_Asians - don't know flag

                   Numeric Variable - no categories

                   answered        : 4000
===============================================================================
Name:              feeling_Feminists
Description:       Feeling thermometer -- Feminists

                   Numeric Variable - no categories

                   answered        : 4000
                   don't know      : 131
===============================================================================
Name:              feeling_Feminists_dk_flag
Description:       feeling_Feminists - don't know flag

                   Numeric Variable - no categories

                   answered        : 4000
===============================================================================
Name:              q4_1
Description:       Whether agree with the statement (racial resentment) -- Irish,
                   Italians, Jewish and many other minorities overcame prejudice
                   and worked their way up. Blacks should do the same without any
                   special favors.

         Count   Code   Label
         -----   ----   -----
          1518     1    Strongly agree
           630     2    Somewhat agree
           566     3    Neither agree nor disagree
           484     4    Somewhat disagree
           802     5    Strongly disagree

===============================================================================
Name:              q4_2
Description:       Whether agree with the statement (racial resentment) --
                   Generations of slavery and discrimination have created
                   conditions that make it difficult for blacks to work their way
                   out of the lower class.

         Count   Code   Label
         -----   ----   -----
          1101     1    Strongly agree
           601     2    Somewhat agree
           358     3    Neither agree nor disagree
           422     4    Somewhat disagree
          1518     5    Strongly disagree
```

```
================================================================================
Name:            q4_3
Description:     Whether agree with the statement (racial resentment) -- Over the
                 past few years, blacks have gotten less than they deserve.

        Count   Code    Label
        -----   ----    -----
          922    1      Strongly agree
          612    2      Somewhat agree
          611    3      Neither agree nor disagree
          477    4      Somewhat disagree
         1378    5      Strongly disagree

================================================================================
Name:            q4_4
Description:     Whether agree with the statement (racial resentment) -- It's
                 really a matter of some people not trying hard enough, if blacks
                 would only try harder they could be just as well off as whites.

        Count   Code    Label
        -----   ----    -----
         1039    1      Strongly agree
          718    2      Somewhat agree
          645    3      Neither agree nor disagree
          500    4      Somewhat disagree
         1098    5      Strongly disagree

================================================================================
Name:            q4_5
Description:     Whether agree with the statement (racial resentment) -- White
                 people in the U.S. have certain advantages because of the color
                 of their skin.

        Count   Code    Label
        -----   ----    -----
         1337    1      Strongly agree
          622    2      Somewhat agree
          478    3      Neither agree nor disagree
          412    4      Somewhat disagree
         1151    5      Strongly disagree

================================================================================
Name:            q4_6
Description:     Whether agree with the statement (racial resentment) -- Racial
                 problems in the U.S. are rare, isolated situations.

        Count   Code    Label
        -----   ----    -----
          635    1      Strongly agree
          657    2      Somewhat agree
```

```
                         579        3    Neither agree nor disagree
                         682        4    Somewhat disagree
                        1447        5    Strongly disagree

===============================================================================
Name:             q4_7
Description:      Whether agree with the statement (feminists) -- When women lose
                  to men in a fair competition, they typically complain about
                  being discriminated against.

                 Count     Code    Label
                 -----     ----    -----
                   802        1    Strongly agree
                   882        2    Somewhat agree
                   961        3    Neither agree nor disagree
                   664        4    Somewhat disagree
                   691        5    Strongly disagree

===============================================================================
Name:             q4_8
Description:      Whether agree with the statement (feminists) -- Feminists are
                  making entirely reasonable demands of men.

                 Count     Code    Label
                 -----     ----    -----
                   867        1    Strongly agree
                   674        2    Somewhat agree
                   802        3    Neither agree nor disagree
                   567        4    Somewhat disagree
                  1090        5    Strongly disagree

===============================================================================
Name:             q5
Description:      How often play video games

                 Count     Code    Label
                 -----     ----    -----
                   844        1    Often
                   947        2    Sometimes
                   695        3    Hardly ever
                  1502        4    Never
                    12        5    Prefer not to answer

===============================================================================
Name:             social_isolation_1
Description:      How often the statement is descriptive -- How often do you feel
                  that you lack companionship?

                 Count     Code    Label
                 -----     ----    -----
                  1315        1    Never
```

```
            1181     2    Rarely
            1010     3    Sometimes
             492     4    Often
               2     8    skipped


==============================================================================
Name:              social_isolation_2
Description:       How often the statement is descriptive -- How often do you feel
                   left out?

            Count  Code  Label
            -----  ----  -----
            1035     1    Never
            1453     2    Rarely
            1082     3    Sometimes
             428     4    Often
               2     8    skipped


==============================================================================
Name:              social_isolation_3
Description:       How often the statement is descriptive -- How often do you feel
                   isolated from others?

            Count  Code  Label
            -----  ----  -----
            1139     1    Never
            1272     2    Rarely
            1070     3    Sometimes
             517     4    Often
               2     8    skipped


==============================================================================
Name:              q6_2
Description:       Whether the statement is True -- Given enough provocation, I may
                   hit a person

            Count  Code  Label
            -----  ----  -----
             226     1    Completely true for me
             242     2    Mostly true for me
             578     3    Slightly true for me
             379     4    Slightly false for me
             827     5    Mostly false for me
            1748     6    Completely false for me


==============================================================================
Name:              q6_4
Description:       Whether the statement is True -- I often find myself disagreeing
                   with people

            Count  Code  Label
```

|  Count | Code | Label |
|---:|---:|---|
|  187 | 1 | Completely true for me |
|  444 | 2 | Mostly true for me |
| 1376 | 3 | Slightly true for me |
|  879 | 4 | Slightly false for me |
|  761 | 5 | Mostly false for me |
|  352 | 6 | Completely false for me |
|    1 | 8 | skipped |

==============================================================================
Name: q6_5
Description: Whether the statement is True -- I can't help getting into arguments when people disagree with me

|  Count | Code | Label |
|---:|---:|---|
|   81 | 1 | Completely true for me |
|  225 | 2 | Mostly true for me |
|  698 | 3 | Slightly true for me |
|  785 | 4 | Slightly false for me |
| 1220 | 5 | Mostly false for me |
|  990 | 6 | Completely false for me |
|    1 | 8 | skipped |

==============================================================================
Name: q6_7
Description: Whether the statement is True -- I have trouble controlling my temper

|  Count | Code | Label |
|---:|---:|---|
|   71 | 1 | Completely true for me |
|  158 | 2 | Mostly true for me |
|  625 | 3 | Slightly true for me |
|  520 | 4 | Slightly false for me |
| 1237 | 5 | Mostly false for me |
| 1388 | 6 | Completely false for me |
|    1 | 8 | skipped |

==============================================================================
Name: q6_9
Description: Whether the statement is True -- I flare up quickly but get over it quickly

|  Count | Code | Label |
|---:|---:|---|
|  190 | 1 | Completely true for me |
|  529 | 2 | Mostly true for me |
| 1042 | 3 | Slightly true for me |
|  615 | 4 | Slightly false for me |
|  867 | 5 | Mostly false for me |

```
       756     6    Completely false for me
         1     8    skipped
```

===============================================================================

```
Name:           q6_10
Description:    Whether the statement is True -- At times I feel I have gotten a
                raw deal out of life

        Count   Code   Label
        -----   ----   -----
          296    1     Completely true for me
          357    2     Mostly true for me
          939    3     Slightly true for me
          530    4     Slightly false for me
          861    5     Mostly false for me
         1017    6     Completely false for me
```

===============================================================================

```
Name:           q6_1
Description:    Whether the statement is True -- There are people who have
                pushed me so far that we have come to blows

        Count   Code   Label
        -----   ----   -----
            0    1     Completely true for me
            0    2     Mostly true for me
            1    3     Slightly true for me
            0    4     Slightly false for me
            1    5     Mostly false for me
            3    6     Completely false for me
         3995    9     not asked
```

===============================================================================

```
Name:           q6_3
Description:    Whether the statement is True -- I have threatened people I know

        Count   Code   Label
        -----   ----   -----
            0    1     Completely true for me
            0    2     Mostly true for me
            1    3     Slightly true for me
            0    4     Slightly false for me
            1    5     Mostly false for me
            3    6     Completely false for me
         3995    9     not asked
```

===============================================================================

```
Name:           q6_6
Description:    Whether the statement is True -- My friends say I'm somewhat
                argumentative
```

```
               Count   Code   Label
               -----   ----   -----
                   0      1   Completely true for me
                   0      2   Mostly true for me
                   0      3   Slightly true for me
                   0      4   Slightly false for me
                   3      5   Mostly false for me
                   2      6   Completely false for me
                3995      9   not asked

==============================================================================
Name:              q6_8
Description:       Whether the statement is True -- Sometimes I fly off the handle
                   for no good reason

               Count   Code   Label
               -----   ----   -----
                   0      1   Completely true for me
                   0      2   Mostly true for me
                   0      3   Slightly true for me
                   1      4   Slightly false for me
                   1      5   Mostly false for me
                   3      6   Completely false for me
                3995      9   not asked

==============================================================================
Name:              q6_11
Description:       Whether the statement is True -- Other people always seem to get
                   the breaks

               Count   Code   Label
               -----   ----   -----
                   0      1   Completely true for me
                   0      2   Mostly true for me
                   3      3   Slightly true for me
                   2      4   Slightly false for me
                   0      5   Mostly false for me
                   0      6   Completely false for me
                3995      9   not asked

==============================================================================
Name:              q6_12
Description:       Whether the statement is True -- I wonder why sometimes I feel
                   so bitter about things

               Count   Code   Label
               -----   ----   -----
                   1      1   Completely true for me
                   1      2   Mostly true for me
                   0      3   Slightly true for me
                   0      4   Slightly false for me
```

```
             2     5   Mostly false for me
             1     6   Completely false for me
          3995     9   not asked

==============================================================================
Name:            q7_1
Description:     Whether agree with the statement (conspiracy predispositions) --
                 Much of our lives are being controlled by plots hatched in
                 secret places.

         Count   Code   Label
         -----   ----   -----
           384    1     Strongly agree
           729    2     Somewhat agree
           873    3     Neither agree nor disagree
           610    4     Somewhat disagree
          1404    5     Strongly disagree

==============================================================================
Name:            q7_2
Description:     Whether agree with the statement (conspiracy predispositions) --
                 Even though we live in a democracy, a few people will always run
                 things anyway.

         Count   Code   Label
         -----   ----   -----
           869    1     Strongly agree
          1753    2     Somewhat agree
           714    3     Neither agree nor disagree
           417    4     Somewhat disagree
           247    5     Strongly disagree

==============================================================================
Name:            q7_3
Description:     Whether agree with the statement (conspiracy predispositions) --
                 The people who really "run" the country are not known to the
                 voter.

         Count   Code   Label
         -----   ----   -----
           832    1     Strongly agree
          1257    2     Somewhat agree
           867    3     Neither agree nor disagree
           582    4     Somewhat disagree
           462    5     Strongly disagree

==============================================================================
Name:            q7_4
Description:     Whether agree with the statement (conspiracy predispositions) --
                 Big events like wars, recessions, and the outcomes of elections
                 are controlled by small groups of people who are working in
```

```
                         secret against the rest of us.

             Count    Code    Label
             -----    ----    -----
               516      1     Strongly agree
               881      2     Somewhat agree
              1004      3     Neither agree nor disagree
               653      4     Somewhat disagree
               945      5     Strongly disagree
                 1      8     skipped


===============================================================================
Name:             q8
Description:   How much trust you have in the mass media

             Count    Code    Label
             -----    ----    -----
               352      1     A great deal
              1253      2     A fair amount
              1054      3     Not very much
              1340      4     None at all
                 1      8     skipped


===============================================================================
Name:             q9
Description:   How accurate is the news posted online

             Count    Code    Label
             -----    ----    -----
               396      1     Very accurate
              1544      2     Somewhat accurate
              1165      3     Not too accurate
               895      4     Not at all accurate


===============================================================================
Name:             q10a_1
Description:   Fox News

             Count    Code    Label
             -----    ----    -----
                 5      1     selected
                 0      2     not selected
              3995      9     not asked


===============================================================================
Name:             q10a_2
Description:   The New York Times

             Count    Code    Label
             -----    ----    -----
                 5      1     selected
```

```
         0     2  not selected
      3995     9  not asked

================================================================================
Name:          q10a_3
Description:   CNN

         Count  Code  Label
         -----  ----  -----
             5     1  selected
             0     2  not selected
          3995     9  not asked

================================================================================
Name:          q10a_4
Description:   The Washington Post

         Count  Code  Label
         -----  ----  -----
             5     1  selected
             0     2  not selected
          3995     9  not asked

================================================================================
Name:          q10a_5
Description:   MSNBC

         Count  Code  Label
         -----  ----  -----
             5     1  selected
             0     2  not selected
          3995     9  not asked

================================================================================
Name:          q10a_6
Description:   Breitbart

         Count  Code  Label
         -----  ----  -----
             3     1  selected
             2     2  not selected
          3995     9  not asked

================================================================================
Name:          q10a_7
Description:   InfoWars

         Count  Code  Label
         -----  ----  -----
             2     1  selected
             3     2  not selected
```

```
         3995       9    not asked


===============================================================================
Name:            q11_1
Description:     How much you trust you have in this news source -- Fox News

         Count   Code   Label
         -----   ----   -----
             0      1   A great deal
             1      2   A fair amount
             3      3   Not very much
             1      4   None at all
          3995      9   not asked


===============================================================================
Name:            q11_2
Description:     How much you trust you have in this news source -- The New York
                 Times

         Count   Code   Label
         -----   ----   -----
             2      1   A great deal
             3      2   A fair amount
             0      3   Not very much
             0      4   None at all
          3995      9   not asked


===============================================================================
Name:            q11_3
Description:     How much you trust you have in this news source -- CNN

         Count   Code   Label
         -----   ----   -----
             1      1   A great deal
             3      2   A fair amount
             1      3   Not very much
             0      4   None at all
          3995      9   not asked


===============================================================================
Name:            q11_4
Description:     How much you trust you have in this news source -- The
                 Washington Post

         Count   Code   Label
         -----   ----   -----
             1      1   A great deal
             4      2   A fair amount
             0      3   Not very much
             0      4   None at all
          3995      9   not asked
```

```
================================================================================
Name:              q11_5
Description:    How much you trust you have in this news source -- MSNBC

           Count   Code   Label
           -----   ----   -----
               1      1   A great deal
               3      2   A fair amount
               1      3   Not very much
               0      4   None at all
            3995      9   not asked

================================================================================
Name:              q11_6
Description:    How much you trust you have in this news source -- Breitbart

           Count   Code   Label
           -----   ----   -----
               0      1   A great deal
               0      2   A fair amount
               0      3   Not very much
               3      4   None at all
            3997      9   not asked

================================================================================
Name:              q11_7
Description:    How much you trust you have in this news source -- InfoWars

           Count   Code   Label
           -----   ----   -----
               0      1   A great deal
               0      2   A fair amount
               1      3   Not very much
               1      4   None at all
            3998      9   not asked

================================================================================
Name:              q12_1
Description:    How much trust you have in information from the following source
               -- Organizations you follow on YouTube or social media platforms
               (Twitter, Facebook, Instagram, Snapchat, etc.)

           Count   Code   Label
           -----   ----   -----
             210      1   A great deal
            1461      2   A fair amount
            1367      3   Not very much
             955      4   None at all
               7      8   skipped
```

```
================================================================================
Name:             q12_2
Description:      How much trust you have in information from the following source
                  -- Celebrities you follow on YouTube or social media platforms
                  (Twitter, Facebook, Instagram, Snapchat, etc.)

        Count    Code    Label
        -----    ----    -----
          107      1     A great deal
          541      2     A fair amount
         1325      3     Not very much
         2023      4     None at all
            4      8     skipped

================================================================================
Name:             q12_3
Description:      How much trust you have in information from the following source
                  -- People you follow but do not personally know on YouTube or
                  social media platforms (Twitter, Facebook, Instagram, Snapchat,
                  etc.)

        Count    Code    Label
        -----    ----    -----
          167      1     A great deal
         1071      2     A fair amount
         1634      3     Not very much
         1123      4     None at all
            5      8     skipped

================================================================================
Name:             q12_4
Description:      How much trust you have in information from the following source
                  -- People you follow and personally know on YouTube or social
                  media platforms (Twitter, Facebook, Instagram, Snapchat, etc.)

        Count    Code    Label
        -----    ----    -----
          301      1     A great deal
         1638      2     A fair amount
         1265      3     Not very much
          792      4     None at all
            4      8     skipped

================================================================================
Name:             q12_5
Description:      How much trust you have in information from the following source
                  -- People you personally know and talk to offline

        Count    Code    Label
        -----    ----    -----
          753      1     A great deal
```

```
        2352     2    A fair amount
         718     3    Not very much
         173     4    None at all
           4     8    skipped
```

===============================================================================
```
Name:            q12_6
Description:     How much trust you have in information from the following source
                 -- The mass media (such as newspapers, TV and radio)

         Count   Code  Label
         -----   ----  -----
           401     1    A great deal
          1284     2    A fair amount
          1020     3    Not very much
          1291     4    None at all
             4     8    skipped
```

===============================================================================
```
Name:            q13
Description:     How frequently you use Google

         Count   Code  Label
         -----   ----  -----
           819     1    Almost constantly
          1561     2    Several times a day
           401     3    About once a day
           482     4    A few times a week
            82     5    About once a week
           175     6    A few times a month
            41     7    Once a month
           214     8    Less often than once a month
           225     9    Never
```

===============================================================================
```
Name:            q14
Description:     How much of the information you find using google is accurate

         Count   Code  Label
         -----   ----  -----
           165     1    All or almost all
           874     2    Most
          1415     3    Some
           839     4    Very little
           401     5    None at all
           306     6    Don't know
```

===============================================================================
```
Name:            q15
Description:     Google personalizes the search results
```

```
           Count   Code   Label
           -----   ----   -----
            1386    1     Very accurate
            1815    2     Somewhat accurate
             515    3     Not very accurate
             282    4     Not at all accurate
               2    8     skipped
```

================================================================================
Name:           q17
Description:    How much of the information you find using YouTube is accurate

```
           Count   Code   Label
           -----   ----   -----
              82    1     All or almost all
             399    2     Most
            1556    3     Some
             994    4     Very little
             416    5     None at all
             552    6     Don't know
               1    8     skipped
```

================================================================================
Name:           q18
Description:    YouTube personalizes the videos

```
           Count   Code   Label
           -----   ----   -----
            1279    1     Very accurate
            1791    2     Somewhat accurate
             583    3     Not very accurate
             344    4     Not at all accurate
               3    8     skipped
```

================================================================================
Name:           q28
Description:    How satisfied you are with the search result quality on Google

```
           Count   Code   Label
           -----   ----   -----
             854    1     Very satisfied
            2056    2     Somewhat satisfied
             492    3     Not very satisfied
             159    4     Not at all satisfied
             439    9     not asked
```

================================================================================
Name:           q29
Description:    How much trust you have in information on Google

```
           Count   Code   Label
```

|  Count | Code | Label |
| --- | --- | --- |
|   551 |  1 | A great deal |
|  2184 |  2 | A moderate amount |
|   708 |  3 | Not much |
|   117 |  4 | Not at all |
|     1 |  8 | skipped |
|   439 |  9 | not asked |

===============================================================================
Name:          q30
Description:   What Google search results favor – Liberals or conservatives

|  Count | Code | Label |
| --- | --- | --- |
|  1418 |  1 | Favor liberals |
|   201 |  2 | Favor conservatives |
|  1798 |  3 | Neither |
|   143 |  4 | Other |
|     1 |  8 | skipped |
|   439 |  9 | not asked |

===============================================================================
Name:          q31
Description:   How satisfied you are with the video quality on YouTube

|  Count | Code | Label |
| --- | --- | --- |
|   611 |  1 | Very satisfied |
|  2119 |  2 | Somewhat satisfied |
|   578 |  3 | Not very satisfied |
|   153 |  4 | Not at all satisfied |
|     2 |  8 | skipped |
|   537 |  9 | not asked |

===============================================================================
Name:          q32
Description:   How much trust you have in YouTube videos

|  Count | Code | Label |
| --- | --- | --- |
|   333 |  1 | A great deal |
|  2120 |  2 | A moderate amount |
|   898 |  3 | Not much |
|   111 |  4 | Not at all |
|     1 |  8 | skipped |
|   537 |  9 | not asked |

===============================================================================
Name:          q33
Description:   What YouTube videos favor – Liberals or conservatives

```
                Count   Code   Label
                -----   ----   -----
                 1163    1     Favor liberals
                  279    2     Favor conservatives
                 1846    3     Neither
                  172    4     Other
                    3    8     skipped
                  537    9     not asked


================================================================================
Name:              q34_1
Description:    How concerned you feel about the following -- Getting
                coronavirus yourself

                Count   Code   Label
                -----   ----   -----
                  725    1     Not at all concerned
                  901    2     Not very concerned
                 1283    3     Somewhat concerned
                 1015    4     Very concerned
                   13    5     Not applicable to me
                   63    6     Already contracted coronavirus


================================================================================
Name:              q34_2
Description:    How concerned you feel about the following -- Family members
                getting coronavirus

                Count   Code   Label
                -----   ----   -----
                    0    1     Not at all concerned
                    0    2     Not very concerned
                    1    3     Somewhat concerned
                    4    4     Very concerned
                    0    5     Not applicable to me
                    0    6     Already contracted coronavirus
                 3995    9     not asked


================================================================================
Name:              q35
Description:    How much of a threat is the coronavirus for US people

                Count   Code   Label
                -----   ----   -----
                 2208    1     A major threat
                 1294    2     A minor threat
                  497    3     Not a threat
                    1    8     skipped


================================================================================
Name:              q36_2
```

Description:     Believe the following or not -- Avoiding larger gatherings of
                 people can help prevent the spread of the coronavirus

        Count   Code    Label
        -----   ----    -----
          250     1     Not at all accurate
          335     2     Not very accurate
         1008     3     Somewhat accurate
         2407     4     Very accurate

================================================================================
Name:            q36_3
Description:     Believe the following or not -- Masks are an effective way to
                 prevent the spread of the coronavirus

        Count   Code    Label
        -----   ----    -----
          600     1     Not at all accurate
          559     2     Not very accurate
         1068     3     Somewhat accurate
         1773     4     Very accurate

================================================================================
Name:            q36_4
Description:     Believe the following or not -- Coronavirus can be spread by
                 people who do not show symptoms

        Count   Code    Label
        -----   ----    -----
          168     1     Not at all accurate
          303     2     Not very accurate
          938     3     Somewhat accurate
         2590     4     Very accurate
            1     8     skipped

================================================================================
Name:            q36_6
Description:     Believe the following or not -- The medication
                 hydroxychloroquine is proven to cure or prevent COVID-19, the
                 illness caused by the novel coronavirus

        Count   Code    Label
        -----   ----    -----
         1697     1     Not at all accurate
          675     2     Not very accurate
          923     3     Somewhat accurate
          704     4     Very accurate
            1     8     skipped

================================================================================
Name:            q36_8

```
Description:     Believe the following or not -- The Chinese government created
                 the coronavirus that causes COVID-19 as a bioweapon

          Count   Code   Label
          -----   ----   -----
           1462    1     Not at all accurate
            760    2     Not very accurate
            826    3     Somewhat accurate
            949    4     Very accurate
              3    8     skipped

================================================================================
Name:            q36_10
Description:     Believe the following or not -- A group funded by Bill Gates
                 patented the coronavirus that causes COVID-19

          Count   Code   Label
          -----   ----   -----
           2372    1     Not at all accurate
            745    2     Not very accurate
            543    3     Somewhat accurate
            339    4     Very accurate
              1    8     skipped

================================================================================
Name:            q36_1
Description:     Believe the following or not -- Frequent handwashing is a way to
                 protect against the coronavirus

          Count   Code   Label
          -----   ----   -----
              0    1     Not at all accurate
              0    2     Not very accurate
              1    3     Somewhat accurate
              4    4     Very accurate
           3995    9     not asked

================================================================================
Name:            q36_5
Description:     Believe the following or not -- A new loss of taste or smell is
                 a symptom of the coronavirus

          Count   Code   Label
          -----   ----   -----
              0    1     Not at all accurate
              0    2     Not very accurate
              2    3     Somewhat accurate
              3    4     Very accurate
           3995    9     not asked

================================================================================
```

```
Name:            q36_7
Description:     Believe the following or not -- The coronavirus is being spread
                 by 5G cell phone technology

        Count   Code   Label
        -----   ----   -----
            5      1   Not at all accurate
            0      2   Not very accurate
            0      3   Somewhat accurate
            0      4   Very accurate
         3995      9   not asked

=============================================================================
Name:            q36_9
Description:     Believe the following or not -- The media is exaggerating the
                 threat from the coronavirus to damage President Trump

        Count   Code   Label
        -----   ----   -----
            4      1   Not at all accurate
            1      2   Not very accurate
            0      3   Somewhat accurate
            0      4   Very accurate
         3995      9   not asked

=============================================================================
Name:            q37_1
Description:     Agree with the following or not -- Getting vaccines is a good
                 way to protect children from disease

        Count   Code   Label
        -----   ----   -----
         2431      1   Strongly agree
          782      2   Somewhat agree
          404      3   Neither agree nor disagree
          142      4   Somewhat disagree
          241      5   Strongly disagree

=============================================================================
Name:            q37_2
Description:     Agree with the following or not -- Generally I do what my doctor
                 recommends about vaccines

        Count   Code   Label
        -----   ----   -----
         1978      1   Strongly agree
          967      2   Somewhat agree
          537      3   Neither agree nor disagree
          255      4   Somewhat disagree
          263      5   Strongly disagree
```

```
================================================================================
Name:              q37_3
Description:       Agree with the following or not -- New vaccines are recommended
                   only if they are safe

           Count   Code   Label
           -----   ----   -----
            1305    1     Strongly agree
            1205    2     Somewhat agree
             755    3     Neither agree nor disagree
             401    4     Somewhat disagree
             333    5     Strongly disagree
               1    8     skipped

================================================================================
Name:              q37_4
Description:       Agree with the following or not -- Children do not need vaccines
                   for diseases that are not common anymore

           Count   Code   Label
           -----   ----   -----
             201    1     Strongly agree
             245    2     Somewhat agree
             567    3     Neither agree nor disagree
             824    4     Somewhat disagree
            2163    5     Strongly disagree

================================================================================
Name:              q37_5
Description:       Agree with the following or not -- I am concerned about serious
                   side effects of vaccines

           Count   Code   Label
           -----   ----   -----
             842    1     Strongly agree
             978    2     Somewhat agree
             764    3     Neither agree nor disagree
             696    4     Somewhat disagree
             720    5     Strongly disagree

================================================================================
Name:              q37_6
Description:       Agree with the following or not -- Some vaccines cause autism in
                   healthy children

           Count   Code   Label
           -----   ----   -----
             412    1     Strongly agree
             409    2     Somewhat agree
             977    3     Neither agree nor disagree
             461    4     Somewhat disagree
```

```
          1740       5    Strongly disagree
             1       8    skipped


===============================================================================
Name:             q37_7
Description:      Agree with the following or not -- Vaccinations are one of the
                  most significant achievements in improving public health

          Count   Code   Label
          -----   ----   -----
           2252      1   Strongly agree
            897      2   Somewhat agree
            527      3   Neither agree nor disagree
            141      4   Somewhat disagree
            183      5   Strongly disagree


===============================================================================
Name:             q38
Description:      How often you don't take surveys seriously

          Count   Code   Label
          -----   ----   -----
           3300      1   Never
            457      2   Rarely
            163      3   Some of the time
             42      4   Most of the time
             38      5   Always


===============================================================================
Name:             q39
Description:      Did you make an effort to look up information

          Count   Code   Label
          -----   ----   -----
            193      1   Yes, I looked up information
           3807      2   No, I did not look up information


===============================================================================
Name:             platform_1
Description:      Desktop or laptop computer

          Count   Code   Label
          -----   ----   -----
           3850      1   selected
            150      2   not selected


===============================================================================
Name:             platform_2
Description:      Tablet

          Count   Code   Label
```

```
             -----    ----    -----
              1344      1     selected
              2656      2     not selected

===============================================================================
Name:           platform_3
Description:    Smartphone

             Count    Code    Label
             -----    ----    -----
              2372      1     selected
              1628      2     not selected

===============================================================================
Name:           browsers_1
Description:    Chrome

             Count    Code    Label
             -----    ----    -----
              2626      1     selected
              1225      2     not selected
               149      9     not asked

===============================================================================
Name:           browsers_2
Description:    Firefox

             Count    Code    Label
             -----    ----    -----
              1092      1     selected
              2759      2     not selected
               149      9     not asked

===============================================================================
Name:           browsers_3
Description:    Safari

             Count    Code    Label
             -----    ----    -----
               524      1     selected
              3327      2     not selected
               149      9     not asked

===============================================================================
Name:           browsers_4
Description:    Microsoft Edge

             Count    Code    Label
             -----    ----    -----
               903      1     selected
              2948      2     not selected
```

```
               149      9    not asked


================================================================================
Name:           browsers_5
Description:    Internet Explorer

              Count    Code    Label
              -----    ----    -----
                595      1    selected
               3256      2    not selected
                149      9    not asked


================================================================================
Name:           browsers_6
Description:    None of the above

              Count    Code    Label
              -----    ----    -----
                 82      1    selected
               3769      2    not selected
                149      9    not asked


================================================================================
Name:           browser_top
Description:    browser_top

              Count    Code    Label
              -----    ----    -----
                733      1    Chrome
                280      2    Firefox
                 79      3    Safari
                208      4    Microsoft Edge
                 79      5    Internet Explorer
               2621      9    not asked


================================================================================
Name:           elig_extension
Description:    elig_extension

              Count    Code    Label
              -----    ----    -----
               2887      1    Yes
               1113      2    No


================================================================================
Name:           extension_install
Description:    Agree to install extension

              Count    Code    Label
              -----    ----    -----
               1473      1    Yes
```

```
            1415      2   No
            1112      9   not asked

================================================================================
Name:           birthyr
Description:    Birth Year

                Numeric Variable - no categories

                answered        : 4000
================================================================================
Name:           gender
Description:    Gender

            Count   Code   Label
            -----   ----   -----
            2175      1    Male
            1825      2    Female

================================================================================
Name:           race
Description:    Race

            Count   Code   Label
            -----   ----   -----
            3035      1    White
             321      2    Black
             278      3    Hispanic
             144      4    Asian
              38      5    Native American
              67      6    Two or more races
             113      7    Other
               4      8    Middle Eastern

================================================================================
Name:           educ
Description:    Education

            Count   Code   Label
            -----   ----   -----
              57      1    No HS
             762      2    High school graduate
             963      3    Some college
             477      4    2-year
            1023      5    4-year
             718      6    Post-grad

================================================================================
Name:           marstat
Description:    Marital Status
```

```
         Count   Code   Label
         -----   ----   -----
          2106     1    Married
            51     2    Separated
           467     3    Divorced
           190     4    Widowed
          1021     5    Never married
           165     6    Domestic / civil partnership


==============================================================================
Name:            employ
Description:     Employment Status

         Count   Code   Label
         -----   ----   -----
          1563     1    Full-time
           388     2    Part-time
           120     3    Temporarily laid off
           261     4    Unemployed
           958     5    Retired
           293     6    Permanently disabled
           192     7    Homemaker
           145     8    Student
            80     9    Other


==============================================================================
Name:            faminc_new
Description:     Family income

         Count   Code   Label
         -----   ----   -----
           163     1    Less than $10,000
           256     2    $10,000 - $19,999
           332     3    $20,000 - $29,999
           389     4    $30,000 - $39,999
           296     5    $40,000 - $49,999
           311     6    $50,000 - $59,999
           286     7    $60,000 - $69,999
           311     8    $70,000 - $79,999
           389     9    $80,000 - $99,999
           255    10    $100,000 - $119,999
           265    11    $120,000 - $149,999
           188    12    $150,000 - $199,999
            65    13    $200,000 - $249,999
            49    14    $250,000 - $349,999
            17    15    $350,000 - $499,999
            17    16    $500,000 or more
           411    97    Prefer not to say


==============================================================================
Name:            presvote16post
```

Description:    2016 President Vote Post Election

```
           Count   Code   Label
           -----   ----   -----
            1234     1    Hillary Clinton
            1614     2    Donald Trump
             114     3    Gary Johnson
              66     4    Jill Stein
              20     5    Evan McMullin
              84     6    Other
             868     7    Did not vote for President
```

==============================================================================
Name:           inputstate
Description:    State of Residence

```
           Count   Code   Label
           -----   ----   -----
              71     1    Alabama
              10     2    Alaska
              86     4    Arizona
              39     5    Arkansas
             342     6    California
              59     8    Colorado
              36     9    Connecticut
              11    10    Delaware
              15    11    District of Columbia
             314    12    Florida
             129    13    Georgia
               9    15    Hawaii
              29    16    Idaho
             155    17    Illinois
              76    18    Indiana
              40    19    Iowa
              30    20    Kansas
              86    21    Kentucky
              43    22    Louisiana
              25    23    Maine
              51    24    Maryland
              82    25    Massachusetts
             117    26    Michigan
              68    27    Minnesota
              24    28    Mississippi
              92    29    Missouri
              17    30    Montana
              23    31    Nebraska
              49    32    Nevada
              21    33    New Hampshire
             118    34    New Jersey
              27    35    New Mexico
             269    36    New York
```

```
       126      37   North Carolina
         6      38   North Dakota
       152      39   Ohio
        30      40   Oklahoma
        76      41   Oregon
       216      42   Pennsylvania
        12      44   Rhode Island
        47      45   South Carolina
        17      46   South Dakota
        77      47   Tennessee
       304      48   Texas
        37      49   Utah
        10      50   Vermont
       113      51   Virginia
        99      53   Washington
        29      54   West Virginia
        75      55   Wisconsin
        11      56   Wyoming
         0      60   American Samoa
         0      64   Federated States of Micronesia
         0      66   Guam
         0      68   Marshall Islands
         0      69   Northern Mariana Islands
         0      70   Palau
         0      72   Puerto Rico
         0      74   U.S. Minor Outlying Islands
         0      78   Virgin Islands
         0      81   Alberta
         0      82   British Columbia
         0      83   Manitoba
         0      84   New Brunswick
         0      85   Newfoundland
         0      86   Northwest Territories
         0      87   Nova Scotia
         0      88   Nunavut
         0      89   Ontario
         0      90   Prince Edward Island
         0      91   Quebec
         0      92   Saskatchewan
         0      93   Yukon Territory
         0      99   Not in the U.S. or Canada


===============================================================================
Name:           votereg
Description:    Voter Registration Status

        Count   Code   Label
        -----   ----   -----
         3796      1   Yes
          165      2   No
           39      3   Don't know
```

```
================================================================================
Name:           ideo5
Description:    Ideology (1)

           Count   Code   Label
           -----   ----   -----
             582      1   Very liberal
             625      2   Liberal
            1049      3   Moderate
             772      4   Conservative
             844      5   Very conservative
             128      6   Not sure


================================================================================
Name:           newsint
Description:    Political Interest

           Count   Code   Label
           -----   ----   -----
            2640      1   Most of the time
             864      2   Some of the time
             277      3   Only now and then
             160      4   Hardly at all
              59      7   Don't know


================================================================================
Name:           religpew
Description:    Religion

           Count   Code   Label
           -----   ----   -----
            1353      1   Protestant
             729      2   Roman Catholic
              57      3   Mormon
              36      4   Eastern or Greek Orthodox
             100      5   Jewish
              24      6   Muslim
              31      7   Buddhist
              20      8   Hindu
             357      9   Atheist
             303     10   Agnostic
             722     11   Nothing in particular
             268     12   Something else


================================================================================
Name:           pew_churatd
Description:    Church attendance (Pew version)

           Count   Code   Label
           -----   ----   -----
```

```
       321    1   More than once a week
       698    2   Once a week
       267    3   Once or twice a month
       431    4   A few times a year
       846    5   Seldom
      1378    6   Never
        59    7   Don't know
```

================================================================================
Name:          pew_bornagain
Description:   Born Again (Pew version)

```
      Count  Code  Label
      -----  ----  -----
       1124    1   Yes
       2876    2   No
```

================================================================================
Name:          pew_religimp
Description:   Importance of religion (Pew version)

```
      Count  Code  Label
      -----  ----  -----
       1565    1   Very important
        816    2   Somewhat important
        556    3   Not too important
       1063    4   Not at all important
```

================================================================================
Name:          pew_prayer
Description:   Frequency of Prayer (Pew version)

```
      Count  Code  Label
      -----  ----  -----
       1157    1   Several times a day
        522    2   Once a day
        438    3   A few times a week
         97    4   Once a week
        220    5   A few times a month
        506    6   Seldom
        948    7   Never
        112    8   Don't know
```

                          Date format variables
================================================================================
Name:          starttime
Description:   Questionnaire Start Time
         DateTime variable – no categories


================================================================================
Name:          endtime

Description:   Questionnaire End Time
              DateTime variable — no categories



\end{document}